\documentclass[pra,twocolumn,floats,superscriptaddress,longbibliography]{revtex4-1}

\usepackage[colorlinks=true,urlcolor=blue,citecolor=blue,linkcolor=blue]{hyperref}

\usepackage{amsmath, amsthm, amssymb}
\usepackage{graphicx}
\usepackage{natbib}
\usepackage{color}
\definecolor{Red}{rgb}{0.95,0.05,0.05}

\usepackage{multirow}
\usepackage{verbatim}

\usepackage[sort&compress]{natbib}
\setcitestyle{numbers,square}
\usepackage{hyperref}
\hypersetup{
        colorlinks=true,
}

\newcommand{\braket}[2]{\langle #1 | #2 \rangle}

\newcommand{\Liquid}{LIQ$Ui\ket{}$\ }

\newtheorem{theorem}{Theorem}
\newtheorem{lemma}{Lemma}

\newtheorem{corollary}{Corollary}

\newcommand{\structOne}{1}
\newcommand{\structTwo}{2}

\usepackage{graphicx}
\usepackage{anyfontsize}
\usepackage{tikz}

\usetikzlibrary{
arrows,shapes.misc,shapes.arrows,matrix,
positioning,scopes,decorations.pathmorphing,shadows,
shapes.geometric,decorations.pathreplacing,shapes.gates.logic.US,calc}

\tikzset{
fontStyle/.style={font=\fontsize{17}{17}},
matStyle/.style={column sep=.4mm,row sep=2mm,font=\fontsize{17}{17},execute at empty cell=\node{\phantom{.}};},
measStyle/.style={and gate US,text centered,minimum size=8mm,thick,draw=black,
    top color=white,bottom color=green!10!white},
meterStyle/.style={rectangle,text centered,minimum size=8mm,thick,draw=black,
    top color=white,bottom color=green!10!white},
lblStyle/.style={rectangle,minimum size=8mm,thick,draw=white,
	top color=white,bottom color=white},
rawStyle/.style={},
prepStyle/.style={rectangle,minimum size=8mm,thick,draw=white,text=red!60!black,
	top color=white,bottom color=white},
ctrlStyle/.style={circle,text centered,inner sep=-1pt,minimum size=1mm,thick,draw=black,
	top color=black,bottom color=black},
ctrloStyle/.style={circle,text centered,inner sep=-1pt,minimum size=1mm,thick,draw=black,
	top color=white,bottom color=white},
targStyle/.style={circle,text centered,inner sep=-1pt,minimum size=3mm,thick,draw=black,
	top color=white,bottom color=white,append after command={
        [shorten >=\pgflinewidth, shorten <=\pgflinewidth, thick,]
        (\tikzlastnode.north) edge[thick] (\tikzlastnode.south)
        (\tikzlastnode.east) edge[thick] (\tikzlastnode.west)
        }},
qswapStyle/.style={circle,text centered,inner sep=0pt,minimum size=3mm,thin,draw=white,opacity=0,
	top color=white,bottom color=white,append after command={
        [shorten >=\pgflinewidth, shorten <=\pgflinewidth, thick,color=black]
        (\tikzlastnode.north east) edge[thick] (\tikzlastnode.south west)
        (\tikzlastnode.north west) edge[thick] (\tikzlastnode.south east)
        }},
gateStyle/.style={rectangle,text centered,anchor=center,yshift=-1mm,minimum size=6mm,inner xsep=2mm,inner ysep=3mm,thick,draw=black,
	top color=white,bottom color=blue!10!white},
mgStyle/.style={rectangle,minimum size=8mm,transparent},
mgBoxStyle/.style={rectangle,minimum size=8mm,thick,draw=black,
	top color=white,bottom color=blue!10!white},
qwStyle/.style={thick,draw=black,font=\ttfamily},
qwxStyle/.style={thick,draw=black,font=\ttfamily},
>=latex,thick,
/pgf/every decoration/.style={/tikz/sharp corners},
pointStyle/.style={coordinate},
>=stealth',thick,draw=black,tip/.style={->,shorten >=0.007pt},
every join/.style={rounded corners},
hv path/.style={to path={-| (\tikztotarget)}},
vh path/.style={to path={|- (\tikztotarget)}},
text height=1.5ex,text depth=.25ex
}

\pgfdeclarelayer{background}		
\pgfdeclarelayer{foreground}		
\pgfsetlayers{background,main,foreground}

\newcommand{\gate}[2][] {\node (#1) [gateStyle] {\phantom{$#2$}} node[fontStyle,yshift=-5pt] {$#2$};}
\newcommand{\lbl}[2][]{\node (#1) [lblStyle] {$#2$};}
\newcommand{\raw}[2][]{\node (#1) [rawStyle] {$#2$};}
\newcommand{\prep}[2][]{\node (#1) [prepStyle] {$#2$};}
\newcommand{\mg}[2][]{\node (#1) [mgStyle] {\phantom{$#2$}};}
\newcommand{\point}[1][]{ \node (#1) [pointStyle] {};}
\newcommand{\targ}[1][]{ \node (#1) [targStyle] {};}
\newcommand{\qswap}[1][]{ \node (#1) [qswapStyle] {};}
\newcommand{\ctrl}[1][]{ \node (#1) [ctrlStyle] {};}
\newcommand{\ctrlo}[1][]{ \node (#1) [ctrloStyle] {};}
\newcommand{\meas}[2][]{\node (#1) [measStyle] {$#2$};}

\newcommand{\meter}[1][] {
	\node (#1) [meterStyle] {\phantom{$X$}};
	\draw[thick] ++(.25,-.2) arc (0:180:.25 and .1);
	\draw[thick] (0,-.14) -- +(.1,.2)
	;}

\newcommand{\multigate}[4][]{
    \draw[gateStyle,name=#1] (#2.north west) rectangle (#3.south east);
    \draw[draw=black] ($0.5*(#2)+0.5*(#3)$) node[fontStyle] {$#4$};
}

\newcommand{\qw}[2]{\draw[qwStyle] (#1) -- (#2);}
\newcommand{\cw}[2]{\draw[qwStyle, double] (#1) -- (#2);}

\newcommand{\qwx}[2]{\draw[qwxStyle] (#1) -- (#2);}
\newcommand{\cwx}[2]{\draw[qwxStyle, double] (#1) -| (#2);}
\newcommand{\dwx}[2]{\draw[qwxStyle, dotted] (#1) -- (#2);}
\newcommand{\fwx}[2]{\draw[qwxStyle] (#1) -- (#2) node[midway,left] {$\mathcal{F}$};}

 \newcommand{\qwxB}[2]{\draw[qwxStyle] (#1)  to [bend left=4]  (#2);}
 \newcommand{\dwxB}[2]{\draw[qwxStyle,dotted] (#1)  to [bend left=4]  (#2);}
 \newcommand{\cwxB}[2]{\draw[qwxStyle,double] (#1)  to [bend left=4]  (#2);}

\newcommand{\bra}[1]{{\left\langle{#1}\right\vert}}
\newcommand{\ket}[1]{{\left\vert{#1}\right\rangle}}

\begin{document}

\title{Elucidating Reaction Mechanisms on Quantum Computers}

\author{Markus Reiher}
\affiliation{Laboratorium f\"ur Physikalische Chemie, ETH Zurich, Valdimir-Prelog-Weg 2, 8093 Zurich, Switzerland}

\author{Nathan Wiebe}
\author{Krysta M.~Svore}
\author{Dave Wecker}
\affiliation{Quantum Architectures and Computation Group, Microsoft Research, Redmond, WA 98052, USA}

\author{Matthias Troyer}
\affiliation{Theoretische Physik and Station Q Zurich, ETH Zurich, 8093 Zurich, Switzerland}
\affiliation{Quantum Architectures and Computation Group, Microsoft Research, Redmond, WA 98052, USA}
\affiliation{Station Q, Microsoft Research, Santa Barbara, CA 93106-6105, USA}

\begin{abstract}
 We show how a quantum computer can be employed to elucidate reaction mechanisms in complex chemical systems, using the open problem of biological nitrogen fixation in nitrogenase as an example. We discuss how quantum computers can augment classical-computer simulations for such problems, to significantly increase their accuracy and enable hitherto intractable simulations. Detailed resource estimates show that, even when taking into account the substantial overhead of quantum error correction, and the need to compile into discrete gate sets, the necessary computations can be performed in reasonable time on small quantum computers. This demonstrates that quantum computers will realistically be able to tackle important problems in chemistry that are both scientifically and economically significant.
\end{abstract}

\maketitle

Chemical reaction mechanisms are networks of molecular structures representing short- or long-lived intermediates connected by transition structures. 
At its core, this surface is determined by the nuclear coordinate dependent electronic energy, i.e., by the Born--Oppenheimer surface, which can be supplemented by nuclear-motion corrections. The relative energies of all stable structures on this surface determine the relative thermodynamical stability. Differences of the energies of these local minima to those of the connecting transition strucures determine the rates of interconversion, i.e., the chemical kinetics of the process. Reaction rates depend on rate constants that are evaluated from these latter energy differences entering the argument of exponential functions. Hence, very accurate energies are required for the reliable evaluation of the rate constants.

The detailed understanding and prediction of complex reaction mechanisms such as transition-metal catalyzed chemical transformations therefore requires highly accurate electronic structure methods. However, the electron correlation problem remains, despite decades of progress \cite{scuseriafrenking}, one of the most vexing problems in quantum chemistry.  While approximate approaches, such as density functional theory \cite{phys_chem_chem_phys_2009_11_10757},
are very popular and have successfully been applied to all kinds of correlated electronic structures, their accuracy is too low for truly reliable quantitative predictions (see, e.g., Refs.\ \cite{angelawilson,wccr10}).  On classical computers, molecules with much less than a hundred strongly correlated electrons are already out of reach for systematically improvable ab initio methods that could achieve the required accuracy.  

The apparent intractability of accurate simulations for such quantum systems led Richard Feynmann to propose quantum computers (see Appendix~\ref{sec:QC} for a brief review of quantum computing). The promise of exponential speedups for quantum simulation on quantum computers was first investigated by Lloyd ~\cite{Llo96} and Zalka \cite{Zal98} and was directly applied to quantum chemistry by Lidar, Aspuru--Guzik and others~\cite{LW99,WKA+08,LWG+10,KWP+11,WBA11}.  Quantum chemistry simulation has remained an active area within quantum algorithm development, with ever more sophisticated methods being employed to reduce the costs of quantum chemistry simulation \cite{JWM+12,PMS+14,WBC+14,MBL+14,HWB15,PHW+15,WHT15,BMW+15,BWM+15}.

The promise of exponential speedups for the electronic structure problem has led many to suspect that quantum computers will one day revolutionize chemistry and materials science.  However, a number of important questions remain.  Not the least of these is the question of how exactly to use a quantum computer to solve an important problem in chemistry.  The inability to point to a clear use case complete with resource and cost estimates is a major drawback; after all, even an exponential speedup may not lead to a useful algorithm if a typical, practical application requires an amount of time and memory so large that it would still be beyond the reach of even a quantum computer.

Here, we demonstrate for an important prototypical chemical system how a quantum computer would be used in practice to address an open problem and to estimate how large and how fast a quantum computer would have to be to perform such calculations within a reasonable amount of time.  This sets a target for the type and size of quantum device that we would like to emerge from existing research and further gives confidence that quantum simulation will be able to provide answers to problems that are both scientifically and economically impactful.

\begin{figure*}
\begin{center}
\includegraphics[width=\textwidth]{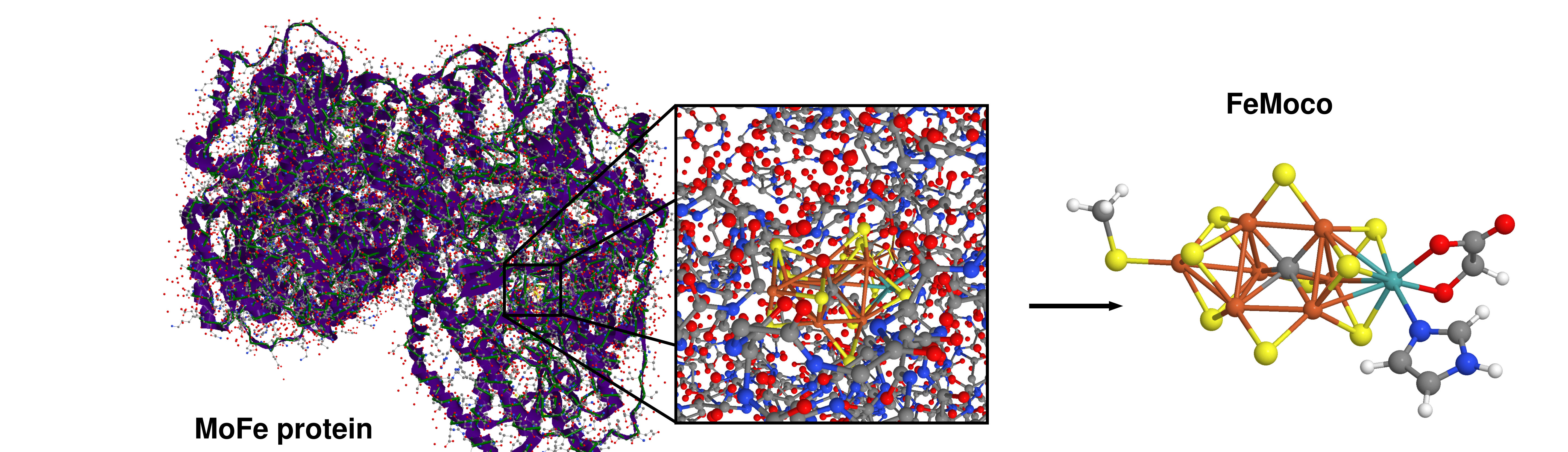}
\end{center}
\caption{X-ray crystal structure 4WES \cite{4wes} of the nitrogenase MoFe protein from {\it Clostridium pasteurianum} taken
from the protein data base (left; the backbone is colored in green and hydrogen atoms are not shown), the close protein environment of the FeMoco (center), and the structural model of FeMoco considered in this work (right; C in gray, O in red, H in white, S in yellow, N in blue, Fe in brown, and Mo in cyan).
\label{nitro}}
\end{figure*}

The chemical process that we consider in this work is that of biological nitrogen fixation by the enzyme nitrogenase \cite{review}. This enzyme accomplishes the remarkable transformation of turning a dinitrogen molecule into two ammonia molecules (and one dihydrogen molecule) under ambient conditions. Whereas the industrial Haber-Bosch catalyst requires high temperatures and pressures and is therefore energy intensive, the active site of Mo-dependent nitrogenase, the iron molybdenum cofactor (FeMoco) \cite{science1,science2}, can split the dinitrogen triple bond at room temperature and standard pressure. Mo-dependent nitrogenase consists of two subunits, the Fe protein, a homodimer, and the MoFe protein, an $\alpha_2\beta_2$ tetramer. Figure \ref{nitro} shows the MoFe protein of nitrogenase on the left and the FeMoco buried in this protein on the right. Despite the importance of this process for fertilizer production that makes nitrogen from air accessible to plants, the mechanism of nitrogen fixation at FeMoco is not known. Experiment has not yet been able to provide sufficient details on the chemical mechanism and theoretical attempts are hampered by intrinsic methodological limitations of traditional quantum chemical methods. Nitrogen fixation also turned out to be a true challenge for synthetic approaches and only three homogeneous catalysts working under ambient conditions have been discovered so far \cite{schrock,nishiba,peters}. All of them decompose under reaction conditions as highlighted by an embarassingly low turnover number of less than a dozen.

\section{Quantum chemical methods for mechanistic studies}

\subsection*{Standard concepts}
At the heart of any chemical process is its mechanism, the elucidation of which requires the identification of all relevant stable intermediates and transition states and the calculation of their properties. The latter allow for experimental verification, whereas energies associated with these structures provide access to thermodynamic stability and kinetic accessibility.  In general, a multitude of charge and spin states need to be explicitly calculated in search for the relevant ones that make the whole chemical process viable. This can lead to thousands of elementary reaction steps \cite{heurex} whose reaction energies must be reliably calculated. 

In the case of nitrogenase, numerous protonated intermediates of dinitrogen-coordinating FeMoco and subsequently reduced intermediates in different charge and spin states are feasible and must be assessed with respect to their relative energy. Especially kinetic modelling poses tight limits on the accuracy of activation energies entering the argument of exponentials in rate expressions.

Although most chemical processes are local and therefore involve only a rather small number of atoms (usually one or two bonds are broken or formed at a time), they can be modulated by environmental effects (e.g., a protein matrix or a solvent). For the consideration of such environment effects, many different embedding methods \cite{kong,jone,warshel,weso,garnet} are available. They incorporate environment terms, ranging from classical electrostatics to full quantum descriptions, into the Hamiltonian to be diagonalized. 

For nitrogenase, an electrostatic quantum-mechanical/molecular-mechanical (QM/MM) model that captures the embedding of FeMoco into the protein pocket of nitrogenase can properly account for the energy modulation of the chemical transformation steps at FeMoco. Accordingly, we consider a structural model for the active site of nitrogenase (Fig.\ \ref{nitro} left) carrying only models of the anchoring groups of the protein, which represents a suitable QM part in such calculations. To study this bare model is no limitation as it does not at all affect our feasibility analysis (because electrostatic QM/MM embedding does not affect the number of orbitals considered for the wave function construction). We carried out (full) molecular structure optimizations with density functional theory (DFT) methods of this FeMoco model in different charge and spin states in order to not base our analysis on a single electronic structure. While our FeMoco model describes the resting state, binding of a small molecule such as dinitrogen, dihydrogen, diazene, or ammonia will not decisively increase or reduce the complexity of its electronic structure.  See Appendix~\ref{sec:Elucidating} for more details.

\begin{figure*}
\begin{center}
\includegraphics[width=1.5\columnwidth]{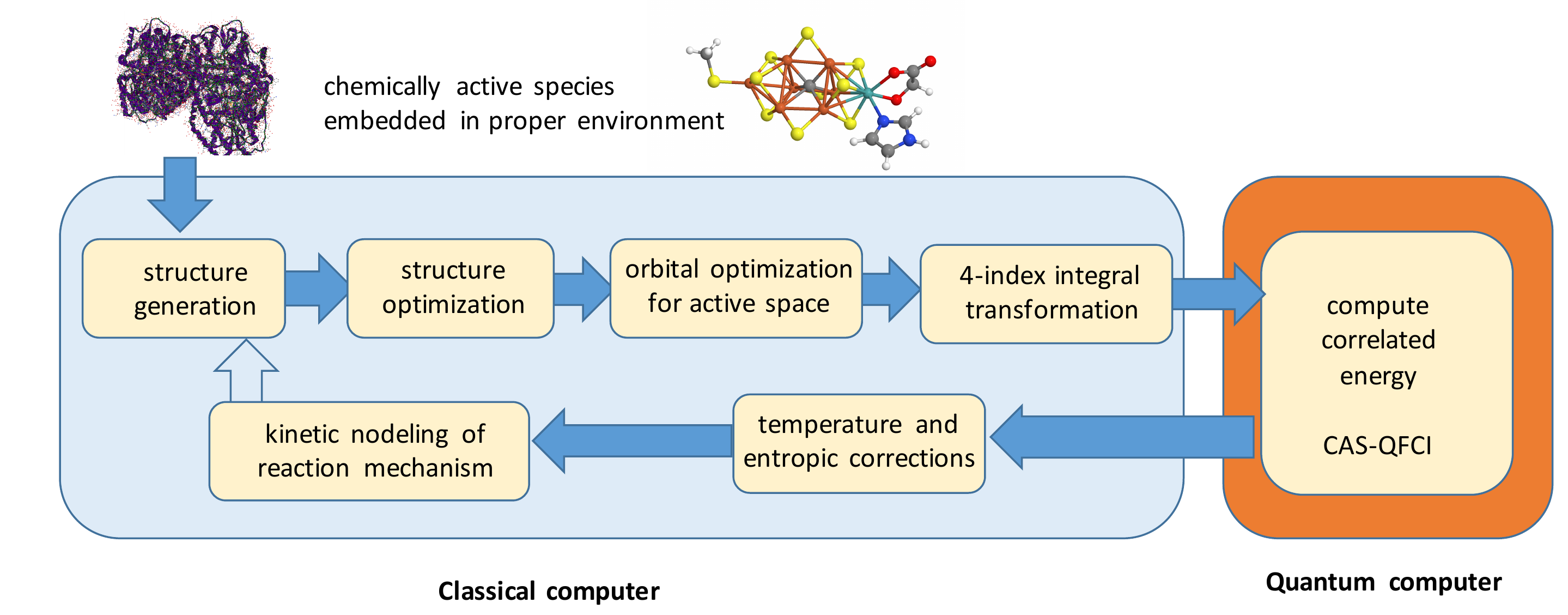}
\end{center}
\caption{
Generic flowchart of a computational reaction-mechanism elucidation with a quantum computer part that delivers
a quantum full configuration interaction (QFCI) energy in a (restricted) complete active orbital space (CAS). Once a structural
model of the active chemical species (here FeMoco, top right) embedded in a suitable environment (the metalloprotein,
top left) is chosen, structures of potential intermediates can be set up and optimized. 
Molecular orbitals are then optimized for a suitably chosen Fock operator. A four-index transformation from the atomic 
orbital to the molecular basis produces all integrals required for the second-quantized Hamiltonian. Once the quantum
computer produces the (ground-state) energy of this Hamiltonian, it can be supplemented by corrections that consider
nuclear motion effects to yield enthalpic and entropic quantities at a given temperature according to standard protocols (e.g., from DFT calculations). 
The temperature-corrected energy differences between stable intermediates and transition structures then enter rate 
expressions for kinetic modeling. For complex chemical mechanisms, this modeling might point to the exploration of 
additional structures.
\label{qcflowchart}}
\end{figure*}

\subsection*{Elucidating reaction mechanisms }

Understanding a complex chemical mechanism requires the exploration of a potential energy landscape that results from the Born--Oppenheimer approximation. This approximation assigns an electronic energy to every molecular structure. The accurate calculation of this energy is the pivotal challenge, here considered by quantum computing. Characteristic molecular structures are optimized to provide local minimum structures indicating stable intermediates and first-order saddle points representing transition structures. 
The electronic energy differences for elementary steps that connect two minima through a transition structure enter expressions for rate constants by virtue of Eyring's absolute rate theory (ART) which is a detailed quantum- and statistical-mechanical interpretation of the phenomenological rate expression by Arrhenius. Although more information on the potential energy surface as well as dynamic and quantum effects may be taken into account, ART is accurate even for large molecular systems such as enzymes \cite{warshel2,ockham}. These rate constants then enter a kinetic description of all elementary steps that ultimately provide a complete picture of the chemical mechanism under consideration.

\subsection*{Exact diagonalization methods in chemistry}
If the frontier orbital region around the Fermi level of a given molecular structure is dense, as is the case in $\pi$-conjugated molecules and open-shell transition metal complexes, then so-called strong {\it static} electron correlation plays a decisive role already in the ground state.  This is even more pronounced for electronically excited states relevant in photophysical and photochemical processes such as light harvesting for clean energy applications.  Such situations require multi-configurational methods of which the complete-active-space self-consistent-field (CASSCF) approach has been established as a well-defined model that also serves as the basis for more advanced approaches \cite{helgaker}.  

CAS-type approaches require the selection of orbitals for the CAS, usually from those around the Fermi energy. This CAS selection is an art that, however, may be automatized \cite{pulay,stein}. 
While CAS-type methods well account for static electron correlation, the remaining dynamic correlation is decisive for quantitative results.
A remaining major drawback of exact-diagonalization schemes therefore is to include the contribution of all neglected virtual orbitals. 

CASSCF is traditionally implemented as an exact diagonalization method, which limits its applicability to 18 electrons in 18 (spatial) orbitals because of the steep scaling of many-electron basis states with the number of electrons and orbitals \cite{molcas8}.  The polynomially scaling density matrix renormalization group (DMRG) algorithm \cite{white} can push this limit to about 100 spatial orbitals. This, however, also comes at the cost of an iterative procedure whose convergence for strongly correlated molecules is, due to the matrix product state representation of the electronic wave function, neither easy to achieve nor is it guaranteed. The groundstate energies output by quantum computer simulations are not affected by these issues, but present other accuracy considerations that we quantify below.

\subsection*{Ways quantum computers will help solve these problems}
As long as molecular structure optimizations cannot be efficiently implemented on a quantum computer, this task can be accomplished with standard DFT approaches. DFT optimized molecular structures are in general reliable, even if the corresponding energies are affected by large uncontrollable errors. The latter problem can then be solved by a quantum computer that implements a multi-configurational (CAS-CI) wave function model to access truly large active orbital spaces. The orbitals for this model do not necessarily need to be optimized as natural orbitals can be taken from an unrestricted Hartree-Fock \cite{pulay2} or small-CAS CASSCF calculation. Clearly, the missing dynamic correlation needs to be implemented. This can be done in a 'perturb-then-digonalize' fashion before the quantum computations start or in a 'diagonalize-then-perturb' fashion, where the quantum computer is used to produce the higher-order reduced density matrices required. The former approach, i.e., built-in dynamic electron correlation, is considerably more advantageous as no wave-function-derived quantities need to be calculated. One option for this approach is, for example, to consider dynamic correlation through DFT that avoids any double counting effects by virtue of range separation as has already been successfully studied for CASSCF and DMRG \cite{jensen,hedegard}. Fig.\ \ref{qcflowchart} presents a flowchart that describes the steps of a quantum-computer assisted chemical mechanism exploration. Moreover, the quantum computer results can be used for the validation and improvement of parametrized approaches such as DFT in order to improve on the latter for the massive pre-screening of structures and energies.

\section{Quantum Simulation of Quantum Chemical Systems}
From among a number of different approaches to determine energies of ground states on a quantum computer~\cite{AL97,PMS+14}, we focus on quantum phase estimation (QPE).  If we take the time evolution of an eigenstate to be 
$e^{-iHt}|n\rangle = e^{-iE_n t} \ket{n}$ 
then the task of QPE is to learn the phase $\phi=E_nt$ in a manner analogous to a Mach--Zehnder interferometer. 
If QPE is applied to a trial state that is a superposition of eigenvectors, it collapses the state to one of the eigenstates and will collapse it to the ground state with high probability if the trial state has a large overlap with the ground state. See Appendix~\ref{sec:prep} for a discussion of state preparation approaches in quantum simulation.

In practice,  $e^{-iHt}$ must be decomposed in QPE into elementary operations.  There are several methods known for achieving this~\cite{BACS07,CW12,berry2015simulating}; here we focus on the Trotter--Suzuki (TS) decompositions and specifically the second-order TS formula, which for the second-quantized Coulomb Hamiltonian
takes the form 
$
e^{-iHt} = \left(\prod_{j=1}^M e^{-iH_jt/2r} \prod_{j=M}^1 e^{-iH_j t/2r}\right)^r + O(t^3/r^2).
$
In the limit $r\rightarrow\infty$ this approximation becomes exact. 
Well-known methods exist for implementing each $e^{-i H_jt/2r}$ arising in a second-quantized formulation using a sequence of elementary gates and single-qubit rotations~\cite{WBA11}.  These rotations dominate the cost of the quantum simulation.

\subsection*{Circuit synthesis and quantum error correction}
While existing estimates of the costs of quantum simulation have taken these rotations to be intrinsic quantum operations, doing so belies the cost of performing the computation fault tolerantly.  To achieve reliable results, fault tolerant implementation of the algorithm is crucial.  Fault tolerance is achieved by encoding a single \emph{logical qubit} in a number of \emph{physical qubits} with a quantum error-correcting code, such as the surface code \cite{fowler2012surface}.  This redundancy  protects the logical qubit against decoherence and other experimental imperfections. 

Quantum error correction cannot directly protect any arbitrary quantum operation, such as arbitrary rotations. Fortunately, however, it can protect a discrete set of gates, from which any continuous quantum operation can be approximated to within arbitrarily small error \cite{NC00}.  

 Approximation takes two steps.  First the exponentials in the Trotter formula are decomposed into single-qubit rotations and so-called Clifford gates. In the surface code, which we consider here, Clifford gates can be implemented fault tolerantly. The single-qubit rotations, however, require approximation by a discrete set of gates consisting of Clifford operations and at least one non-Clifford operation, usually taken to be the $\tt T$ gate, a rotation by $\pi/8$ about the $z$-axis. 
Each $\tt T$ gate requires a procedure called magic state distillation, which consumes a host of noisy quantum states to output a single accurate magic state.  The magic state is then used to teleport a $\tt T$ gate into the computation~\cite{bravyi2005universal}.  The space and time overheads of state distillation render it by far the most costly aspect of quantum error correction, leading to a large multiplicative overhead for each single-qubit rotation. The number of $\tt T$ gates therefore typically dominates the cost when implementing a fault tolerant algorithm.

\section{Resource Estimates}
We now show that the resources required to elucidate reaction mechanisms on quantum computers not only scale polynomially, but also that the calculations are feasible on a relatively small quantum computer in reasonable time. These resource estimates are based on extrapolations from the costs found from simulations of smaller molecules (for loose upper bounds see Appendix~\ref{sec:trotter}) and focus on two protoypical structures of FeMoco that are typical of the complexity of those that naturally would arise when probing the potential energy landscape of the complex to find the reaction rates.  Our work differs from previous studies in that we not only look at an important target for simulation but we also examine it within a basis set that is reasonable match for the target accuracy required (see Appendix~\ref{sec:methodology}).

We first estimate the runtime of the computation assuming a quantum computer can perform a {\em logical} {\tt T} gate every $10$ ns, Clifford circuits require negligible time and that a good trial state for the ground state is available.
We then determine the cost of performing this simulation fault tolerantly using the surface code, such that each {\em physical} gate takes $10$ ns.

\begin{table}[t!]
\begin{tabular}{|c|c|c|c|c|}
\hline
\multicolumn{5}{|c|}{\bf Quantitatively accurate simulation (0.1 mHa)} \\
\hline
Struct. {\bf 1}& {\tt T}-Gates & Clifford Gates& Time &Log. Qubits\\
\hline
\hline
Serial & $1.1\times 10^{15} $ &$1.7\times 10^{15}$ & $130$ days &$111$\\
Nesting & $3.5\times 10^{15}$ & $5.7 \times 10^{15}$ & $15$ days & $135$ \\
PAR & $3.1\times 10^{16} $ & $3.1\times 10^{16}$ & $110$ hours&$1982$\\
\hline
\end{tabular}
\begin{tabular}{|c|c|c|c|c|}
\hline
Struct. {\bf 2} & {\tt T}-Gates & Clifford Gates& Time &Log. Qubits\\
\hline
\hline
Serial & $2.0\times 10^{15} $& $3.1\times 10^{15}$ & $240$ days&$117$\\
Nesting & $6.5\times 10^{15}$ & $1.0 \times 10^{16}$ & $27$ days & $142$ \\
PAR & $6.0\times 10^{16} $&$6.0\times 10^{16}$ & $204$ hours&$2024$\\
\hline
\hline
\multicolumn{5}{|c|}{\bf Qualitatively accurate simulation (1 mHa)} \\
\hline
Struct. {\bf 1} & {\tt T}-Gates & Clifford Gates& Time &Log. Qubits\\
\hline
\hline
Serial & $1.0\times 10^{14} $ &$1.6\times 10^{14}$ & $12$ days &$111$\\
Nesting & $3.3\times 10^{14}$ & $5.6 \times 10^{14}$ & $1.4$ days & $135$ \\
PAR & $3.0\times 10^{15} $ & $3.0\times 10^{15}$ & $11$ hours&$1982$\\
\hline
\hline
Struct. {\bf 2} & {\tt T}-Gates & Clifford Gates& Time &Log. Qubits\\
\hline
\hline
Serial & $1.9\times 10^{14} $& $3.0\times 10^{14}$ & $22$ days&$117$\\
Nesting & $6.0\times 10^{14}$ & $9.9 \times 10^{14}$ & $2.5$ days & $142$ \\
PAR & $5.5\times 10^{15} $&$5.5\times 10^{15}$ & $20$ hours&$2024$\\
\hline
\end{tabular}

\caption{Simulation time estimates.  Listed are the number of logical qubits and gate operations and an estimate of the runtime required to obtain energies within $0.1$ mHa or $1$ mHa for two different structures of
FeMoco on a quantum computer and the number of logical qubits needed for the simulations (width). 
Structure 1 is for spin state $S=0$ and charge $+3$ elementary charges with 54 electrons in 54 spatial orbitals. Structure 2 is for spin state $S=1/2$ and charge $0$ with 65 electrons in 57 spatial orbitals (see Supporting Material for further details).
 These runtimes and gate counts are likely to exceed the actual requirements.}
\label{tab:costs1}
\end{table}

For chemical significance, we aim to compute the energies with a total error of at most $0.1$ mHartree, adding up all sources of systematic and statistical error (see Appendix~\ref{sec:costest}). These errors include the error in phase estimation, the error in the Trotter--Suzuki deocomposition and errors from decomposing the Trotter--Suzuki approximation into $\tt H$ and $\tt T$ gates.  We then optimize over these three costs to find the most efficient way to divide the error budget up over these three sources. We also consider a larger error of $1$ mHartree that would still put our approach into the accuracy range of standard state-of-the-art quantum chemical methods for simple (mono-nuclear) transition metal complexes.

Three concrete implementations of the quantum algorithm are considered here.
In the first approach, called Serial, the single-qubit rotations are constrained to occur serially.  
In a second approach, called Nesting, Hamiltonian terms that affect disjoint sets of spin orbitals are executed in parallel, which substantially reduces the runtime of the calculations.  
In the third approach, programmable ancilla rotations (PAR), rotations are cached in qubits, which are then teleported into the circuit as needed~\cite{JWM+12}.  
For each approach, the overall cost is found by decomposing the rotations into Clifford and $\tt T$ gates using~\cite{bocharov2015efficient} for the serial case and~\cite{selinger2015efficient} in the other two cases.  

The costs at a logical level, shown in Table~\ref{tab:costs1} for simulation of two different structures of FeMoco in different charge and spin states (see Appendix~\ref{sec:methodology} for details) are within reach of a small scale quantum computer.  
If all gates are executed in series then we estimate that the simulation will complete in under a year and use a small number of logical qubits.  PAR can reduce the time required to several days at the price of requiring nearly $20$ times as many logical qubits.  Nesting gives a reasonable tradeoff between the two and may be the preferred approach.  Finally, if we focus on learning qualitatively accurate reaction rates then the costs drop by roughly a factor of $10$ in all examples.
Further details can be found in the appendix.

\begin{figure}[t!]
\includegraphics[width=\linewidth]{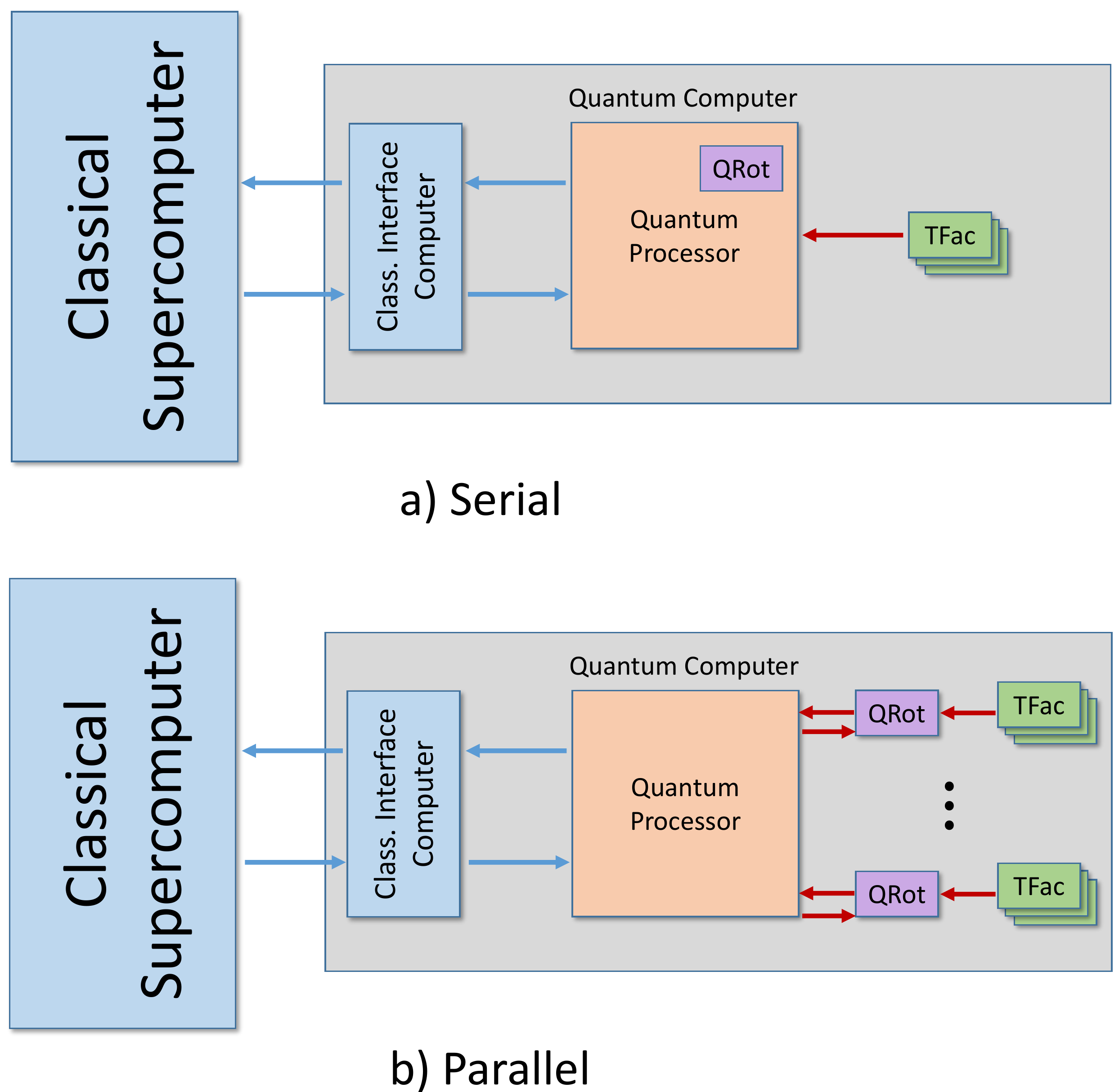}
\caption{  We show the architecture of a hybrid classical/quantum computer for quantum chemistry type calculations. The quantum computer acts as an accelerator to the classical supercomputer. It consists of a classical control frontend, a main quantum processor and a number of auxiliary processing units. The devices labeled QRot build single qubit rotations using $\pi/8$ rotations created in  the $\tt T$ gate factories labeled Tfac.  Red arrows denote quantum communication and blue represent classical communication.}
\label{fig:arch}
\end{figure}

\subsection*{Resource requirements with quantum error correction}
We next add the overheads required to perform the simulation fault tolerantly and summarize the costs incurred by performing the simulation of structure 1 using the surface code in Table~\ref{tab:ftcosts}.   The underlying resource calculations follow~\cite{fowler2012surface}.  These costs naturally depend on the physical error rates assumed in the hardware, and we consider three cases that correspond to (a) a near-term error rate of $10^{-3}$, (b) an error rate of $10^{-6}$ that might one day be feasible in superconducting qubits, but is expected early on for topological quantum bits \cite{topo} and (c) an error rate of  $10^{-9}$ that may be achievable for topological quantum computers.

Rather than focusing solely on the number of physical qubits 
we 
consider a fine-grained modular architecture for a quantum computer, sketched in Fig. \ref{fig:arch}, consisting of a classical supercomputer interfacing with a quantum computer that consists of a main quantum processor and dedicated separate $\tt T$ factories and rotation factories to produce $\tt T$ gates and to synthesize rotations, respectively. Only the main quantum processor is a general purpose device, while the other factories are special purpose hardware that is expected to be easier to construct. 

We first observe in Table~\ref{tab:ftcosts} that the number of logical qubits in the {\em main quantum processor} is only of the order of a hundred, which translates into tens of thousands to millions of physical qubits -- which is challenging but not out of reach for early devices. Most of the qubits are used in the $\tt T$ factories, which each need even fewer physical qubits than the main quantum processor. Even the number of $\tt T$ factories needed to perform the serial calculation is reasonable, with only $202$ such factories required in this case for an error rate of $10^{-3}$, and fewer for better qubits.  After factoring in the number of qubits needed per $\tt T$ factory, we arrive at the conclusion that only $10^5$ to $10^6$ physical qubits are needed for the $10^{-6}$ and $10^{-9}$ error rates.  

If we consider the parallelizing with the PAR approach then these costs become much more daunting.  The number of $\tt T$ factories required increases by a factor of roughly $1000$ although the number of qubits required in each factory remains comparable.  This explosion is due to the fact that PAR  perform about one thousand times as many rotations as the serial code. The costs required by the nesting approach, on the other hand, are much more modest and lead to only an order of magnitude increase in the number of qubits with a comparable reduction in runtimes.

At low error rates, the total number of qubits required  in both the serial and nested cases are of the order of a million, which could even be envisioned as a single quantum processor.  The case of $10^{-3}$ errors or the PAR approach require hundreds of millions to nearly a trillion physical qubits, which seems more daunting. However, considering  a modular design where each factory requires  about one million physical qubits this is nevertheless certainly within the realm of possibility for future generations of quantum devices and the potential economic benefits of solving chemical catalysis problems of industrial importance may offset the cost.

\begin{widetext}
\begin{table*}[t!]
\begin{center}
\begin{tabular}{||l||c|c|c||c|c|c||c|c|c||}
\hline
\hline
& \multicolumn{3}{c||}{\bf Serial rotations} & \multicolumn{3}{c||}{\bf PAR rotations}& \multicolumn{3}{c||}{\bf Nested rotations} \\
\hline 
{\bf Error Rate} & $\mathbf{10^{-3}}$ & $\mathbf{10^{-6}}$ & $\mathbf{10^{-9}}$ & $\mathbf{10^{-3}}$ & $\mathbf{10^{-6}}$ & $\mathbf{10^{-9}}$& $\mathbf{10^{-3}}$ & $\mathbf{10^{-6}}$ & $\mathbf{10^{-9}}$ \\
\hline
Required code distance & 35,17 & 9 & 5 &37,19 & 9,5 & 5 &37,17 &9 &5 \\
\hline
\hline
&\multicolumn{9}{c||}{\bf Quantum processor} \\
\hline
Logical qubits & \multicolumn{3}{c||}{111}  & \multicolumn{3}{c||}{110} & \multicolumn{3}{c||}{109} \\
\hline
Physical qubits per logical qubit & 15313&1013 &313 &17113 &1013 &313 &17113 &1013 &313 \\
\hline
Total physical qubits for processor & $1.7\times 10^6$ &$1.1\times 10^5$  &$3.5\times 10^{4}$  & $1.9\times 10^6$ & $1.1\times 10^5$ & $3.4\times 10^4$ &$1.9\times 10^6$ &$1.1\times 10^5$ & $3.4\times 10^4$  \\
\hline
\hline
&\multicolumn{9}{c||}{\bf Discrete Rotation factories} \\
\hline
Number &  \multicolumn{3}{c||}{0}   & \multicolumn{3}{c||}{1872}  & \multicolumn{3}{c||}{26}   \\
\hline
Physical qubits per factory &-- &-- &-- & 17113 & 1013 &313 &17113 &1013 &313\\
\hline
Total physical qubits for rotations & -- & -- & -- & $3.2\times 10^7$ &$1.9\times 10^6$ & $5.9 \times 10^5$ & $4.5\times 10^5$ & $2.6\times 10^4$ & $8.1\times 10^3$\\
\hline
\hline
&\multicolumn{9}{c||}{\bf $\mathbf{T}$ factories} \\
\hline
Number & 202 & 68 & 38 & 166462 &  41110 & 29659 & 5845 &1842 &1029\\
\hline
Physical qubits per factory & $8.7\times 10^5$ & $1.7\times 10^4$ & $5.0\times 10^3$  & $1.1\times 10^6$ & $7.5\times 10^4$ & $5.0\times 10^3$ & $8.7\times 10^5$ & $1.7\times 10^4$ & $5.0\times 10^3$ \\
\hline
Total physical qubits for T factories & $1.8\times 10^8$ & $1.1\times 10^6$ & $1.9\times 10^5$ & $1.8\times 10^{11}$ & $3.1 \times 10^9$ & $1.5\times 10^8$ & $5.1\times 10^9$ &$3.0\times 10^7$ & $5.2\times 10^6$ \\
\hline
\hline
Total physical qubits & $1.8 \times 10^{8}$	& $1.2 \times 10^6$ &$2.3\times 10^5$ & $1.8 \times 10^{11}$ & $3.1 \times 10^{9}$ &	$1.5 \times 10^8$ & $5.1\times 10^9$	&$3.0\times 10^7$&	$5.2 \times 10^6$  \\
\hline
\hline
\end{tabular}
\end{center}
\caption{  This table gives the resource requirements including error correction for simulations of nitrogenase's FeMoco in structure 1 within the times quoted in Table~\ref{tab:costs1} using physical gates operating at $100$ MHz and taking the error target to be $0.1$ mHa. While the total number of qubits seems large, the quantum computer is composed of many small and weakly coupled devices that can be mass produced. \label{tab:ftcosts}}
\end{table*}
\end{widetext}

\section{Discussion}
While at present a quantitative understanding of chemical processes involving complex open-shell species such as FeMoco in biological nitrogen fixation remains beyond the  capability of classical-computer simulations, our work shows that quantum computers used as accelerators to classical computers could be used to elucidate this mechanism using a manageable amount of memory and time. In this context a quantum computer would be used to obtain, validate, or correct the energies of intermediates and transition states and thus give accurate activation energies for various transitions.

We  note that the required quantum computing resources are comparable to that needed for Shor's factoring algorithm for interesting  $4096$ bit numbers; both in terms of number of gates and also physical qubits~\cite{fowler2012surface}.  The complexity of these simulations is thus typical of that required for other targets for quantum computing.

We also observe that the resources required depend sensitively on gate error rates.  Therefore robust qubits, such as topologically protected qubits~\cite{topo}, may be invaluable for using a quantum computer to elucidate reaction mechanisms for important chemical processes such as the mode of action of nitrogenase.

Chemical reactions that involve strongly correlated species that are hard to describe by traditional multi-configuration approaches are not just limited to nitrogen fixation: they are ubiquitous. They range from small molecule activating catalysts for C-H bond breaking, to those for  hydrogen and oxygen production, carbon dioxide fixation and transformation to industrially useful compounds, and to photochemical processes.  Given the economic and societal impact of chemical processes ranging from fertilizer production to polymerization catalysis and clean energy processes, the importance of a versatile, reliable, and fast quantum chemical approach powered by quantum computing can hardly be overemphasized. Parallelising the quantum computation of the energy landscape will be crucial to providing answers within a timeframe of several days instead of several years.  Quantum computers therefore must be designed such that the components can be mass produced and clusters of quantum computers can be built in order to scan over the many structures that need to be examined to identify and estimate all important reaction rates accurately~\cite{heurex}.

\section{Acknowledgements}

We acknowledge support by the Swiss National Science Foundation, the Swiss National Competence Center in Research NCCR QSIT, and hospitality of the Aspen Center for Physics, supported by NSF grant  \# PHY-1066293.

\appendix

\section{Introduction to Quantum Computing}\label{sec:QC}
Here we provide a brief review of quantum computing for those that are not experts in quantum computing.  The aim of this review is to introduce the basic concepts and notations
that are used in quantum computing as well as introduce quantum algorithms for simulating chemistry and also quantum error correction.  For a more detailed exposition see~\cite{NC00}.  These concepts will be used in section~\ref{sec:PE} where we discuss quantum processes such as phase estimation.

\subsection{Qubits and Quantum Gates}
In quantum computation, quantum information is stored in a quantum bit, or \textit{qubit}.
Whereas a classical bit has a state value $s\in \{0,1\}$, a qubit state $\ket{\psi}$ is a \textit{linear superposition} of states:
\begin{equation}
\ket{\psi} = \alpha\ket{0} + \beta\ket{1} = \left[
    \begin{smallmatrix}
    \alpha \\
    \beta \end{smallmatrix} \right] \label{eqn:qubit},
\end{equation}
where the $\{0,1\}$ basis state vectors are represented in Dirac notation (called \textit{ket} vectors) as
$\ket{0} = \begin{bmatrix}\setlength{\arraycolsep}{3pt} 1 & 0 \end{bmatrix}^T$, and
$\ket{1} = \begin{bmatrix} 0 & 1 \end{bmatrix}^T$, respectively.  The {\it amplitudes} $\alpha$ and $\beta$ are complex numbers that satisfy the normalization condition: $|\alpha|^2 + |\beta|^2 = 1$.
Upon {\it measurement} of the quantum state $\ket{\psi}$, either state $\ket{0}$ or $\ket{1}$ is observed with probability $|\alpha|^2$ or $|\beta|^2$, respectively.

Dirac notation is used largely because it contains an implicit tensor product structure that makes expressing qubit  states much easier.  For example, that the four-qubit state $\ket{0000}$ is equivalent to writing the tensor product of the four states: 
$\ket{0}\otimes\ket{0}\otimes\ket{0}\otimes\ket{0} = \ket{0}^{\otimes 4} = \left[\begin{smallmatrix} 1 & 0 & 0 & 0 & 0 & 0 & 0 & 0& 0 & 0 & 0 & 0& 0 & 0 & 0 & 0 \end{smallmatrix}\right]^T$.

A general $n$-qubit quantum state lives in a $2^n$-dimensional Hilbert space and is represented by a $2^n \times 1$-dimensional state vector whose entries represent the amplitudes of the basis states.
A superposition over $2^n$ states is given by:
\begin{equation}
\ket{\psi} = \sum_{i=0}^{2^n - 1} \alpha_i \ket{i}\mbox{, such that }\sum_i |\alpha_i|^2 = 1 \label{eqn:ket},
\end{equation}
where $\alpha_i$ are complex amplitudes and $i$ is the binary representation of integer $i$.
The ability to represent a superposition over exponentially many states with only a linear number of qubits is one of the essential ingredients of a quantum algorithm --- an innate massive parallelism.

In a quantum computation, unitary transformations are used to transform quantum states into other quantum states.  
In particular, the quantum state $\ket{\psi_1}$ of the system at time $t_1$ is related to the quantum state $\ket{\psi_2}$ at time $t_2$ by a unitary  $\ket{\psi_2} = U\ket{\psi_1}$.

In general we cannot expect that a quantum computer can implement every unitary transformation on $n$ qubits exactly because there are an infinite number of such transformations.  Instead, gate model quantum computers use a discrete set of quantum gates (which can be represented as $2^n\times 2^n$ unitary matrices) to approximate these continuous transformations.  Any such set of gates is known as universal if any unitary transformation can be expressed, within arbitrarily small error, as a sequence of such gates.

The gate set that we consider in this work includes the following gates:
\begin{itemize}
\item the {\tt X} gate which is the classical NOT gate that maps
$\ket{0}\rightarrow \ket{1}$ and
$\ket{1}\rightarrow \ket{0}$.
\item the {\it Hadamard} gate {\tt H} maps
$\ket{0} \rightarrow \frac{1}{\sqrt{2}}\left( \ket{0} + \ket{1}\right)$ and
$\ket{1} \rightarrow \frac{1}{\sqrt{2}}\left( \ket{0} - \ket{1}\right)$.
\item the {\tt Z} gate maps $\ket{1}\rightarrow -\ket{1}$ and ${\tt T}$ is the fourth--root of $Z$,
and can be used interconvert $Z$ and $X$ gates via ${\tt H} {\tt Z} {\tt H} = {\tt X}$.
\item the {\tt Y} gate maps $\ket{1}\rightarrow -i\ket{0}$ and $\ket{0} \rightarrow i\ket{1}$.
\item the identity gate is represented by \openone.
\item the two-qubit {\it controlled-NOT} gate, {\tt CNOT}, maps $\ket{x,y}\rightarrow\ket{x, x\oplus y}$.
The corresponding unitary matrices are:
 \[
\mbox{H} =
  \left[\begin{smallmatrix}1&1\\1&\textrm{-}1\end{smallmatrix}\right],
  ~ \mbox{\tt X} =
  \left[\begin{smallmatrix}0&1\\1&0\end{smallmatrix}\right], 
~ \mbox{\tt Y} =
  \left[\begin{smallmatrix}0&i\\-i&0\end{smallmatrix}\right], 
~ \mbox{\tt Z}  =
  \left[\begin{smallmatrix}1&0\\0&\textrm{-}1\end{smallmatrix}\right],
\]
\[
\mbox{\tt T}=\left[\begin{smallmatrix}1&0\\ 0&e^{i\pi/4} \end{smallmatrix}\right], \openone = \left[\begin{smallmatrix}1&0\\0&1\end{smallmatrix}\right],
  \mbox{\tt CNOT} = \left[\begin{smallmatrix}1 & 0 & 0 & 0 \\ 0 & 1 & 0 & 0 \\ 0 & 0 & 0 & 1 \\ 0 & 0 & 1 & 0\end{smallmatrix}\right].
\]
\end{itemize}

Measurement is an exception to this rule;
it collapses the quantum state to the observed value, thereby erasing any information about the amplitudes $\alpha$ and $\beta$.  This means that
extracting information from a quantum system irreversibly damages a state.  Consequently, although the exponential parallelism that quantum computers
naturally possess is mitigated by the fact that most quantum computations will have to be repeated many times to extract the necessary information about the output
of the quantum algorithm.

\subsection{Quantum Circuit Synthesis}
In order to understand how the cost estimates of our algorithms are found it is important to understand how arbitrary unitaries can be converted into discrete gate sequences.  For our purposes, we consider the Clifford + ${\tt T}$ gate library discussed above, but
compilation into other gate sets is also possible~\cite{bocharov2013efficient,forest2015exact}.  
The simplest case to consider is that of single qubit unitary synthesis where $U\in \mathbb{C}^{2\times 2}$.  In such cases, an Euler angle decomposition exists such that, up to an irrelevant overall phase,
\begin{equation}
U=e^{i \alpha Z} {\tt H} e^{i \beta Z} {\tt H} e^{i \gamma Z}.
\end{equation}
Thus the problem of implementing a single qubit transformation reduces to the problem of performing the rotation $e^{i \theta Z} = cos(\theta)\openone + i \sin(\theta) Z$.  The general case of $U\in \mathbb{C}^{2^n \times 2^n}$ similarly reduces to instances of single qubit synthesis interspersed with entangling gates such as {\tt CNOT}.  Thus implementing single qubit rotations can be seen as an atomic
operation on which compilation process for $U$ is based.

There are a host of methods known for decomposing rotations into Clifford and ${\tt T}$ gates.  Perhaps the earliest such algorithm is the Solovay--Kitaev algorithm~\cite{dawson2005solovay}, which provided an efficient method for performing this decomposition based on Lie--algebraic techniques.  In recent years, much more efficient methods for decomposition have been discovered that are based on number theory.  These methods require near--quartically fewer operations than the Solovay--Kitaev algorithm would need~\cite{dawson2005solovay} and are useful, if not necessary, for reducing the gate counts of the simulation algorithm to a palatable level.

In the absence of ancillae, the cost required to approximate an axial rotation within error $\epsilon$ for the worst possible input angle, $C(\epsilon)$, is known to lie within the interval~\cite{selinger2015efficient}
\begin{equation}
4\log_2(1/\epsilon) -9 \le C(\epsilon)\le 4\log_2(1/\epsilon) +11.\label{eq:synthbd}
\end{equation}
Subsequent work \cite{kliuchnikov2013fast} has provided a method that has an average case complexity roughly $3/4$ of this worst case bound and also has made the compilation process efficient~\cite{ross2014optimal}.  

More recent methods have introduced the use of ancillae to reduce the costs of synthesis~\cite{bocharov2015efficient}.  These methods can substantially reduce the number of ${\tt T}$ gates required to perform
the synthesis.  This approach can reduce the scaling of the average number of ${\tt T}$ gates needed to implement an axial rotation
\begin{equation}
C(\epsilon) \approx 1.15 \log_2(1/\epsilon)+9.2.
\end{equation}
This approach requires one additional ancilla qubit.

Asymptotically better methods exist, such as repeat-until-success (RUS) synthesis with fallback~\cite{bocharov2015efficient}, but these methods do not outperform this method given the range of $\epsilon$ that we require for the simulation to reach chemical accuracy.  Further methods such as PAR rotations~\cite{JWM+12} or gearbox circuits~\cite{wiebe2013floating} can be used to substantially reduce the ${\tt T}$--depth of the simulation circuits at the price of requiring greater parallel width.  While we do not consider the impact of gearbox synthesis here, we will examine the costs and benefits of PAR rotations.

An important issue arises when using such methods for parallel execution.  If several rotations are implemented simultaneously on different qubits and RUS synthesis is used then the dominant contribution to the cost is given by the longest sequence of gates used in each block of parallelized rotations.  In particular, while RUS synthesis reduces the expectation value by a factor of roughly $4$ from the worst case bounds, it is clear that when performing many of them in parallel it is very likely that at least one of them will saturate the bound. For this reason we use upper bounds on deterministic synthesis given by~\eqref{eq:synthbd}, which have a better worst case scaling.

\subsection{Simulating Quantum Chemistry}

Our approach to quantum chemistry simulations closely follows the strategy proposed in Ref.~\cite{LWG+10}.  
We begin by representing the quantum chemistry Hamiltonian in a second quantized form, keeping track of the locations of the electrons by using the occupations of a discrete basis of spin-orbitals.  Since a spin orbital can only contain one electron, such states are very natural to express in a quantum computer.  Each spin orbital is assigned a qubit where the state $\ket{1}$ corresponds to an occupied orbital and $\ket{0}$ an unoccupied orbital.  

These states are often described using creation and annihilation operators which obey $a^\dagger \ket{0} = \ket{1}$, $a^\dagger \ket{1} =0$, $a \ket{1} =\ket{0}$ and $a\ket{0} =0$.  Since these operators correspond to creating electrons, they must respect the symmetries appropriate for Fermions.  The most important property is the anti--commutation property $\{a_i^\dagger, a_j\}=\delta_{i,j}$.  This means that while it may be tempting to identify $a^\dagger = ({\tt X} - i {\tt Y})/2$, such a representation does not satisfy the anti--commutation relation for Fermionic operators.  Instead, these creation and annihilation operators can be converted into Pauli operators using  the Jordan--Wigner or Bravyi--Kitaev~\cite{seeley2012bravyi} transformations.

In order to simulate the dynamics of the system we need to emulate the unitary dynamics that the system undergoes using a sequence of gate operations.  This dynamics is of the form $e^{-iHt}$ where 
\begin{equation}
H=\sum_{pq} h_{pq} a^\dagger_p a_q + \frac{1}{2} \sum_{pqrs} h_{pqrs} a^\dagger_pa^\dagger_q a_ra_s,\label{eq:Hamiltonian}
\end{equation}
Once these fermionic creation and annihilation operators are translated into Pauli operators using, for example, the Jordan--Wigner transformation then the Hamiltonian can be translated into a sequence of operations that individually could be performed on a quantum computer.  Well known circuits exist for performing the exponentials yielded by the Trotter--Suzuki formulas~\cite{NC00,WBA11,WBC+14}.

The most direct method for estimating the ground-state energy is phase estimation (see Section~\ref{sec:PE} for more detail).   
Phase estimation uses quantum interference between controlled executions of $\openone, e^{-iHt_1},e^{-iHt_2},\ldots$, to infer the eigenvalues of $e^{-i Ht}$.  If $t$ is smaller than $\pi/\|H\|$ this also directly yields the eigenvalues of $H$.  Apart from the ability to directly sample from the eigenvalues, a further advantage to phase estimation is that it requires $O(1/\epsilon)$ applications of $e^{-i H dt}$ to learn its eigenphase within error $\epsilon$ with high probability.  This is quadratically better than bounds on the variance that would be seen from optimal classical sampling methods using an unbiased estimator.

It is necessary to apply phase estimation on an initial state that has large overlap with the ground state in order to find the ground-state energy with high probability,.  Specifically, if applied to an initial state $\ket{\psi}=a_g \ket{E_G} + \sqrt{1-|a_g|^2} \ket{E^\perp_G}$ then PE returns an estimate of ground-state energy $E_G$ with probability $|a_g|^2$.  The simplest state to use is the Hartree--Fock state, which only requires applying a sequence of NOT gates to the state $\ket{0}^{n}$, however configuration interaction with sufficiently high excitations
may be required to achieve high overlap for systems that have strong correlations in their ground states.  We do not consider the cost of preparing such a state here, since such a cost of preparing a sufficiently accurate approximation to the ground state of FeMoco is difficult to determine in absentia of large scale quantum computers.  We discuss this issue in more detail in Section~\ref{sec:prep}.

\subsection{Quantum error correction}
Quantum hardware is far less  robust to errors as classical hardware.  Quantum error
correction provides a way to reduce the errors in the device without sacrificing the quantum nature of the system.  Quantum error correcting codes require that the physical error rates, say of qubits, quantum gates, and measurements, are less than a given threshold value.  This threshold value depends on the error correcting code being used and the type of noise of the system. If error rates are below the threshold, then errors in the computation, referred to as the logical gates and logical qubits, can be made arbitrarily small at only a polylogarithmic overhead.

The surface code is currently the most popular code due in part to its relatively high threshold of $1\%$ \cite{fowler2009high}.  Much of this enthusiasm has arisen because of evidence that existing superconducting quantum computers may already have error rates near this threshold \cite{barends2014superconducting}.
This raises the hope that a fault-tolerant, scalable quantum computer may be just over the horizon.

While quantum error correction promises the ability to perform arbitrarily long quantum computations on a noisy device, the resources required to execute such a fault-tolerant computation can be large if the system operates close to the threshold.
The majority of the cost of quantum error correction arises from the need to perform a universal set of quantum gates.  While protecting so-called Clifford gates requires very little overhead in the surface code, in comparison protecting a non-Clifford gate, such as a $T$ or Toffoli gate, requires substantial resources.
While several techniques for producing fault-tolerant non-Clifford gates exist \cite{PR13,bombin2009quantum,bravyi2015doubled,jones2013multilevel}, here we focus on the use of magic state distillation in conjunction with the surface code \cite{fowler2009high}.
Magic state distillation \cite{bravyi2005universal} takes as input a set of noisy resource states and outputs a cleaner resource state.  
For the surface code, we must distll the $T$ state as it cannot be implemented directly in the code.
Here we consider the $15-1$ distillation scheme of Bravyi and Kitaev \cite{bravyi2005universal}, where 15 noisy input states with error rate $p$ produce a single magic resource state with error rate roughly $35p^3$.

Magic state distillation consists only of Clifford operations, which can be implemented easily within the surface code.

\section{Implementing Phase Estimation}
In this section, we discuss how to implement the phase estimation protocol in quantum computers.  This is important to our subsequent estimates of the complexity of simulating nitrogenase's FeMoco because phase estimation constitutes the outer most loop of the quantum simulation and hence is a major driver of the cost of the simulation.
\subsection{Phase Estimation}\label{sec:PE}
Phase estimation is one of the most critical components of the quantum simulation algorithm.  Without phase estimation, the amount of simulation time
needed to estimate the ground-state energy would grow quadratically as $\epsilon^{-2}$ with the precision $\epsilon$ required.  Since $\epsilon$ is on the order of $0.1$mHartree,
the phase must be estimated for one time step of the evolution operator within an error $0.1/r$ mHartree where $r$ is the number of Trotter steps required.  If statistical sampling,
rather than phase estimation, were used to estimate the phase then on the order of $10^{12}$ experiments would be needed to make the variance in the estimate
sufficiently small.  This would be impractical and so phase estimation is crucial for most large scale applications in quantum chemistry.

The standard quantum phase estimation algorithm~\cite{NC00} allows eigenvalues to be learned within error $\epsilon$ with probability at least $1/2$ uses a number of applications of the unitary circuit, $M(\epsilon)$, that is bounded above by
\begin{equation}
M(\epsilon) \le  \frac{16\pi}{\epsilon}.
\end{equation}
This value is far from optimal time scaling of $\pi/\epsilon$, but has the advantage of requiring neither measurements of the quantum system nor a classical computer to infer the most likely eigenvalue.  We will also use this result below because it provides an upper bound on the cost of the optimal phase estimation algorithm.

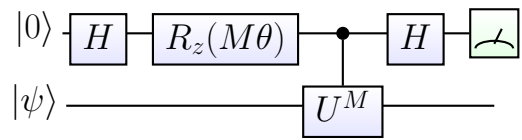
\begin{figure}[t!]
\centering
\begin{tikzpicture}[scale=1.00,every node/.style={scale=0.80}]
\matrix[matStyle] {
  \lbl[0-0]{\ket{0}} 		& \gate[0-1]{H}	   & 			 & \gate[0-2]{R_z(M\theta)}		&  	\ctrl[0-3]		&\gate[0-4]{H} 	&		&\meter[0-8]    \\
  \lbl[1-0]{\ket{\psi}} 	& \point[1-1]		   & \point[1-3]	 &  \point[1-4]				& \gate[1-5]{U^M}		&  	&\point[1-7]	&\point[1-8]   \\
 
   &  &  &  &  &  &  &  &  &  &   \\
};
\begin{pgfonlayer}{background}
\qw{0-0}{0-8} \qw{1-0}{1-8}  
 \qwx{0-3}{1-5} 
\end{pgfonlayer}
\end{tikzpicture}
\caption{  This circuit is used in iterative phase estimation algorithms wherein the eigenvalues are inferred from measurement statistics. \label{fig:PE}}
\end{figure}

An alternative approach is to use iterative phase estimation.  Iterative phase estimation forgoes storing the
phase in a quantum register and instead uses a classical inference algorithm to learn the eigenphase from
measurements of an auxillary probe qubit that is iteratively measured and re-entangled with the system.
Kitaev proposed the first variant of iterative phase estimation.  The circuit used in this process is given in Fig~\ref{fig:PE}.

The ultimate limit that can be achieved, in terms of the number of times the unitary is applied as a function of desired error tolerance, is given by~\cite{wiseman1997adaptive,van2007optimal,berry2009perform}
\begin{equation}
M(\epsilon) \approx \frac{\pi}{\epsilon}.\label{eq:Mopt}
\end{equation}
Furthermore, Gaussian strategies such as RFPE~\cite{wiebe2015efficient} yield Bayesian Cramer--Rao bounds that come close to saturating this: $3.3/\epsilon$.  As these bounds are often saturated for likelihood functions of this form~\cite{wiebe2014hamiltonian} and because the Gaussian assumption causes the user to throw away substantial information from higher moments in the posterior distribution, we expect that~\eqref{eq:Mopt} represents the ultimate limit for phase estimation and that based on previous studies the use of optimized policies~\cite{hentschel2010machine,granade2012robust,bonato2015optimized} will enable performance that comes close to saturating this limit.  As a result we use $M(\epsilon) \approx \pi/(2\epsilon)$ as a surrogate for the expected performance of these optimized approaches, where the factor of $2$ difference follows from an important optimization that holds for the case of quantum simulation algorithms that we discuss in detail in the following section.

\subsection{Controlled rotations}
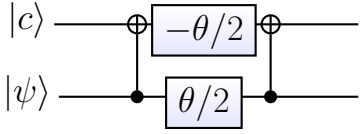
\begin{figure}
\centering
\begin{tikzpicture}[scale=1.00,every node/.style={scale=0.80}]
\matrix[matStyle] {
  \lbl[0-0]{\ket{c}} 		& \point[0-1] &  & 			 & \targ[2-4]		&  	\gate[0-5]{-\theta/2}		&\targ[2-6] 	&		&&&\point[0-8]    \\
  \lbl[1-0]{\ket{\psi}} 	& \point[1-1] &  & \point[1-3]	 &  \ctrl[0-4]		& \gate[1-5]{\theta/2} 			&  \ctrl[0-6] 		&\point[1-7]	&&&\point[1-8]   \\
 
   &  &  &  &  &  &  &  &  &  &   \\
};
\begin{pgfonlayer}{background}
\qw{0-0}{0-8} \qw{1-0}{1-8}  \qwx{0-4}{2-4}
 \qwx{0-6}{2-6} 
\end{pgfonlayer}
\end{tikzpicture}
\caption{  Low depth circuit for controlled ${\tt Z}$--rotations where the $\theta$ gate represents $e^{i\theta Z}$. \label{fig:parallel}}
\end{figure}

While the previous discussion only showed how to perform a ${\tt Z}$--rotation, we need to perform controlled ${\tt Z}$--rotations to perform the phase estimation algorithm.  Fortunately, there are well known methods that can be used to perform such controlled rotations using a pair of rotations and two {\tt CNOT} gates.  A better approach, previously shown in unpublished work by Tsuyoshi Ito in 2012, is given in Figure~\ref{fig:parallel} and the validity of the circuit is proven in the following lemma.
\begin{lemma}
The circuit of~Fig. \ref{fig:parallel} implements the controlled operation $\Lambda(R_Z(2\theta))$ where the top most qubit is the control.
\end{lemma}
\begin{proof}
Assume that $\ket{c} = \ket{0}$ or $\ket{1}$ then the {\tt CNOT} gate performs
\begin{align}
\ket{c}\ket{\psi} \mapsto ({\tt X}^c\otimes \openone) (a\ket{00}+b\ket{11}).
\end{align}
The rotation gates then prepare the state
\begin{align}
(ae^{i(1-(-1)^c)\theta/2}\ket{c}\ket{0}+be^{-i(1-(-1)^c)\theta/2}\ket{c\oplus 1}\ket{1}).
\end{align}
Finally the {\tt CNOT} gate yields the state
\begin{align}
&\ket{c}(ae^{i(1-(-1)^c)\theta/2}\ket{0}+be^{-i(1-(-1)^c)\theta/2}\ket{1})\nonumber\\
&\qquad=\left(\Lambda(R_z(2\theta)) \ket{c}\ket{\psi}\right).
\end{align}
Therefore the circuit functions properly for either $\ket{c}=\ket{0}$ or $\ket{c}=\ket{1}$.  By linearity it is also valid for any input.
\end{proof}

This form of a controlled rotation is well suited for many quantum simulation applications because it allows the controlled evolutions to be replaced with evolutions that control only on the single qubit rotations in the circuits for implementing the individual terms in the Hamiltonian. However, it is not optimal here because we can implement a variant of a controlled rotation gate to in effect double the impact that the eigenphase has on the measurement probabilities.

\begin{figure}[t!]
\centering
\begin{tikzpicture}[scale=1.00,every node/.style={scale=0.80}]
\matrix[matStyle] {
  \lbl[0-0]{\ket{c}} 		& \point[0-1] &  & 			 & \ctrl[2-4]		&  			&\ctrl[2-6] 	&		&&&\point[0-8]    \\
  \lbl[1-0]{\ket{\psi}} 	& \point[1-1] &  & \point[1-3]	 &  \targ[0-4]		& \gate[1-5]{\theta/2} 			&  \targ[0-6] 		&\point[1-7]	&&&\point[1-8]   \\
 
   &  &  &  &  &  &  &  &  &  &   \\
};
\begin{pgfonlayer}{background}
\qw{0-0}{0-8} \qw{1-0}{1-8}  \qwx{0-4}{2-4}
 \qwx{0-6}{2-6} 
\end{pgfonlayer}
\end{tikzpicture}
\caption{  Circuit used to implement analogue of controlled ${\tt Z}$--rotations used in Lemma~\ref{lem:rot}, which is a rotation whose sign is conditioned on the control qubit. \label{fig:parallel2}}
\end{figure}
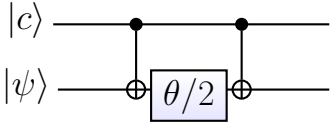

\begin{lemma}\label{lem:rot}
Let $U = \prod_{j=1}^{N_{\exp}} B_j (\openone\otimes R_z(\phi_j)) B_j^\dagger$ where $B_j$ are unitary transformations and assume that it is chosen such that $U^{\dagger} = \prod_{j=1}^{N_{\exp}} B_j (\openone\otimes R_z(-\phi_j)) B_j^\dagger$  then if $U\ket{\phi}=e^{i\phi}$ then there exists a quantum protocol that uses an $M$ rotations and samples from a Bernoulli distribution such that the probability of the protocol outputting $0$ is $$P(0|\phi;\theta,M) = \frac{1+\cos(2M[\phi +\theta])}{2}.$$
\end{lemma}
\begin{proof}
Our proof is constructive and follows directly from the arguments presented informally in~\cite{WHW+15}.  The idea behind the protocol is to replace every controlled rotation used in the circuit  in Fig.~\ref{fig:PE} with the circuit in Fig.~\ref{fig:parallel2}.  Formally we denote this by replacing the controlled operation $\Lambda(U^M)$ with  $W^M$, which is defined such that $W\ket{0}\ket{\psi} = \ket{0}U \ket{\psi}$ and $W\ket{1}\ket{\psi} = \ket{1}U^\dagger\ket{\psi}$.  Since $XR_z(\theta_j)X= R_z(-\theta_j)$, the operation $W$ can be written as
\begin{equation}
W=\prod_{j=1}^{N_{\exp}} B_j \Lambda(X)(\openone\otimes R_z(-\phi_j))\Lambda(X) B_j^\dagger,
\end{equation}
where the controlled--not operation $\Lambda(X)$ uses the control qubit as its control and the qubit that $R_z$ acts on as its target.  This circuit is shown in Fig.~\ref{fig:parallel2}.

Applying $({\tt H}\otimes \openone)W({\tt H} \otimes \openone)$ to the eigenstate $\ket{0}\ket{\phi}$, such that $U\ket{\phi} = e^{i\phi} \ket{\phi}$, yields
\begin{equation}
\frac{1}{2}\left(\ket{0}\ket{\phi} (1+ e^{2i(\theta-\phi) M})+ \ket{1}\ket{\phi}(1- e^{2i(\theta-\phi) M})  \right),
\end{equation}
up to a global phase.
The probability of measuring the ancilla qubit to be zero is $(1+\cos(2M(\phi-\theta)))/2$ as claimed.
\end{proof}

Lemma~\ref{lem:rot} shows that the number of rotations needed to perform phase estimation is $1/4$ the value that would be expected if circuits such as that of Fig.~\ref{fig:parallel} were used (or 1/2 the cost in parallel settings).  As an example, for the case of RFPE numerical experiments show that we can learn the eigenphase within error $\epsilon$ with probability $1/2$ using
\begin{equation}
M(\epsilon) \approx \frac{2.3}{\epsilon}.
\end{equation}
In comparison, the ultimate lower limit given by previous studies becomes $\pi/(2\epsilon)$ and the Bayesian Cramer--Rao bound gives a lower limit of approximately $1.6/\epsilon$ under the assumptions of Gaussian priors made in RFPE~\cite{wiebe2015efficient}.  Similarly the upper bounds on the cost of traditional QPE in~\cite{NC00} become $8\pi/\epsilon$ rather than $16\pi/\epsilon$ if these circuits are used.

While the assumption that the underlying Trotter--Suzuki formula can be inverted by simply inverting the sign of the evolution time.  Specifically, this assumptions applies for the $(2k+1)^{\rm th}$ order Trotter-Suzuki formulas for all $k\ge 1$ but it does not apply to the second order formula unless further assumptions are made.  We can see this from the fact that $$\left(\prod_{j=1}^N e^{-i H_j t}\right)\left(\prod_{j=1}^N e^{iH_j t} \right)= 1+O(t^2).$$
If we assume that we are interested only in the ground-state energy and the Hamiltonian is real valued (like in the quantum chemistry applications that we consider) then the error in assuming that $U^\dagger$ can be formed by simply flipping the signs is $O(t^3)$~\cite{HWB15}, which is by no means fatal but it could potentially contribute to the error in Trotter--Suzuki decompositions.  Whereas if the third (or higher) order Trotter--Suzuki formula is used then no such danger exists.  This, along with the superior bounds proven for the error in the third order formula, provide the justification for using this formula in preference to the asymptotically equivalent second--order formula.

\section{Trotter errors}\label{sec:trotter}
Here we discuss the issue of how errors from the use Trotter Suzuki formulas lead to systematic errors in the ground-state energy estimates output by phase estimation.  We further discuss the methodology that we use to upper bound these errors and also give empirical estimates of the scaling of such errors.

\subsection{Rigorous bounds}
A major source of error in most quantum simulation algorithms arises from the use of Trotter formula expansions.
Such errors can be made arbitrarily small but the need to make such errors smaller then chemical accuracy
mean that the cost of doing so can substantially impact the time required to perform the simulation.  In this
appendix we will provide a detailed discussion about how to bound the error in low--order Trotter formulas.

Trotter errors arise from the fact that the terms in the Hamiltonian used in the expansion do not commute.
In principle, the Zassenhaus formula provides everything that is needed in order to understand the scaling of these errors:
\begin{equation}
e^{(A+B)t}=e^{At}e^{Bt}e^{\frac{1}{2}[A,B]t^2} + O(t^3),
\end{equation}
and thus
\begin{equation}
\|e^{(A+B)t}-e^{At}e^{Bt}\| \in O(\|[A,B]\|t^2).
\end{equation}

Although this expression is useful in estimating the scaling of simulation errors, the question that we're interested
in is somewhat orthogonal to this.  Instead, we are interested in the errors in estimated eigenvalues.  The Baker--Campbell--Hausdorff formula, which
is the dual to the Zassenhaus formula, can be used to estimate these errors.  If $H=\sum_{\alpha=1}^L H_\alpha$ where the $H_\alpha$ correspond to terms in~\eqref{eq:Hamiltonian} then the second order Trotter--Suzuki formula (also known as the Straang splitting) gives 
\begin{align}
\prod_{\alpha=1}^L e^{-i H_\alpha t/2}\prod_{\alpha=L}^1 e^{-i H_\alpha t/2} = e^{-iH_{\rm eff} t},
\end{align}
where $H-H_{\rm eff}$ is 
\begin{equation}
-\frac{1}{12} \sum_{\alpha \le \beta} \sum_{\beta} \sum_{\alpha' < \beta} [H_{\alpha}(1-\frac{\delta_{\alpha,\beta}}{2}),[H_\beta,H_{\alpha'}]]t^2 +O(t^3).\label{eq:errbd1}
\end{equation}
This shows that, to leading order, the error in the Strang splitting can be estimated by the ground--state expectation value of a double commutator sum.  Furthermore, since $L\in O(N^4)$ it is clear from this form that the error in the Trotter formula must scale at most as $O(N^{10} t^2)$.  It is not $O(N^{12}t^2)$ because the commutator structure restricts two of the orbitals $H_{\alpha}$ and $H_{\beta}$ act on.  Although this estimate can be computed in polynomial time, there are too many terms for this to be reliably estimated for molecules on the scale of nitrogenase even using Monte Carlo methods~\cite{BMW+15}.

An upper bound on the asymptotic scaling can be trivially found by applying the triangle inequality to~\eqref{eq:errbd1}.  This approach is not appropriate for our purposes because we do not
know apriori how large $t$ must be for the leading order term to be dominant.  As a result, we use the following result, which is provably an upper bound on the error in the energy of the evolution:
\begin{align}
\Delta E_{\rm TS}/t^2\le &
4\!\!\sum_{\alpha,\beta,\alpha'}\!\! \|H_{\alpha}\|\|H_\beta\| \|H_{\alpha'}\| \nonumber\\
&\times (\delta_{\alpha>\beta}\delta_{\alpha'>\beta} + \delta_{\beta>\alpha}\delta_{\alpha',\alpha})W(\alpha,\beta,\alpha') ,\label{eq:bound2}
\end{align}
where $W(\alpha,\beta,\alpha')$ is an indicator function that takes the value $1$ if and only if the corresponding double commutator is non--zero.  Note that this is slightly tighter than the bound used in (16) of~\cite{PHW+15}.  

There are several criteria that we know a priori lead to a double commutator vanishing:
\begin{enumerate}
\item $H_{\alpha'}$ and $H_\beta$ act on disjoint sets of qubits.
\item $H_{\alpha}$ acts on a disjoint set of qubits from the set of qubits that $H_{\alpha'}$ and $H_{\beta}$ act on.
\item $H_{\alpha'}$ and $H_{\beta}$ correspond to PP or PQQP terms.
\item $H_{\alpha'}$ and $H_{\beta}$ correspond to PR and PQQR terms with the same P and R.
\item Only one of $[H_{\alpha},[H_\beta,H_{\alpha'}]]$, $[H_{\beta},[H_{\alpha'},H_{\alpha}]]$ and $[H_{\alpha'},[H_{\alpha},H_{\beta}]]$ is non--zero according to the prior rules (Jacobi identity).
\end{enumerate}
There are other symmetries to the terms that can be used to argue that even more terms are necessarily zero.  Since we do not consider these properties, we overcount the contribution
of any such terms and so our estimate remains an upper bound.

This bound is much more computable than the original expression, but is computationally challenging to compute exactly owing to the $O(N^{10})$ terms in the double commutator sum.
The Cauchy--Schwarz inequality can be used to convert this expression into one that can be computed in $O(N^4)$ operations, but doing so can over-represent the influence of large terms in the Hamiltonian~\cite{PHW+15}.  
\subsection{Empirical estimates of Trotter Error}
Given that computation of the matrix elements of the error operator in~\eqref{eq:errbd1} is computationally challenging, we rely on Monte Carlo sampling to estimate the upper bound in~\eqref{eq:bound2}.  Monte Carlo sampling is much more effective here because the use of the triangle inequality removes the alternating sign that we see when summing the original series.
We achieve this by drawing $M$ samples uniformly at random for $\{\alpha_j:j=1\ldots M\}$, $\{\beta_j:j=1\ldots M\}$ and $\{\beta_j:j=1\ldots M\}$.  We then reject the sample if $(\delta_{\alpha>\beta}\delta_{\alpha'>\beta} + \delta_{\beta>\alpha}\delta_{\alpha',\alpha})W(\alpha,\beta,\alpha')=0$, and otherwise compute the product of the products of the norms of the three corresponding Hamiltonian terms.  If there are $L$ terms in the Hamiltonian and define
\begin{align}
\Gamma(\alpha,\beta,\alpha'):=&4\|H_{\alpha}\|\|H_\beta\| \|H_{\alpha'}\| \nonumber\\
&\times(\delta_{\alpha>\beta}\delta_{\alpha'>\beta} + \delta_{\beta>\alpha}\delta_{\alpha',\alpha})W(\alpha,\beta,\alpha')
\end{align}
then
\begin{equation}
\Delta E_{\rm TS} \lessapprox \frac{L^3}{M} \sum_{j=1}^M \Gamma(\alpha_j,\beta_j,\alpha'_j)t^2:=ht^2.
\end{equation}
This implies that, for a fixed $h$, if we wish to achieve error $\epsilon$ in the eigenvalues of the Trotter--Suzuki expansion then it suffices to pick
\begin{equation}
t=\sqrt{\frac{\epsilon}{h}}.
\end{equation}
The variance in this estimator is
\begin{equation}
\frac{L^6\mathbb{V}_{\alpha,\beta,\alpha'}(\Gamma(\alpha,\beta,\alpha'))t^4}{M},
\end{equation}
which implies using Chebyshev's inequality that with probability greater than $75\%$ the sample error is less than $2\epsilon$ if
\begin{equation}
M\ge \frac{L^6\mathbb{V}_{\alpha,\beta,\alpha'}(\Gamma(\alpha,\beta,\alpha'))t^4}{\epsilon^2}=\frac{L^6\mathbb{V}_{\alpha,\beta,\alpha'}(\Gamma(\alpha,\beta,\alpha'))}{h^2}.
\end{equation}

In practice, uniform sampling is not necessarily the best option because the importance of the different terms can vary
wildly within a class.  For example, the one-body terms tend to be much larger than the two-body terms but the 
two-body terms are far more numerous.  This means that uniform sampling can underestimate the contributions of
such terms because of their relative scarcity in the sample space.

\begin{figure*}[t!]
\centering
\includegraphics[width=\linewidth]{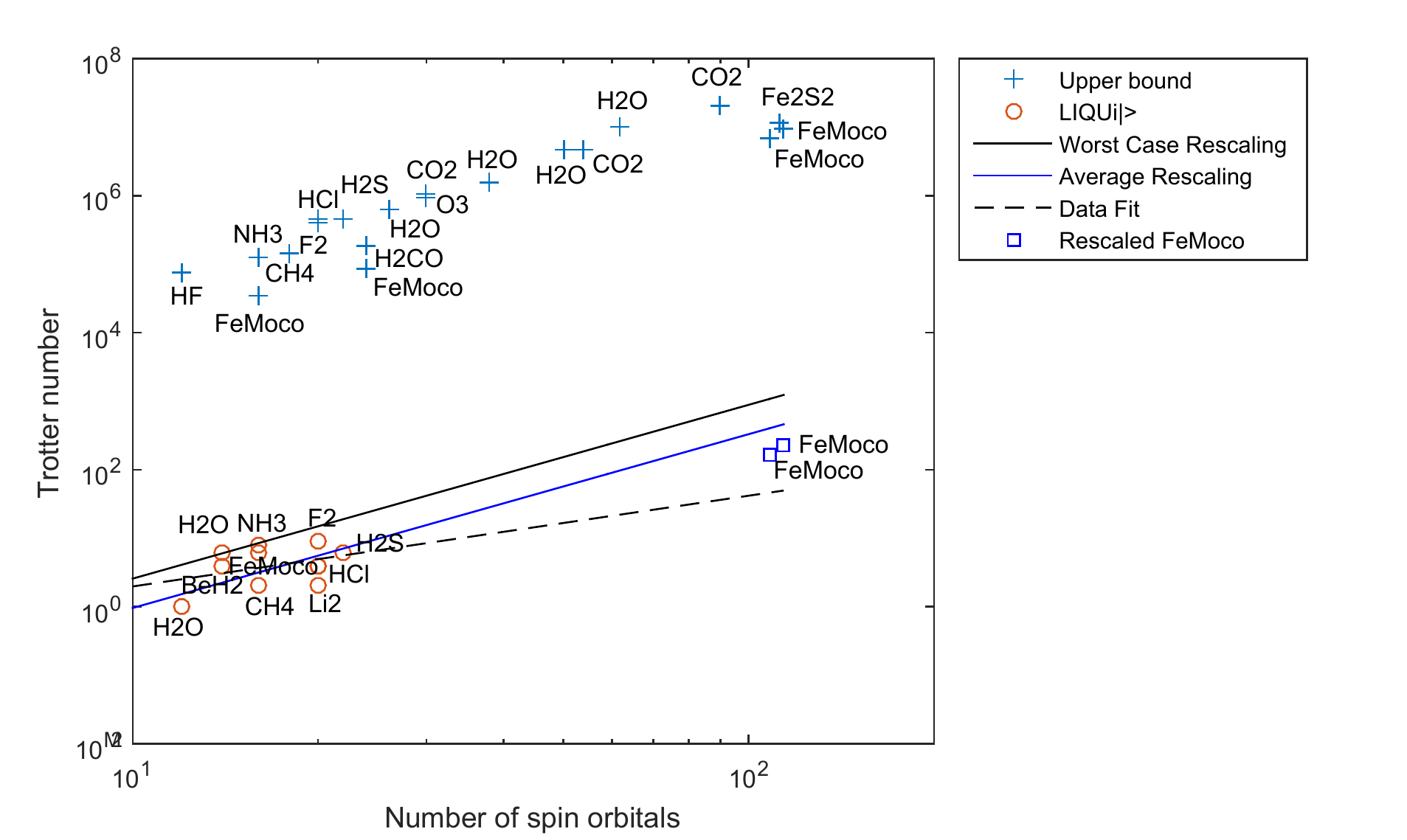}
\caption{Trotter number ($1/dt$) needed to reach $0.1$ mHartree of accuracy assuming no errors from synthesis or phase estimation.  Lines represent projected upper bounds based on current data.
\label{fig:trotter}
}
\end{figure*}

\begin{table*}
\begin{tabular}{|c|c|c||c|c|c|}
\hline
\multicolumn{3}{|c||}{Upper Bound}&\multicolumn{3}{c|}{\text{LIQUi$|\rangle$}}\\
\hline
Molecule & Spin Orbitals &Basis&Molecule & Spin Orbitals &Basis\\
\hline
HF & 12 &sto6g&H2O (frozen core)& 12 & sto6g \\
FeMoco & 16 &tzvp&BeH2&14&sto6g\\
NH3 & 16 &sto6g&H2O&14&sto6g\\
CH4 & 18 &sto6g&CH4 (frozen core)&16& sto6g \\
F2 & 20 &sto6g&FeMoco & 16&tzvp\\
HCl & 20 &sto6g&NH3&16&sto6g\\
H2S & 22 & sto6g&Li2&20&sto6g\\
FeMoco & 24 &tzvp&HCl&20&sto6g\\
H2CO & 24 &tzvp&F2&20&sto6g\\
H2O & 26 &p321&CH4&20&sto6g\\
CO2 &30 &sto3g&H2S&22&sto6g\\
O3 & 30 & sto6g&&&\\
H2O &38&dzvp&&&\\
H2O&50&p6311ss&&&\\
CO2&54&p321&&&\\
H2O&62&p6311ss&&&\\
CO2 &90&dzvp&&&\\
FeMoco&108&tzvp&&&\\
Fe2S2 & 112 & sto3g&&&\\
FeMoco&114 &tzvp&&&\\
\hline
\end{tabular}
\caption{Table contains the identities of each molecule sorted first by the number of {\it spin} orbitals, which is twice the number of spatial orbitals, and then by the actual, or upper bounded, Trotter number. \label{fig:trotterTable}}
\end{table*}

We combat this by sampling from each~\emph{type} of double commutator.  In particular, we sample over a class of double
commutators such as $[{\rm PQ},[{\rm PP},{\rm PQ}]]$ and uniformly draw PQ, PP and PQ terms to estimate those terms contribution to the
overall error.  The total error is then the sum of the estimates over all such classes.  We use a minimum of $10^8$ samples per
class, which renders the sample standard deviation in our estimates of the error less than $1\%$.

The resultant bounds can be seen in Fig.~\ref{fig:trotter} wherein we examine the predicted Trotter numbers and the empirically observed Trotter numbers for small molecules.  The rough scaling of the upper bound on the Trotter number that we see corresponds to $N^{2.5}$ which was noted in previous numerical studies~\cite{PHW+15}, but owing to the scatter of the data due to the widely varying chemical properties of the molecules, this scaling should not be seen as definitive.  We observe that the upper bounds seem to be roughly a factor of $10~000$ times too loose for small molecules.  For this reason we plot three reasonable extrapolations of the scaling based on the numerical results for small molecules.  The most pessimistic bound rescales the scaling extracted from the upper bound such that all the data remains beneath the curve.  The middle one rescales the upper bound data by the average discrepancy between the upper bound and the numerically computed examples.  The most optimistic curve is simply a polynomial fit to the numerical data that ignores the upper bound.  We expect the middle curve to be the most realistic estimates, but provide resource estimates for these three cases below as well as results that follow from using the upper bound.

\section{Error propagation}\label{sec:errorProp}
Here we provide proofs of some basic results that we will use to propagate these errors through the quantum simulation.  These results are crucial for the cost estimates in the subsequent section because they show how large the worst case errors can be in the eigenvalue estimation given errors of these magnitudes.  It is worth noting that we expect these results to yield substantial overestimates of the error because they do not consider the natural cancellations that are likely to occur in practical eigenvalue estimation problems.
\begin{lemma}
Let $A$ and $B$ be Hermitian operators acting on finite dimensional Hilbert spaces such that $\|A-B\|_2 \le \epsilon$ and $A\ket{\psi_A} = E_A \ket{\psi_A}$ and $B\ket{\psi_B}=E_B\ket{\psi_B}$ where $E_A$ and $E_B$ are the smallest eigenvalues of either operator then $|E_A -E_B|\le \epsilon$.\label{lem:1d}
\end{lemma}
\begin{proof}
Because $A$ and $B$ are Hermitian they satisfy the variational property meaning that
\begin{equation}
E_B \le \bra{\psi_A} B \ket{\psi_A}.
\end{equation}
Since $\|A-B\|_2 \le \epsilon$ it follows that there exists $C$ such that $\|C\|_2 \le 1$ and $A= B+\epsilon C$.  This implies that
\begin{equation}
E_B \le \bra{\psi_A} A+\epsilon C \ket{\psi_A}=E_A + \epsilon \bra{\psi_A} C \ket{\psi_A}.
\end{equation}
Thus
\begin{equation}
E_B - E_A \le \epsilon \bra{\psi_A} C \ket{\psi_A}.
\end{equation}
If $E_A \le E_B$ then we have from the definition of $\|\cdot \|_2$ that
\begin{equation}
|E_B - E_A| \le \epsilon |\bra{\psi_A} C \ket{\psi_A}|\le \epsilon.
\end{equation}
Now assume that $E_B > E_A$.  We then have 
\begin{equation}
E_A \le \bra{\psi_B} A \ket{\psi_B}.
\end{equation}
Then by repeating the same argument we conclude that $|E_B - E_A| \le \epsilon$ regardless of the sign of $E_A -E_B$.
\end{proof}

Now we will go beyond this bound to show that the error scaling in the eigenvalues of the unitary evolutions generated by two similar Hamiltonians is no more pathological than the scaling of errors in the ground-state energies.

\begin{lemma}
Assume that for Hermitian bounded operators $A$ and $B$ acting on a finite dimensional Hilbert space $\|e^{-i A t} -e^{-iBt}\|_2\le t\gamma(t)$ for $\gamma(t)$ a non--decreasing continuous function of $t$ on $[0,\infty)$ then $\|A-B\|_2 \le \gamma(t)$.\label{lem:gamma}
\end{lemma}
\begin{proof}
Using standard bounds~\cite{NC00}, we have that $\|e^{-i A t} - e^{-i Bt}\|_2\le \|A-B\|_2 t$ and furthermore from Taylor's theorem $\|e^{-i A t} - e^{-i Bt}\|_2 = \|A-B\|_2t + O([\|A-B\|_2 t]^2)$.  Therefore the former upper bound is tight in the limit as $t\rightarrow 0$.  By assumption $\|e^{-i A t} - e^{-i Bt}\|_2 \le t\gamma(t)$ for all $t$ in a compact subinterval containing $0$.  Assume that $\lim_{t\rightarrow 0} \gamma(t)/\|A-B\|_2<1$.  This implies that there exists a compact interval containing $0$ such that for all $t$ in this interval $\|e^{-i A t} -e^{-iBt}\|_2 > \|A-B\|_2t$, which leads to a contradiction because we have already demonstrated that $\|A-B\|_2t$ is a tight bound on the error in this limit.  Therefore 
$\gamma(t) \ge \lim_{t\rightarrow 0} \gamma(t) \ge \|A-B\|_2$ under the assumptions of the lemma.
\end{proof}

\begin{theorem}
Let $H=\sum_{j=1}^M H_j$ where each $H_j$ is a bounded Hermitian operator acting on a finite dimensional Hilbert space.  Furthermore, let $\tilde{H} = \sum_{j=1}^M \tilde{H}_j$ be a similar sum of bounded operators such that $\|H_j - \tilde{H}_j\|_2\le \delta$.  Finally, let $\|e^{-iH t} - \prod_{j=1}^M e^{-iH_j t/2} \prod_{j=M}^1 e^{-iH_j t/2}\|_2 \le \Delta E_{TS}(t)t$.  Then the difference in ground-state energies between $H(t)$ and $\tilde{H}(t):=i\log(\prod_{j=1}^M e^{-i\tilde H_j t/2} \prod_{j=M}^1 e^{-i\tilde H_j t/2})/t$ is at most $\Delta E_{TS}(t) + (2M-1)\delta$.\label{thm:gstate}
\end{theorem}
\begin{proof}
First, since $\tilde{H}_j$ is Hermitian $$\prod_{j=1}^M e^{-i\tilde H_j t/2} \prod_{j=M}^1 e^{-i\tilde H_j t/2}$$ is a unitary operator.  The matrix logarithm is defined if and only if the matrix in question is invertible and hence the matrix logarithm exists because unitary matrices are invertible.  The logarithm is then clearly an anti--Hermitian operator and hence $\tilde H(t)$ is Hermitian.  This implies that $\prod_{j=1}^M e^{-iH_j t/2} \prod_{j=M}^1 e^{-iH_j t/2} \equiv e^{-i\tilde{H}(t) t}.$

The triangle inequality implies that
\begin{align}
\|e^{-i H t}& - e^{-i \tilde{H}(t) t} \|_2 \le \nonumber\\
&\left\|e^{-i Ht} -\prod_{j=1}^M e^{-iH_j t/2} \prod_{j=M}^1 e^{-iH_j t/2}\right\|_2\nonumber\\
&+\left\|\prod_{j=1}^M e^{-iH_j t/2} \prod_{j=M}^1 e^{-iH_j t/2} - e^{-i\tilde{H}(t)t}\right\|_2.
\end{align}
Then using our assumptions and standard inequalities for the errors in unitary operations~\cite{NC00} this error is at most
\begin{equation}
[\Delta E_{TS}(t)   + (2M-1)\delta] t.
\end{equation}
Then applying Lemma~\ref{lem:gamma} we have that $\|H-\tilde{H}(t)\|_2 \le \Delta E_{TS}(t)  + (2M-1)\delta$.  The result then follows from applying Lemma~\ref{lem:1d}.
\end{proof}

These error bounds apply for generic quantum simulation based on the second order Trotter formula (also known as the Strang splitting), however they apply in particular to the case of simulating quantum chemistry.  We have already discussed the Trotter--Suzuki error for quantum chemistry in~\eqref{eq:bound2}.  Even in the absence of decoherence, a source of error inevitably comes into our analysis from approximating the single qubit rotations.  Now we need to show a result that bounds the impact of such eigenvalue estimation.
\begin{lemma}
For Hermitian $H$ and $t\ge 0$ let $U(t)$ be a family of unitary operations such that $\|e^{-iHt} -U(t)\|_2\le \epsilon t$ for all $t\ge 0$, then for every $t$ there exists a Hermitian operator $\tilde H(t)$ such that $U=e^{-i \tilde H t}$ and $\|H-\tilde H(t)\|_2\le \epsilon$.\label{lem:Heff}
\end{lemma}
\begin{proof}
By taking the principal matrix logarithm of $U(t)$ it is clear that there exists $\tilde H(t)$ such that $U=e^{-i \tilde H(t) t}$.  Furthermore, using standard inequalities it is easy to see that $\|e^{-i H t} - e^{-i \tilde H(t) t}\|_2\le \|H-\tilde H(t)\|_2 t$.  Because this bound is tight in the limit as $t\rightarrow 0$, if $\epsilon< \|H-\tilde H(t)\|_2$ then a contradiction is reached for $t$ sufficiently small.  Therefore since we require that the inequality hold for all $t\ge 0$, $\|H-\tilde H(t)\|_2\le \epsilon$.
\end{proof}

It is easy to see from the fact that there is a one to one mapping between the Hamiltonian terms and the rotations performed in the circuit.  Thus if the error in each such rotation is $\delta t$ then the error in each term of the effective Hamiltonian is at most $\delta$.  This means that the total shift in energy from circuit synthesis is at most $M\delta$.

\begin{corollary}
Assume $H=\sum_{j=1}^{M} h_j P_j$ for $h_j \in \mathbb{R}$, Pauli operators $P_j$ and
$\|e^{-i Ht} - \prod_{j=1}^{M} e^{-i h_{j} P_{j} t/2}\prod_{j=M}^{1} e^{-i h_{j} P_{j} t/2}\|_2\le \Delta E_{TS}(t)t$.  Also let each $e^{-iH_j t/2}$ be implemented using a unitary $U_j(t)$ such that $\|U_j(t) - e^{-i H_j t/2}\|_2 \le \Delta_{\rm synth}$ for all $t$.  The error in the ground-state energy that results from such a simulation is at most $\Delta E_{TS} +(2M-1)\Delta_{\rm synth}/t$.\label{cor:err}
\end{corollary}
\begin{proof}
The result then follows directly from Theorem~\ref{thm:gstate} and Lemma~\ref{lem:Heff}.
\end{proof}
This shows that it suffices to choose synthesis error that shrinks linearly with the timestep used in the Trotter decomposition.  Since the cost of circuit synthesis of rotations in the the Clifford + $T$ gate library scales logarithmically with $\Delta_{\rm synth}/t$~\cite{selinger2015efficient}.

\section{Cost estimates for nitrogenase}\label{sec:costest}
Fundamentally, two factors contribute to the cost of the quantum simulation (assuming that the user can prepare an exact copy of the ground state at negligible cost).  The first is the cost of implementing the Trotter decomposition of the Hamiltonian and the second is the number of times that the Trotter circuit must be repeated in the phase estimation algorithm.

One might object that the number of time steps required in the Trotter--Suzuki decomposition also is a driving factor in the cost.  Of course the Trotter decomposition is a major driver of the cost, but it comes in only indirectly through the cost of phase estimation.  This is because, in principle, the phase estimation algorithm learns the eigenphases of a single Trotter
step.  The Trotter error can be made arbitrarily small by choosing shorter evolution times, but this in turn requires the phase estimation algorithm to take more steps.  As the cost of phase estimation scales inversely with the desired uncertainty, this causes the cost to scale inversely with the time step used in the Trotter--Suzuki decomposition.  Thus the cost
of the simulation can be thought of as arising from only two sources, the cost of each depending on the error tolerances allowed for all three contributions to the error.

If we then define $\epsilon_1$ to be the error in phase estimation, $\epsilon_2:=\Delta E_{\rm TS}$ to be the error in the Trotter--Suzuki expansion and $\epsilon_3:=(2M-1)\Delta_{\rm synth}/t$ to be the error in circuit synthesis then it follows from the triangle inequality and Corollary~\ref{cor:err} that the error in the ground-state energy is at most $0.1$ mHartree if
\begin{equation}
\epsilon_1 + \epsilon_2 +\epsilon_3 \le \epsilon:= 10^{-4} {\rm Ha}.\label{eq:epsilonConst}
\end{equation}
In this section we will focus on the target of $0.1$ mHa level of accuracy, which is appropriate for quantitative calculations of the reaction rates.

We estimate the cost of the circuit using the number of ${\tt T}$ gates required in the algorithm, which is a function of the form
\begin{equation}
C = 2M\left\lceil\frac{\alpha}{\epsilon_1}\right\rceil \left\lceil \beta \sqrt{\frac{\epsilon}{\epsilon_2}}\right\rceil \biggr(\gamma \log_2\left(\frac{2M}{\epsilon_3} \left\lceil\beta\sqrt{\frac{\epsilon}{\epsilon_2}}\right\rceil\right)+\delta\biggr),\label{eq:C}
\end{equation}
and then optimize over $\epsilon_1, \epsilon_2$ and $\epsilon_3$ to minimize $C$ subject to the constraint in~\eqref{eq:epsilonConst}.  Here $\alpha$ is the scaling constant for the phase estimation algorithm used, $\beta$ is the Trotter number (or multiplicative factor by which $t$ is decreased from $1~{\rm Ha}^{-1}$) needed to achieve an error of $0.1~{\rm mHartree}$ in the ground-state energy estimate and $\gamma$ and $\delta$ are the constants used in the quantum circuit synthesis algorithm.  Here $M=6.1\times 10^6$ for nitrogenase in the $54$ orbital basis and $M=8.2\times 10^6$ for nitrogenase in the $57$ orbital basis using the circuits of~\cite{LWG+10}.

The true optima of~\eqref{eq:C} are difficult to find because of the factor of $\sqrt{\epsilon/\epsilon_2}$ in the logarithm.  In order to simplify our optimization we instead choose $\epsilon_1,\epsilon_2$ and $\epsilon_3$ to minimize
\begin{equation}
\tilde C = 2M\left(\frac{\alpha}{\epsilon_1}\right) \left( \beta \sqrt{\frac{\epsilon}{\epsilon_2}}\right) \biggr(\gamma \log_2\left(\frac{2M}{\epsilon_3} \beta\right)+\delta\biggr),\label{eq:Capprox}
\end{equation}
subject to the same constraint. The global optimum of~\eqref{eq:Capprox} can be found directly from calculus, which allows near optimal parameters for~\eqref{eq:C} to be found easily.
The value of~\eqref{eq:C} at these parameters is then an upper bound on the minimum of~\eqref{eq:C} and so the estimates provided remain upper bounds (modulo assumptions about the Trotter error).  

\begin{table}[t!]
\begin{tabular}{|c|c|c|}
\hline
Case & Gates & Time ($100$ MHz ${\tt T}$ gates)\\
\hline
\hline
Rigorous bound & $1.0\times 10^{21}$ & $3.2\times 10^5$ years.\\
Clifford & $1.4\times 10^{21}$ & --\\
\hline
Rigorous + PAR & $3.2 \times 10^{22}$ & $8300$ years.\\
Clifford & $3.3\times 10^{22}$ & --\\
\hline
\hline
Pessimistic bound & $7.9 \times 10^{15}$ & $30$ months\\
Clifford & $1.2\times 10^{16}$ & --\\
\hline
Pessimistic + PAR & $2.3 \times 10^{17}$ & $31$ days\\
Clifford & $2.3\times 10^{17}$ & --\\
\hline
\hline
Rescaled bound & $1.2\times 10^{15} $ & $135$ days\\
Clifford & $1.8\times 10^{15}$ & --\\
\hline
Rescaled + PAR & $3.5\times 10^{16} $ & $120$ hours\\
Clifford & $3.5\times 10^{16}$ & --\\
\hline
\hline
Optimistic bound& $ 1.6 \times 10^{14}$ & $19 $ days\\
Clifford & $2.4\times 10^{14}$ & --\\
\hline
Optimistic + PAR & $4.7 \times 10^{15}$ & $17$ hours\\
Clifford & $4.7\times 10^{15}$ & --\\
\hline
\end{tabular}

\vspace{5mm}

\begin{tabular}{|c|c|c|c|c|}
\hline
Case & $\alpha$ & $\beta$ & $\gamma$ & $\delta$\\
\hline
Rigorous & $8\pi$ & $7\times 10^6$ & $4$ & $11$\\
Pessimistic & $\pi/2$ & $1075$ & $1.15$ & $9.2$\\
Rescaled & $\pi/2$ & $166$ & $1.15$ & $9.2$\\
Optimistic & $\pi/2$ & $24$ & $1.15$ & $9.2$\\
\hline
\end{tabular}
\caption{Resource estimates for simulation of nitrogenase's FeMoco in structure \structOne~ which requires a small basis consisting of $108$ spin orbitals.  PAR uses $\gamma=4$ and $\delta=11$. }\label{tab:cost}
\end{table}

The ``worst'' case assumptions in Table~\ref{tab:cost} correspond to only using rigorously proven upper bounds on the cost.  These lead to estimates that are clearly extremely pessimistic.  Even given our optimistic assumptions about the target computer, the worst case bounds suggest between millions and tens of thousands of years depending on whether parallelism is used.  

The ``pessimistic'' assumptions use empirical scalings for circuit synthesis and phase estimation and use the worst scaling supported from our bounds on the Trotter error, but rescaled by the ratios observed between the actual Trotter numbers required and the theoretically predicted ones.

The ``rescaled'' case takes the rigorous upper bound for nitrogenase and divides it by the average ratio observed between the exact Trotter numbers and their upper bounds for the tractable molecules molecules.  Rescaling the upper bound by a constant and applying least squares fitting to find the most consistent constant yields similar results.  We suspect these rescaled estimates may provide the most realistic estimate of the Trotter number required to simulate nitrogenase.

\begin{table}[t!]
\begin{tabular}{|c|c|c|}
\hline
Case & Gates & Time ($100$ MHz ${\tt T}$ gates)\\
\hline
\hline
Rigorous bound & $1.8\times 10^{21}$ & $5.5\times 10^5$ years.\\
Clifford & $2.5 \times 10^{21}$ & --\\
\hline
Rigorous + PAR & $5.9 \times 10^{22}$ & $1.5\times 10^4$ years.\\
Clifford & $6.0 \times 10^{22}$ & --\\
\hline
\hline
Pessimistic bound& $1.2 \times 10^{16}$ & $3.8$ years\\
Clifford & $1.8 \times 10^{16}$ & --\\
\hline
Pessimistic + PAR & $3.5 \times 10^{17}$ & $48$ days\\
Clifford & $3.5 \times 10^{17}$ & --\\
\hline
\hline
Rescaled bound& $2.2\times 10^{15} $ & $250$ days\\
Clifford & $3.1 \times 10^{15}$ & --\\
\hline
Rescaled + PAR & $6.3\times 10^{16}$ & $8.8$ days\\
Clifford & $6.3 \times 10^{16}$ & --\\
\hline
\hline
Optimistic bound& $2.3 \times 10^{14}$ & $27$ days\\
Clifford & $3.5 \times 10^{14}$ & --\\
\hline
Optimistic + PAR & $6.6 \times 10^{15}$ & $23$ hours\\
Clifford & $6.6 \times 10^{15}$ & --\\
\hline
\end{tabular}

\vspace{5mm}

\begin{tabular}{|c|c|c|c|c|}
\hline
Case & $\alpha$ & $\beta$ & $\gamma$ & $\delta$\\
\hline
Rigorous & $8\pi$ & $9.5\times 10^6$ & $4$ & $11$\\
Pessimistic & $\pi/2$ & $1233$ & $1.15$ & $9.2$\\
Rescaled & $\pi/2$ & $225$ & $1.15$ & $9.2$\\
Optimistic & $\pi/2$ & $25$ & $1.15$ & $9.2$\\
\hline
\end{tabular}
\caption{Resource estimates for simulation of nitrogenase's FeMoco in structure \structTwo~which requires a small basis consisting of $114$ spin orbitals.  PAR uses $\gamma=4$ and $\delta=11$.  }\label{tab:cost2}
\end{table}

The ``optimistic'' assumptions again use the same empirical scalings for PE and synthesis, but instead extrapolate the average scaling observed for the ensemble of molecules whose Trotter numbers we can compute and scaling up the result so that all of the data lies beneath the curve.  This scaling is optimistic, as there is little evidence for a clear trend in the empirical data and the range provided is insufficient to meaningfully extrapolate out to $108$ spin orbitals ($54$ spatial orbitals) or more.  We provide this estimate because it provides the best scaling that could reasonably be claimed to be supported by the data.

\subsection{Variance-based estimates}

{Such errors arise from three sources:  the systematic error in the TS decomposition $\epsilon_{\rm TS}$, the statistical error tolerance for phase estimation $\epsilon_{\rm QPE}$, and the statistical error in synthesizing rotations from Clifford and $\tt T$ gates $\epsilon_{\rm Rot}$. We then require the total error to be $\epsilon_{\rm TS}+\sqrt{\epsilon_{\rm QPE}^2 + \epsilon_{\rm Rot}^2}=0.1$ mHa.  The three uncertainties are then chosen such that the number of $\tt T$ gates required for the simulation is minimized given the target accuracy. }

The previous analysis for the estimates in the error can be used within this expression for the error under the assumptions that the errors in QPE and synthesis are not adversarial.  This approach was taken with the estimates in the main body, wherein the three dominant costs are optimized against each other to minimize the resources needed to achieve the $0.1$ mHartree target.  The optimization process is exactly the same as that used to minimize the cost given in~\eqref{eq:C}, however a different constraint linking the three errors is used.  This leads to modest reductions in the costs relative to the worst case bounds, which we provide in Tables~\ref{tab:cost} and \ref{tab:cost2}.

\subsection{PAR circuits}
There are several approaches that can be taken to parallelize rotations.  The first, often coined nesting, is discussed in~\cite{HWB15,raeisi2012quantum}.  It involves taking terms that commute with each other in the Hamiltonian and grouping them together so that they can be executed simultaneously.  In principle, this can lead to substantial reductions in the depth but in practice it is difficult to assess the performance of these schemes here because of the size of the molecule and the fact that we have chosen to restrict ourselves to lexicographic ordering.  This means that if we are to estimate the impact that parallelization can bring to these calculations we need to introduce a method that can reduce the $T$--depth without changing the ordering of terms in the Trotter--Suzuki decomposition.

The PAR method gives a way to achieve this goal~\cite{JWM+12}.  It works by teleporting a rotation into a state with probability $1/2$ using only Clifford operations and a pre--rotated ancilla.  In the event that this method fails then instead of performing $R_z(\theta)$ it performs $R_z(-\theta)$.  This can be corrected by teleporting a rotation $R_z(2\theta)$ into the qubit in question.  Should this fail (and it will half the time) the rotation can be corrected by teleporting a rotation of $R_z(4\theta)$ and so on.  This creates a geometric distribution of the number of pre--cached qubits needed to perform a given rotation.  These rotation angles are known before hand and so can be prepared offline in parallel.  Hence a logarithmic multiplicative overhead in space is needed to guarantee that enough ancilla qubits are prepared to perform the rotations with such high probability that it is unlikely that the cache of ancillae will ever be depleted.  

In order to bound the number of ancillae needed in order to parallelize a factor of $M$ rotations with high probability consider the following protocol.
\begin{itemize}
\item Divide the terms in the Trotter expansion of the Hamiltonian into blocks consisting of $M$ sequential terms.
\item For each rotation angle $\theta_j$ in a given bloeck and each $j=0,\ldots, n-1$, prepare the states $R_z(2^j\theta_j)\ket{+}$ in parallel for a predetermined value of $n$.
\item Implement each of the $M$ (possibly sequential operations) using programmable ancilla rotations using at most $n$ attempts, if the PAR circuit fails in each attempt then a failure is said to have occurred.
\item If a failure occurs then implement the correct rotation and proceed to the next precached rotation.
\end{itemize}
This protocol can be used to perform the desired rotation and its performance is summarized below.
\begin{theorem}\label{thm:M}
There exists a protocol for implementing PAR rotations that caches $M$ rotations of the form $R_z(\theta_j), R_z(2\theta_j),\ldots,R_z(2^{n-1}\theta_j)$ for $j=1,\ldots,M$ such that
 the expected number of rotations performed before a failure occurs is
$$
2^n(1-(1-2^{-n})^M).
$$
\end{theorem}
\begin{proof}
The proof is constructive.  To see this consider the protocol discussed above.  Such a protocol fails when all $n$ PAR circuits fail for any of the $M$ rotations cached, and a failure occurs when all $n$ attempts at the rotation that have been precomputed are expended.  The probability of such a failure is clearly $2^{-n}$ because the PAR circuit's success probabilities are independent and each attempt has success probability $1/2$~\cite{JWM+12}.

From the geometric distribution, the probability that no failure occurs in $M$ trials is then simply $(1-2^{-n})^M$.  Similarly, the probability that a failure occurs after precisely $k$ attempts is $2^{-n}(1-2^{-n})^{k-1}$.  Since there are only $M$ rotations in the cache any branch that has more than $M$ successes can only yield $M$ rotations.  This implies that the mean is
\begin{equation}
\sum_{k=1}^M k2^{-n}(1-2^{-n})^k + M(1-2^{-n})^M= 2^n(1-(1-2^{-n})^M).\label{eq:PARmean}
\end{equation} 
\end{proof}
A consequence of Theorem~\ref{thm:M} is that a PAR cache of $M$ rotations that further caches the correction operations for $n$ failures can reduce the $T$--depth by a factor of $2^n(1-(1-2^{-n})^M)$.  We can therefore adjust these parameters to substantially reduce the $T$--depth without adding a prohibitive number of rotations.

Note that as $n\rightarrow \infty$ \eqref{eq:PARmean} approaches $M$ as expected.  This suggests that taking large $n$ allows the PAR rotations to be paralellized more efficiently, but this comes at the price of requiring more $T$ gates and hence increases the overheads of quantum error correction.  

\subsubsection{Factory approach}
A major challenge with PAR facing costing it in a fault tolerant setting is that all the non--Clifford operations can be prepared simultaneously and offline.  This means that if we take the cost model where only $\tt T$ gates are considered then we come to the absurd conclusion that the costs of all quantum simulation algorithms can be reduced to that of synthesizing a single rotation.
In order to prevent such absurd tradeoffs we assume here that a delay of time equal to 1 $\tt T$ gate is included to model the measurement and feed-forward step that is needed for the programmable ancilla rotation.  This fixed cost means that even if all of the $\tt T$--gates are precached before hand then the time required for the remainder of the simulation will never be zero.

The above assumptions lead to an alternative approach to implementing PAR rotations, which we follow in the PAR costs in the following as well as the main body.  Rather than constructing a large cache of rotations offline, it makes sense to produce the rotations just in time.  Specifically if the cost of synthesizing a rotation is $C$ $\tt T$ gates then it makes sense to have $nC$ factories constantly producing rotation states of the form $R_z(2^j\theta_j)\ket{+}$ for $j=1,\ldots,n$.
Each of these $nC$ factories is staggered such that one set of factories finishes with their $n$ states at least $1$ cycle before the next rotation is needed.  As soon as a set of factories finishes it then proceeds on to the next set of rotations needed (excluding those that are currently being generated).  If a failure occurs, then the factory approach halts just like the traditional approach and waits for a rotation to be synthesized that applies the correct rotation online.

Imagine for the moment that the PAR circuit succeeds in $C$ consecutive attempts.  Since each attempt requires time equivalent to a single invocation of a $\tt T$ gate and the cost of synthesis is $C$ $\tt T$ gates, the first set of factories will have finished producing their states before the last set applies theirs.  This means that with $C$ such factories rotation states can be continuously generated even in the ``worst case scenario'' where each PAR circuit succeeds on the first attempt.  This is why in this setting it does not make sense to use more than $nC$ rotation factories given the assumption that the cost of each PAR attempt is $1$ $\tt T$ gate and that the rotations for at most $n$ failures are to be pre-cached.

\begin{theorem}
The average time required per rotation to apply the factory--based PAR strategy, assuming each PAR application requires time at most equal to $1$ $\tt T$ gate and all remaining Clifford operations are free is 
$$
\left(2 -\frac{n+2}{2^n} \right) + \frac{C+n}{2^n}.
$$
\end{theorem}
\begin{proof}
The probability of success in any PAR attempt is $1/2$ therefore the expected number of $\tt T$ gates that need to be applied before a solution is found is
\begin{equation}
\sum_{j=1}^n \frac{j}{2^n} + \frac{C+n}{2^n}.
\end{equation}
The latter term gives the expected impact of a failure, which occurs with probability $1/2^n$, makes $n$ PAR attempts and incurs a cost of $C$ $\tt T$ gates in applying the fallback rotation.  The result then follows from summing the geometric series.
\end{proof}

We use the above theorem to compute the expected time required by PAR under our assumptions of the costs of feed forwarding.  If such costs are neglected then the online cost of performing a PAR rotation is simply $C/2^n$ {\tt T} gates.

\subsubsection{Nesting estimates}

Finally, we would like to reiterate that in both the serial and PAR cases, the costs are found by building the circuit corresponding to the TS formula and counting the gates that compose it.  The estimates for nesting given in the main body are upper bounds based on empirical estimates of the number of terms that can be simultaneously executed.  We estimate this by taking FeMoco and greedily grouping terms that act on distinct sets of spin orbitals.  We specifically find that there exists a grouping that can simultaneously execute $26.43$ terms simultaneously for strucure \structOne~and $27.83$ for structure~\structTwo.  These values roughly correspond to the optimal scaling of $N/4$ that can be achieved through this nesting strategy.  We then assume the quantum computer can simultaneously execute each of these commuting groups and find the corresponding time by dividing the ${\tt T}$ count by these factors.  

Rather than executing the grouping in \Liquid we use a simple upper bound on the number of Clifford gates that could arise from the grouping.  We do this because the grouping strategy breaks the lexicographic ordering that leads to dramatic cancelation of the {\tt CNOT} strings that arise from the Jordan--Wigner decomposition.

For the nested data, the number of timesteps needed is assumed to be the same as for the other cases, which is reasonable given that operator ordering tends to not have a dramatic impact on the error.  Subsequent work will investigate the precise interplay that operator ordering has in nesting.  

\section{State Preparation}\label{sec:prep}

We discuss in this section the issue of state preparation, which is a major unanswered question that impacts the cost of quantum simulations.  While we do not discuss these costs in detail in the rest of the paper, we discuss below the issues that arise when using elementary state preparation methods based on coupled cluster, configuration interaction or Hartree--Fock states as well as adiabatic state preparation.  We also provide numerical results showing that the ground state overlap with Hartree--Fock states scales for small molecules and discuss the costs involved in adiabatic state preparation.

\subsection{Elementary state preparation methods}

Although we cannot rigorously prove that elementary state preparation methods such as Hartree--Fock states, unitary coupled cluster or truncated configuration interaction states (such as CISD or difference dedicated CI (DDCI) \cite{malrieu} states) will suffice for preparing a state with large overlap with the ground state, it is still important to ask how good simple ansatzes perform for numerically tractable cases.  We provide some numerical evidence for small molecules showing that the overlaps of the true ground state with the Hartree--Fock ground state is not necessarily small.  We leave similar studies of the overlap for unitary coupled cluster and truncated configuration interaction ansatzes for subsequent work.

\begin{figure}[t!]
\includegraphics[width=0.8\linewidth]{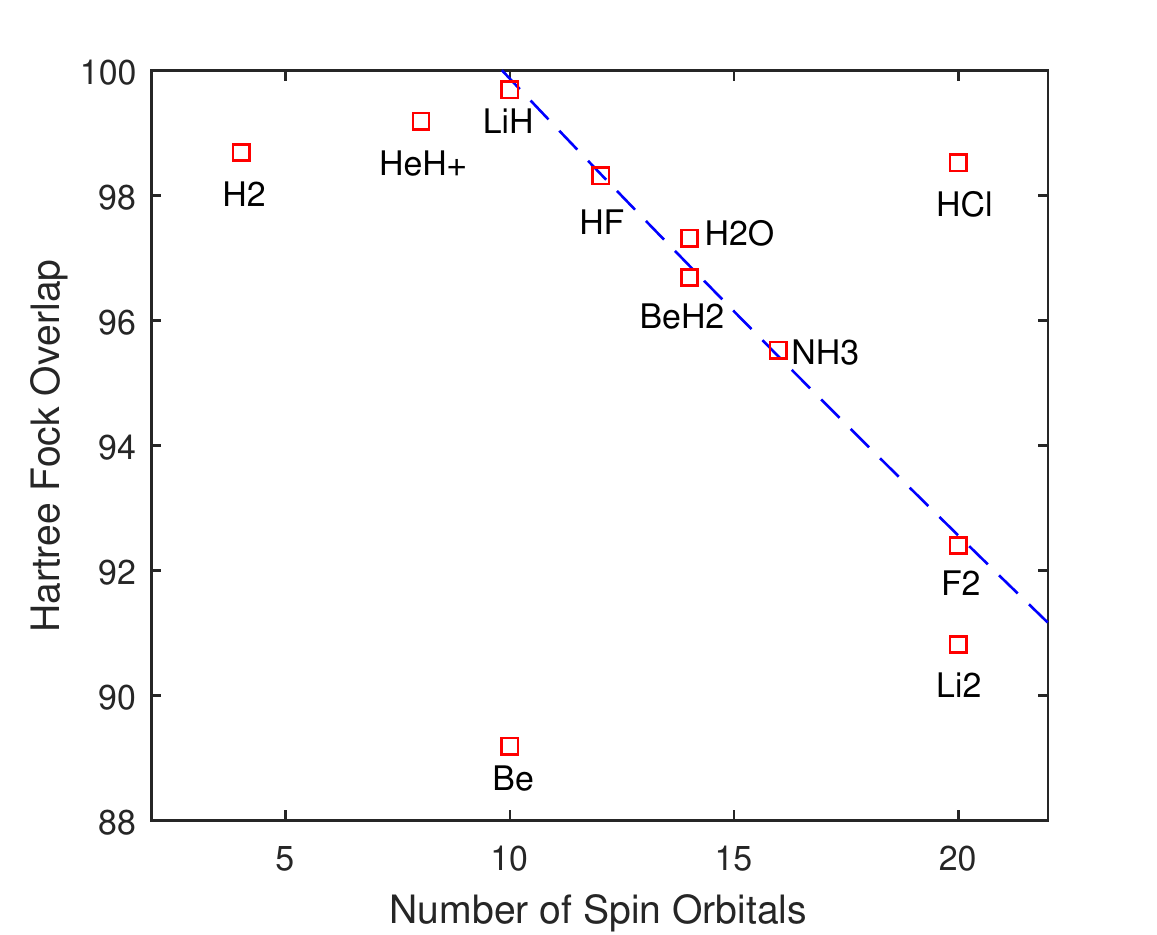}
\caption{Percent overlap $(|\braket{\psi}{\psi_{\rm HF}}|^2\times 100$) of the Hartree--Fock state with the electronic ground state computed by \Liquid.  All integrals are computed in a sto6g basis  except for ${\rm H}_2$ and HeH+.  The integrals for those molecules are computed using sto3g and 3-21g bases respectively.  The blue dashed line shows a possible extrapolated trend from the data.}\label{fig:HF}
\end{figure}

We see in Figure~\ref{fig:HF} that there is substantial overlap between the Hartree--Fock state and the true ground state of the molecules calculated using \Liquid. In particular, the smallest overlap that we see is $89\%$.  Other studies that have looked at chains of hydrogen atoms that are near disassociation show very small overlaps with the Hartree--Fock states.  Thus there are small molecules that can be constructed that are not well described by such ansatzes.

We find weak evidence for a downward trend in the overlap with the Hartree--Fock state as the size of the system grows.  Specifically, we see that roughly $50\%$ of the data points are well approximated by $107.75\%\times e^{-0.0076n}$ where $n$ is the number of spin orbitals.  This scaling would suggest that the overlap with the Hartree--Fock state for a molecule typical of this ensemble of the scale of FeMoco may be roughly $43\%$, we cannot say whether nitrogenase is indeed typical of this ensemble.  Indeed, one may expect stronger correlations to be present for molecules that contain atoms with $d$--electrons than those examined in Figure~\ref{fig:HF}.   Such molecules are frequently outside of our ability to simulate classically so finding an appropriate ensemble of molecules to use for such benchmarks remains an open research problem; however, the success of methods such as CI dynamically extended active space (CI-DEAS) \cite{legeza2004,legeza2010} in DMRG suggests that elementary truncated configuration interaction states may also suffice for quantum simulation~\cite{keller2014determining}. 

We further see no compelling evidence for scaling with the maximum nuclear charge of the constituent atoms for the molecules in this set.  This can clearly be seen from Be${\rm H_2}$ which has substantially better fidelity with the ground state than Be does.  Similarly, HF and HCl have nearly identical overlaps despite the fact that Cl has nearly twice the nuclear charge of F.  More study is needed in order to understand how these overlaps scale for large molecules, however it is strongly suspected that multi--reference states will be required to accurately model highly correlated ground states.

\subsection{Adiabatic State Preparation}

Adiabatic state preparation~\cite{wu2002polynomial} provides an alternative approach state preparation wherein the Hamiltonian to be simulated is replaced with a time--dependent Hamiltonian whose final ground state coincides with the target ground state and whose initial ground state is an easily preparable state.  An example of such a Hamiltonian is
\begin{align}
H(s) &= \sum_{p} h_{pp} a^\dagger_p a_p + \sum_{p,q} h_{pqqp} a^\dagger_p a^\dagger_q a_p a_q\nonumber\\
&~+s\left(H-\sum_{p} h_{pp} a^\dagger_p a_p - \sum_{p,q} h_{pqqp} a^\dagger_p a^\dagger_q a_p a_q \right),
\end{align}
where $s\in [0,1]$ is a dimensionless time that is $0$ at the beginning of the evolution and $1$ at the end.  It is then easy to see that the ground state of $H(0)$ is the Hartree--Fock state and the ground state of $H(1)$ is the full configuration interaction (FCI) ground state.

For such Hamiltonians (or more generally for those that are differentiable at least three times and whose resulting derivatives are $O(1)$~\cite{cheung2011improved}) a sufficiently slow evolution under the Hamiltonian will cause the Hartree--Fock state to be transformed into the FCI ground state of $H$.  In other words if $\mathcal{T}$ is the time--ordering operator which is defined such that
\begin{equation}
\partial_s \mathcal{T}(e^{-i\int_0^s H(s') \mathrm{d}s' T}):= -iH(s)\mathcal{T}(e^{-i\int_0^s H(s') \mathrm{d}s' T})
\end{equation}
and if we define $\mathcal{P}$ to be a projector onto the FCI ground state and $\Delta(s)$ to be the smallest eigenvalue gap between the ground state and the rest of the spectrum of $H(s)$ then
\begin{equation}
|(\openone - \mathcal{P})\mathcal{T}(e^{-i\int_0^1 H(s') \mathrm{d}s' T}) \ket{\psi_{\rm HF}}| \in O\left(\frac{\max_s \|\dot H(s)\|}{\min_s \Delta^2(s)T} \right).
\end{equation}
Here we take $T\gg 1$ and treat the other parameters to bounded above by a constant in this asymptotic expansion.
For such Hamiltonians the triangle inequality clearly shows that $\|\dot H(s)\| \in O(N^4)$ (although on physical grounds we expect $\|\dot{H(s)}\|$ to be in $O(\eta^2)\in O(N^2)$ because the potential energy scales quadratically with the number of constituent particles).  Thus if we want to have $O(1)$ probability of preparing the FCI ground state it suffices to simulate the time--dependent Hamiltonian $H(s)$ for time 
\begin{equation}
T\in O\left(\frac{N^4}{\min_s \Delta^2(s)} \right).
\end{equation}

Prima facie, the best known bounds on the costs of Trotter--Suzuki based simulation give the circuit size for such a simulation to be~\cite{WBH+10,wiebe2011simulating}
\begin{equation}
N_{\rm operations} \in \left(N^4 \left[\frac{hN^8}{\min_s \Delta^2(s)}\right]^{1+o(1)} \right),
\end{equation}
where $h\ge \max\{|h_{pq}|,|h_{pqrs}|\}$.
If we take $h\in O(1)$, this rigorous bound suggests that the cost of adiabatic state preparation may be prohibitively expensive even if the eigenvalue gap is constant, it is important to note that this upper bound on the norm of the derivative of $H$ is expected to be extremely loose and the scaling of the error in the Trotter--Suzuki formulas is expected to be much better than the scaling quoted above.

If we take the depth of second-order Trotter--Suzuki formula simulations of real molecules  to scale as $O(N^{5.5})$, as observed in previous studies~\cite{PHW+15}, and take the norm of the Hamiltonian to scale as $O(N^2)$ as anticipated asymptotically for a local basis~\cite{mcclean2014exploiting} then the scaling that arises from using the second--order Trotter formula for time--dependent Hamiltonians~\cite{WBH+10} would be
\begin{equation}
O\left( \frac{h^{3/2}N^{8.5}}{\min_s \Delta^3(s)} \right).
\end{equation}
If $h\in O(N^{-2})$ then this scaling reduces to $N^{5.5}/\min_s \Delta^3(s)$, which is expected if the two body terms dominate the cost of the Trotter--Suzuki decomposition.
This means that even after making empirical assumptions, highly gapped adiabatic paths are likely to be necessary for adiabatic state preparation to be useful. 

It is worth mentioning that adiabatic state preparation has been investigated for other systems and in these settings it has been found to be a highly practical method of state preparation~\cite{WHW+15,BWM+15}.  Although it should be noted that the paths used from the initial Hamiltonian to the final Hamiltonian are often non-trivial.  These observations suggest that the above complexity analysis may be quite loose.  Further work is needed to better estimate the cost of adiabatic state preparation for realistic molecules and also the cost of learning optimal adiabatic paths from easily preparable Hamiltonians to the FCI Hamiltonian.

\section{Cost estimates for topological qubits}

In this section we will examine the impact topological quantum computing may have on these numbers.  This is important because the fault tolerant overheads quoted in the main body depend sensitively on the error rate.  Topological quantum computing promises to provide physical error rates that are orders of magnitude beyond what is achievable in alternative platforms; however there is a catch.  Much of the current research underway focuses on topological quantum computing using Ising anyons, which provide topological protection for Clifford operations but do not provide protection for {\tt T} gates.  This means that even if we achieve very low error rates using this technology then the costs of magic state distillation may not be reduced dramatically despite the quality of the Clifford gates that topological quantum computing affords.  We examine this by modeling the errors in generating the ${\tt T}$ states to be $10^{-4}$ and then consider the costs of using the surface code to distill the necessary gates.  We provide the data for this scenario in Table~\ref{tab:ftcosts}.

The data in Table~\ref{tab:ftcosts} shows that even in this setting assuming low quality magic states does not remove the benefit of topological protection for the Clifford operations.  In particular, if we look at the savings in physical qubits that occur from going from error rate $10^{-6}$ to $10^{-9}$ we see in the data in the main body that roughly an order of magnitude separates the two numbers of physical qubits.  In contrast, if we assume the lower quality magic states given above then the reductions are more modest: they are roughly a factor of $5$.  This shows that while having high fidelity magic states is ideal, even if such states are low quality then a topological quantum computer with protected Clifford operations can nonetheless see advantages from having topologically protected gates at the level of accuracy required to simulate nitrogenase.

\begin{widetext}

\begin{table}
\begin{center}
\begin{tabular}{||l||c|c||c|c||c|c||}
\hline
\hline
& \multicolumn{2}{c||}{\bf Serial rotations} & \multicolumn{2}{c||}{\bf PAR rotations}& \multicolumn{2}{c||}{\bf Nested rotations} \\
\hline 
{\bf Clifford Error Rate}  & $\mathbf{10^{-6}}$ & $\mathbf{10^{-9}}$  & $\mathbf{10^{-6}}$ & $\mathbf{10^{-9}}$ & $\mathbf{10^{-6}}$ & $\mathbf{10^{-9}}$ \\
\hline
Required code distance  & 9,3 & 5,3  & 9,5 & 5,3  &9,3 &5,3 \\
\hline
\hline
&\multicolumn{6}{c||}{\bf Quantum processor} \\
\hline
Logical qubits & \multicolumn{2}{c||}{111}  & \multicolumn{2}{c||}{110} & \multicolumn{2}{c||}{109} \\
\hline
Physical qubits per logical qubit &1013 &313  &1013 &313  &1013 &313 \\
\hline
Total physical qubits for processor  &$1.1\times 10^5$  &$3.5\times 10^{4}$   & $1.1\times 10^5$ & $3.4\times 10^4$ &$1.1\times 10^5$ & $3.4\times 10^4$  \\
\hline
\hline
&\multicolumn{6}{c||}{\bf Discrete Rotation factories} \\
\hline
Number &  \multicolumn{2}{c||}{0}   & \multicolumn{2}{c||}{1872}  & \multicolumn{2}{c||}{26}   \\
\hline
Physical qubits per factory  &-- &-- &  1013 &313 &1013 &313\\
\hline
Total physical qubits for rotations & -- & --  &$1.9\times 10^6$ & $5.9 \times 10^5$ & $2.6\times 10^4$ & $8.1\times 10^3$\\
\hline
\hline
&\multicolumn{6}{c||}{\bf $\mathbf{T}$ factories} \\
\hline
Number &  64 & 30 &   41110 & 23248 & 1427 &813\\
\hline
Physical qubits per factory & $2.7\times 10^4$ & $2.7\times 10^4$   & $7.5\times 10^4$ & $2.7\times 10^4$ & $7.5\times 10^4$ & $2.7\times 10^4$ \\
\hline
Total physical qubits for T factories  & $1.7\times 10^6$ & $8.1\times 10^5$ &  $3.1 \times 10^9$ & $6.4\times 10^8$  &$1.1\times 10^8$ & $2.2\times 10^7$ \\
\hline
\hline
Total physical qubits 	& $1.8 \times 10^6$ &$8.5\times 10^5$ & $3.1 \times 10^{9}$ &	$6.4 \times 10^8$ &$1.1\times 10^8$&	$2.2 \times 10^7$  \\
\hline
\hline
\end{tabular}
\end{center}
\caption{  This table gives the resource requirements including error correction for simulations of nitrogenase's FeMoco in a $54$ (spatial) orbital basis within the times quoted in Table I in the main body using physical gates operating at $100$ MHz. Here we use error rates that are appropriate for quantum computing with Ising anyons, wherein topological protection is granted to Clifford operations but not to non--Clifford rotations.  The error rate used in the production of the raw magic states is taken to be $10^{-4}$ in all of the above cases. \label{tab:ftcosts}}
\end{table}
\end{widetext}

\section{FeMoco --- the active site of nitrogenase}\label{sec:Elucidating}
In this section we provide more background on the importance of biological dinitrogen fixation and on the active site model of nitrogenase prepared in different charge and spin states applied in the feasibility analysis of this work. 

For decades, a Holy Grail in chemistry has been
the catalytic fixation of molecular nitrogen under ambient conditions. Less than half a dozen synthetic catalysts have
been developed for this purpose \cite{schrock,nishiba,peters} (after decades of fruitless efforts). All of them suffer from
low turnover numbers and the synthetic dinitrogen-fixation problem under ambient conditions can thus be considered largely unsolved. The process is of
tremendous importance for society as fertilizers are produced from ammonia, the final product of dinitrogen fixation.
Industrially, ammonia is produced in the very efficient Haber--Bosch process, which, however, requires high temperature and pressure (and consumes
up to 2\% of the annual energy production) \cite{leig02}. 

While it currently appears unrealistic that this simple heterogeneous process will be replaced
by a sophisticated synthetic homogeneous catalyst, which is likely to be less stable and expensive to produce, a mono- or poly-nuclear iron-based catalyst working under ambient conditions and feeding on an easily accessible source for 'hydrogen' (and dinitrogen from air) could become important for local small-scale fertilizer-production concepts. 

In any case, nitrogen fixation represents a tremendous chemical challenge to activate and break the strong triple bond
in dinitrogen at room temperature and pressure. Nature found an efficient way to achieve this goal. It is accomplished by the enzyme 
nitrogenase whose active site, the iron--molybdenum cofactor FeMoco, consists of seven iron atoms and
one molybdenum atom, which are clamped together by bridging sulfur atoms \cite{science_1992_257_1677}. The complete structure of the FeMoco
was solved only very recently when a central main-group atom was discovered \cite{eins02} that, surprisingly, turned out to be a carbon atom \cite{science1,science2,ribbe}.
The complex electronic structure of this cluster of open-shell iron atoms, the possible charge, spin, and protonation states as well as the
different ligand binding sites to be considered makes this active site a nightmare for electronic structure calculations, which is the basis
of all theoretical approaches toward the elucidation of the mode of action of metalloenzymes such as nitrogenase.

It is thus no surprise that the specific mechanism of dinitrogen reduction
at this active site has been elusive, especially given the fact that the mechanism of nitrogenase is difficult to study 
experimentally. Computational approaches 
suffer from the static electron correlation problem. For two ammonia molecules to be produced, the transfer of six protons and six electrons
is required per dinitrogen molecule (in fact, eight protons and eight electrons are
needed as one dihydrogen molecule is produced stoichiometrically). The transfer of these highly reactive agents leads
to many stable intermediates and side products (see, for example, the analogous discussion of these steps in Ref.\ \cite{heurex} for the first
synthetic dinitrogen-fixating complex by Yandulov and Schrock). To elucidate the mechanism of nitrogenase, which is
important for a better understanding of the activation of inert bonds by synthetic catalysts, therefore requires the consideration of a very large
number of molecular structures. 

While molecular structure of stable intermediates and
transition states may be optimized within unrestricted Kohn--Sham DFT, the calculation of their energies demands an accurate wave-function-based approach.
We therefore optimize molecular structures of a FeMoco model in the 
resting structure (Fig. 1 (right) in the main article)
for varying charge and spin states in order to create different electronic situations that challenge the 
feasibility analysis presented in this work. For these structures, integrals in a molecular orbital basis have been obtained that parametrize
the electronic structure of the cluster in the second-quantized quantum-chemical Hamiltonian of Eq.\ (\ref{eq:Hamiltonian}).

\section{Exact diagonalization techniques in chemistry}

The electronic structure of a molecular structure determines its reactivity. Predicting chemical reactions
requires the solution of the electronic Schr\"odinger equation to obtain the electronic energy and wave function.
Whereas the expansion of the many-electron wave function into a (quasi-) complete many-electron basis will produce the exact solution,
called full configuration interaction in chemistry or exact diagonalization in physics, this approach is
unfeasible for molecules of more than a few atoms. As all standard quantum-chemical solution approaches construct
many-electron basis functions from orbitals, the number of the former is determined by the number of the latter.
Exact diagonalization is therefore limited to about 18 electrons distributed among 18 spatial orbitals due to the
exponential scaling of the many-electron basis states with the number of orbitals \cite{molcas8}.
Unfortunately, the size of an orbital basis is already very large for moderately sized molecules.
As a consequence, a restricted orbital space must be chosen.

Of all approximations developed in quantum chemistry to overcome this problem \cite{helgaker} the complete-active-space
self-consistent-field (CASSCF) approach (and related models) utilizes exact diagonalization, but, because of the exponential scaling,
in a reduced orbital space, the so-called CAS, that selects orbitals around the Fermi energy.
To compensate for this approximation, the orbitals are relaxed self consistently,
hence the name. Still, the CAS is limited by the 18-orbital wall and by the neglect of most of the (virtual) orbitals for the
construction of the wave function. Considering the fact that molecules of
a hundred atoms or more quickly require much more than a thousand one-electron basis functions for an accurate description of their electronic
structure, most of the orbitals constructed from these basis functions are omitted in the construction of a CASSCF wave function.
Even iterative techniques such as the density matrix renormalization group (DMRG), which can be understood as a polynomially
scaling CASSCF approach, can push this wall only to about a hundred spatial orbitals.

As a result, a CASSCF-type wave function solves only the so-called static electron correlation problem. It is therefore particularly well suited
for molecular structures with near-degenerate orbitals. The resulting
electronic structure is only qualitatively well described. However, this feature is maintained throughout a reaction coordinate,
which makes a CASSCF-type approach an appealing universal approach.
For a quantitative description, the many virtual orbitals not considered for the CAS
make a nonnegligible contribution to electronic-energy differences. To account for this so-called dynamic correlation is mandatory and typically
achieved by a subsequent multi-reference perturbation theory calculation. Such perturbation theories to second order require elements of
the three- and four-electron reduced density matrices, which are difficult to calculate and to store.

Instead of this 'diagonalize, then perturb' approach, also 'perturb, then diagonalize' ideas have been studied in
chemistry. These latter approaches produce a 'dressed' many-electron Hamiltonian which is then
better conditioned for a CASSCF-type approach. A new development in this area is the combination of density
functional theory (DFT), known to describe dynamic electron correlation well 
with a CASSCF-type approach by spatial range separation introduced by Savin (see Refs.\ \cite{jensen,hedegard} and references cited therein). 
Range separation is accomplished for the electron--electron interaction matrix elements
and allows one to apply DFT at short range, whereas at long range the full flexibility of a CASSCF-type wave
function can be exploited. This approach is very efficient and does not compromise the efficiency of a CASSCF-type
approach (by contrast to perturbation theory). In combination with the polynomially scaling DMRG, this approach has delivered
very promising results \cite{hedegard} for benchmark reactions involving transition metals \cite{wccr10}.

As a consequence, a chemically sensible exact diagonalization technique is a CASSCF-type approach that captures
dynamic correlation effects in the one-electron states. Such an approach could be directly implemented on a quantum
computer with the specific exact diagonalization technology developed for such a machine.

\begin{table*}[t!]
\begin{tabular}{|l|c|r||c|c|c|c|}
\hline
Structure & \multicolumn{2}{c||}{Structure opt./B3LYP} & \multicolumn{4}{c|}{small-CAS CASSCF orbitals} \\
          & Total spin $S$ & Charge & Act.\ electrons & Act.\ orbitals & Total spin $S$ & Charge \\
\hline
{\bf 1} &  0 & 3 & 54 & 54 & 0 & 3  \\
{\bf 2} &  $1/2$ & 0 & 65 & 57 & $3/2$ & 0\\

\hline
\end{tabular}

\caption{The structures optimized for FeMoco and the settings for the small-CAS CASSCF orbital optimization. 
$S$ is the total spin quantum number and the charge is measured in units of the elementary charge.
}
\label{tab:configs}
\end{table*}
\begin{widetext}
\begin{table}
\begin{tabular}{|c|ccc|ccc|}
\hline
Atom & \multicolumn{3}{c|}{Coordinates Structure {\bf 1}} \\
\hline
S&   0.193509  & -1.756174 & -6.077728 \\
FE&  -0.073029 &  0.147462 & -7.060398 \\
S & -0.155670 & -0.304014 & -9.055299 \\
FE & -1.194342 & -0.633669 & -4.857848 \\
C  & 0.097884 &  0.293121 & -3.822079 \\
FE&   0.330922 &  1.682738 & -2.639449 \\
S  & 0.144858 &  3.427365 & -3.756994 \\
FE  & 0.000866 &  1.806617  &-5.056623 \\
S &  -1.759709  & 1.058486 & -6.097784 \\
FE &  1.393308 & -0.388901 & -4.927653 \\
S  & 2.935549 & -1.083502 & -3.683191 \\
FE  & 1.581418 & -0.451806 & -2.290305 \\
S  & 1.643565  & 1.242708 & -0.922047 \\
MO & -0.046077 & -0.172415  & 0.259925 \\
O  &-0.081763 &  1.100331 &  1.760010 \\
FE & -0.867802 & -0.634367 & -2.449654 \\
S & -1.333714 &  1.160052 & -1.286619 \\
S  &-2.410505 & -1.577384 & -3.437041 \\
S   &1.630797 &  1.240870 & -6.345822 \\
S  & 0.353345 & -1.960563 & -1.213579 \\
N  & 1.652155 & -0.910661 &  1.409953 \\
O & -1.541962 & -0.693648 &  1.167835 \\
C & -2.111338 & -0.028674 &  2.280713 \\
C & -0.152453 &  1.144086 &-10.129068 \\
C & -1.083149 &  1.057101 &  2.683832 \\
H  &-3.060803 &  0.437015 &  2.003587 \\
H & -2.276775 & -0.745389 &  3.087505 \\
O & -1.167081 &  1.740174 &  3.642482 \\
H &  0.625836 &  0.971552 &-10.878592 \\
H & -1.119129 &  1.154099 &-10.642807 \\
H &  0.015059 &  2.077284 & -9.598730 \\
C &  2.339711 & -0.178076 &  2.302234 \\
N &  3.244459 & -0.944382 &  2.902881 \\
C &  3.160159 & -2.227901 &  2.407388 \\
C &  2.176178 & -2.202712 &  1.475167 \\
H &  1.794821 & -3.019854 &  0.890693 \\
H &  3.781670 & -3.031530 &  2.764468 \\
H &  3.882526 & -0.630293 &  3.624464 \\
H &  2.182421 &  0.864873 &  2.517705 \\
\hline
\end{tabular}

\caption{Coordinates for Structure {\bf 1} of FeMoco in \AA.}
\label{tab:coord1}
\end{table}
\begin{table}
\begin{tabular}{|c|ccc|ccc|}
\hline
Atom & \multicolumn{3}{c|}{Coordinates Structure {\bf 2} } \\
\hline
S  & 0.032866 & -2.093214 & -6.099450 \\
FE &  0.017622 & -0.056816 & -7.223052 \\
S &  0.143644 & -0.401241 & -9.315852 \\
FE & -1.471854 & -0.721074 & -4.983076 \\
C & -0.152873 &  0.225677 & -3.664235 \\
FE & -0.118135  & 1.845485&  -2.405214 \\
S & -0.036661 &  3.550618 & -3.771677 \\
FE & -0.049780  & 1.771611&  -5.047025 \\
S & -1.760858 &  1.077437 & -6.479794 \\
FE &  1.306370 & -0.657731 & -4.880018 \\
S &  2.846629 & -1.294232 & -3.368478 \\
FE &  1.196355 & -0.482357&  -2.281465 \\
S &  1.566535 &  1.288016 & -0.916398 \\
MO & -0.186653 & -0.006921 &  0.115165 \\
O & -0.140683  & 1.329614 &  1.698525 \\
FE & -1.529490 & -0.648712 & -2.302651 \\
S & -1.922731  & 1.225686 & -0.982284 \\
S & -3.001156  &-1.537533 & -3.665135 \\
S &  1.731662  & 1.109781 & -6.342619 \\
S & -0.092449 & -1.984660 & -1.157533 \\
N &  1.598971 & -0.845206 &  1.346832 \\
O & -1.365000 & -0.810743 &  1.369699 \\
C & -1.748263 & -0.161994 &  2.555155 \\
C &  0.321553  & 1.195240& -10.180658 \\
C & -0.891249 &  1.088353 &  2.753014 \\
H & -2.802446 &  0.128482 &  2.480632 \\
H & -1.641761 & -0.837543 &  3.407640 \\
O & -0.904893 &  1.741420 &  3.768510 \\
H &  0.360712 &  0.987604& -11.249659 \\
H & -0.529456  & 1.838638 & -9.965894 \\
H &  1.243686  & 1.686635 & -9.874888 \\
C &  2.191192 & -0.163147 &  2.313386 \\
N &  3.212741 & -0.875946 &  2.820353 \\
C &  3.284255 & -2.074995 &  2.147363 \\
C &  2.276098  &-2.042456 &  1.234236 \\
H  & 1.996684  &-2.787079 &  0.511609 \\
H  & 4.022709 & -2.823292 &  2.371900 \\
H &  3.814462 & -0.573873 &  3.568856 \\
H &  1.907357 &  0.820486  & 2.645710 \\
\hline
\end{tabular}

\caption{Coordinates for Structure {\bf 2} of FeMoco in \AA.}
\label{tab:coord2}
\end{table}
\end{widetext}

What chemical problems will benefit from such an implementation? Clearly, these will be problems that are dominated
by strong static electron correlation (rather than by weak dispersion interactions originating from dynamic electron correlation).
The electronic Schr\"odinger equation assigns an energy to
a given molecular structure in the Born--Oppenheimer approximation, and this electronic energy (evaluated at zero
Kelvin without vibrational, temperature, and entropy corrections) should dominate the energy change of a chemical process.
Transition-metal catalysis is a field that presents such situations.

Many important chemical transformations are mediated by complicated electronic
structures featured by transition-metal complexes. Especially late 3$d$ transition metals are of this kind, most
importantly iron, which is cheap and ubiquitous, often yields nontoxic compounds, and therefore represents an ideal catalytic center. Moreover, stable
intermediates and, in particular, transition states of a reaction mechanism often represent typical static electron
correlation problems as bonds are formed and broken on the way to stable products. To reliably predict a chemical
transformation of this kind usually requires to study many more than one elementary reaction step --- especially
when reactive intermediates are involved. The number of stable intermediates also increases due to unwanted side
reactions that need to be inspected.
Therefore, the total number of molecular structures whose electronic energy is required for
an understanding of a reaction mechanism is, in general, very large.

Clearly, all these structures must be optimized with an
efficient quantum chemical method. While the accuracy of electronic energies obtained with present-day DFT approaches
is often not satisfactory for predictive purposes (see, e.g., Refs.\ \cite{wccr10,HallHyd}),
molecular structures can be reproduced with remarkable accuracy (obviously, a geometry gradient would replace DFT structures, too).
Hence, CASSCF-type methods will mostly be required for the validation of electronic energies of DFT-optimized molecular
structures, which is the essential piece of information for establishing a reaction mechanism.

\section{Computational Methodology}\label{sec:methodology}

We optimized the structure of a FeMoco resting-state model that takes those residues of the protein backbone into account which anchor 
the metal cluster in the enzyme (see Fig.~1 (right) in the main article). 
Note that different spin and charge states were considered for the structure optimization
in order to obtain a variety of electronic structures for the assessment of a solution algorithm on a quantum computer. 
These spin and charge states do not necessarily match the one of the resting state of nitrogenase. 

Structure {\bf 1} for three positive excess charges and an equal number of $\alpha$- and $\beta$-spins, and
structure {\bf 2} for an uncharged FeMoco model with one unpaired $\alpha$-spin. 

The Cartesian coordinates of these structures are collected in Tables \ref{tab:coord1} and \ref{tab:coord2}.
In these unrestricted DFT calculations, the spin symmetry
was broken \cite{jaco12} and only $S_z$ remains as a good quantum number. For these structure optimizations we chose the {\sc Turbomole} program package
(V6.4) \cite{turbomole} and employed the B3LYP density functional \cite{lyp,becke88,becke93,stephens} with the def2-TZVP Ahlrichs triple-zeta 
basis set plus polarization functions on all atoms \cite{phys_chem_chem_phys_2005_7_3297}. 
An effective core potential was chosen only for the molybdenum atom \cite{theor_chim_acta_1990_77_123},
which also takes care of all scalar-relativistic effects on this heavy atom. 

Then, integrals in the molecular orbital (MO) basis were produced 
MO integrals for structure {\bf 1} were generated for a CAS of 54 electrons in 54 spatial CASSCF orbitals (108 spin orbitals), which were obtained from a singlet CASSCF calculation with 24 electrons in 16 orbitals.
Accordingly, 
for structure {\bf 2} we generated MO integrals for a CAS of 65 electrons in 57 spatial CASSCF orbitals  (114 spin orbitals) from a quartet CASSCF calculation with 21 electrons in 12 orbitals.

The choice of the large active spaces was based on Pulay's UNO-CAS criterion \cite{pulay,bogus15}, in which
the occupation number of the natural orbitals serves as a selection criterion. However, rather than unrestricted Hartree--Fock natural orbitals,
we selected those small-CAS CASSCF natural orbitals in the occupation intervals [1.98,0.02] and [1.99,0.01], respectively.
The molecular orbital integrals for the second-quantized electronic Hamiltonian in these small-CAS orbital bases were calculated with the
{\sc Molcas} program \cite{molcas8}.

All settings for the calculations are summarized in Table \ref{tab:configs}.


\begin{thebibliography}{99}%
\makeatletter
\providecommand \@ifxundefined [1]{%
 \@ifx{#1\undefined}
}%
\providecommand \@ifnum [1]{%
 \ifnum #1\expandafter \@firstoftwo
 \else \expandafter \@secondoftwo
 \fi
}%
\providecommand \@ifx [1]{%
 \ifx #1\expandafter \@firstoftwo
 \else \expandafter \@secondoftwo
 \fi
}%
\providecommand \natexlab [1]{#1}%
\providecommand \enquote  [1]{``#1''}%
\providecommand \bibnamefont  [1]{#1}%
\providecommand \bibfnamefont [1]{#1}%
\providecommand \citenamefont [1]{#1}%
\providecommand \href@noop [0]{\@secondoftwo}%
\providecommand \href [0]{\begingroup \@sanitize@url \@href}%
\providecommand \@href[1]{\@@startlink{#1}\@@href}%
\providecommand \@@href[1]{\endgroup#1\@@endlink}%
\providecommand \@sanitize@url [0]{\catcode `\\12\catcode `\$12\catcode
  `\&12\catcode `\#12\catcode `\^12\catcode `\_12\catcode `\%12\relax}%
\providecommand \@@startlink[1]{}%
\providecommand \@@endlink[0]{}%
\providecommand \url  [0]{\begingroup\@sanitize@url \@url }%
\providecommand \@url [1]{\endgroup\@href {#1}{\urlprefix }}%
\providecommand \urlprefix  [0]{URL }%
\providecommand \Eprint [0]{\href }%
\providecommand \doibase [0]{http://dx.doi.org/}%
\providecommand \selectlanguage [0]{\@gobble}%
\providecommand \bibinfo  [0]{\@secondoftwo}%
\providecommand \bibfield  [0]{\@secondoftwo}%
\providecommand \translation [1]{[#1]}%
\providecommand \BibitemOpen [0]{}%
\providecommand \bibitemStop [0]{}%
\providecommand \bibitemNoStop [0]{.\EOS\space}%
\providecommand \EOS [0]{\spacefactor3000\relax}%
\providecommand \BibitemShut  [1]{\csname bibitem#1\endcsname}%
\let\auto@bib@innerbib\@empty
\bibitem [{\citenamefont {Dykstra}\ \emph {et~al.}(2005)\citenamefont
  {Dykstra}, \citenamefont {Frenking}, \citenamefont {Kim},\ and\ \citenamefont
  {Scuseria}}]{scuseriafrenking}%
  \BibitemOpen
  \bibfield  {author} {\bibinfo {author} {\bibfnamefont {Clifford}\
  \bibnamefont {Dykstra}}, \bibinfo {author} {\bibfnamefont {Gernot}\
  \bibnamefont {Frenking}}, \bibinfo {author} {\bibfnamefont {Kwang~S.}\
  \bibnamefont {Kim}}, \ and\ \bibinfo {author} {\bibfnamefont {Gustavo~E.}\
  \bibnamefont {Scuseria}},\ }\href@noop {} {\emph {\bibinfo {title} {Theory
  and Applications of Computational Chemistry: The First Forty Years}}}\
  (\bibinfo  {publisher} {Elsevier},\ \bibinfo {address} {Amsterdam},\ \bibinfo
  {year} {2005})\BibitemShut {NoStop}%
\bibitem [{\citenamefont {Cramer}\ and\ \citenamefont
  {Truhlar}(2009)}]{phys_chem_chem_phys_2009_11_10757}%
  \BibitemOpen
  \bibfield  {author} {\bibinfo {author} {\bibfnamefont {Christopher~J.}\
  \bibnamefont {Cramer}}\ and\ \bibinfo {author} {\bibfnamefont {Donald~G.}\
  \bibnamefont {Truhlar}},\ }\bibfield  {title} {\enquote {\bibinfo {title}
  {{Density Functional Theory for Transition Metals and Transition Metal
  Chemistry}},}\ }\href@noop {} {\bibfield  {journal} {\bibinfo  {journal}
  {Phys. Chem. Chem. Phys.}\ }\textbf {\bibinfo {volume} {11}},\ \bibinfo
  {pages} {10757--10816} (\bibinfo {year} {2009})}\BibitemShut {NoStop}%
\bibitem [{\citenamefont {Jiang}\ \emph {et~al.}(2012)\citenamefont {Jiang},
  \citenamefont {DeYonker}, \citenamefont {Determan},\ and\ \citenamefont
  {Wilson}}]{angelawilson}%
  \BibitemOpen
  \bibfield  {author} {\bibinfo {author} {\bibfnamefont {Wanyi}\ \bibnamefont
  {Jiang}}, \bibinfo {author} {\bibfnamefont {Nathan~J.}\ \bibnamefont
  {DeYonker}}, \bibinfo {author} {\bibfnamefont {John~J.}\ \bibnamefont
  {Determan}}, \ and\ \bibinfo {author} {\bibfnamefont {Angela~K.}\
  \bibnamefont {Wilson}},\ }\bibfield  {title} {\enquote {\bibinfo {title}
  {Toward accurate theoretical thermochemistry of first row transition metal
  complexes},}\ }\href@noop {} {\bibfield  {journal} {\bibinfo  {journal} {J.
  Phys. Chem. A}\ }\textbf {\bibinfo {volume} {116}},\ \bibinfo {pages}
  {870--885} (\bibinfo {year} {2012})}\BibitemShut {NoStop}%
\bibitem [{\citenamefont {Weymuth}\ \emph {et~al.}(2014)\citenamefont
  {Weymuth}, \citenamefont {Couzijn}, \citenamefont {Chen},\ and\ \citenamefont
  {Reiher}}]{wccr10}%
  \BibitemOpen
  \bibfield  {author} {\bibinfo {author} {\bibfnamefont {Thomas}\ \bibnamefont
  {Weymuth}}, \bibinfo {author} {\bibfnamefont {Erik P.~A.}\ \bibnamefont
  {Couzijn}}, \bibinfo {author} {\bibfnamefont {Peter}\ \bibnamefont {Chen}}, \
  and\ \bibinfo {author} {\bibfnamefont {Markus}\ \bibnamefont {Reiher}},\
  }\bibfield  {title} {\enquote {\bibinfo {title} {New benchmark set of
  transition-metal coordination reactions for the assessment of density
  functionals},}\ }\href@noop {} {\bibfield  {journal} {\bibinfo  {journal} {J.
  Chem. Theory Comput.}\ }\textbf {\bibinfo {volume} {10}},\ \bibinfo {pages}
  {3092--3103} (\bibinfo {year} {2014})}\BibitemShut {NoStop}%
\bibitem [{\citenamefont {Lloyd}(1996)}]{Llo96}%
  \BibitemOpen
  \bibfield  {author} {\bibinfo {author} {\bibfnamefont {Seth}\ \bibnamefont
  {Lloyd}},\ }\bibfield  {title} {\enquote {\bibinfo {title} {Universal quantum
  simulators},}\ }\href@noop {} {\bibfield  {journal} {\bibinfo  {journal}
  {Science}\ }\textbf {\bibinfo {volume} {273}},\ \bibinfo {pages} {1073}
  (\bibinfo {year} {1996})}\BibitemShut {NoStop}%
\bibitem [{\citenamefont {Zalka}(1998)}]{Zal98}%
  \BibitemOpen
  \bibfield  {author} {\bibinfo {author} {\bibfnamefont {Christof}\
  \bibnamefont {Zalka}},\ }\bibfield  {title} {\enquote {\bibinfo {title}
  {Simulating quantum systems on a quantum computer},}\ }in\ \href@noop {}
  {\emph {\bibinfo {booktitle} {Proceedings of the Royal Society of London A:
  Mathematical, Physical and Engineering Sciences}}},\ Vol.\ \bibinfo {volume}
  {454}\ (\bibinfo {organization} {The Royal Society},\ \bibinfo {year}
  {1998})\ pp.\ \bibinfo {pages} {313--322}\BibitemShut {NoStop}%
\bibitem [{\citenamefont {Lidar}\ and\ \citenamefont {Wang}(1999)}]{LW99}%
  \BibitemOpen
  \bibfield  {author} {\bibinfo {author} {\bibfnamefont {Daniel~A}\
  \bibnamefont {Lidar}}\ and\ \bibinfo {author} {\bibfnamefont {Haobin}\
  \bibnamefont {Wang}},\ }\bibfield  {title} {\enquote {\bibinfo {title}
  {Calculating the thermal rate constant with exponential speedup on a quantum
  computer},}\ }\href@noop {} {\bibfield  {journal} {\bibinfo  {journal}
  {Physical Review E}\ }\textbf {\bibinfo {volume} {59}},\ \bibinfo {pages}
  {2429} (\bibinfo {year} {1999})}\BibitemShut {NoStop}%
\bibitem [{\citenamefont {Wang}\ \emph {et~al.}(2008)\citenamefont {Wang},
  \citenamefont {Kais}, \citenamefont {Aspuru-Guzik},\ and\ \citenamefont
  {Hoffmann}}]{WKA+08}%
  \BibitemOpen
  \bibfield  {author} {\bibinfo {author} {\bibfnamefont {Hefeng}\ \bibnamefont
  {Wang}}, \bibinfo {author} {\bibfnamefont {Sabre}\ \bibnamefont {Kais}},
  \bibinfo {author} {\bibfnamefont {Al{\'a}n}\ \bibnamefont {Aspuru-Guzik}}, \
  and\ \bibinfo {author} {\bibfnamefont {Mark~R}\ \bibnamefont {Hoffmann}},\
  }\bibfield  {title} {\enquote {\bibinfo {title} {Quantum algorithm for
  obtaining the energy spectrum of molecular systems},}\ }\href@noop {}
  {\bibfield  {journal} {\bibinfo  {journal} {Physical Chemistry Chemical
  Physics}\ }\textbf {\bibinfo {volume} {10}},\ \bibinfo {pages} {5388--5393}
  (\bibinfo {year} {2008})}\BibitemShut {NoStop}%
\bibitem [{\citenamefont {Lanyon}\ \emph {et~al.}(2010)\citenamefont {Lanyon},
  \citenamefont {Whitfield}, \citenamefont {Gillett}, \citenamefont {Goggin},
  \citenamefont {Almeida}, \citenamefont {Kassal}, \citenamefont {Biamonte},
  \citenamefont {Mohseni}, \citenamefont {Powell}, \citenamefont {Barbieri},\
  and\ \citenamefont {White}}]{LWG+10}%
  \BibitemOpen
  \bibfield  {author} {\bibinfo {author} {\bibfnamefont {Benjamin~P}\
  \bibnamefont {Lanyon}}, \bibinfo {author} {\bibfnamefont {James~D}\
  \bibnamefont {Whitfield}}, \bibinfo {author} {\bibfnamefont {Geoff~G}\
  \bibnamefont {Gillett}}, \bibinfo {author} {\bibfnamefont {Michael~E}\
  \bibnamefont {Goggin}}, \bibinfo {author} {\bibfnamefont {Marcelo~P}\
  \bibnamefont {Almeida}}, \bibinfo {author} {\bibfnamefont {Ivan}\
  \bibnamefont {Kassal}}, \bibinfo {author} {\bibfnamefont {Jacob~D}\
  \bibnamefont {Biamonte}}, \bibinfo {author} {\bibfnamefont {Masoud}\
  \bibnamefont {Mohseni}}, \bibinfo {author} {\bibfnamefont {Ben~J}\
  \bibnamefont {Powell}}, \bibinfo {author} {\bibfnamefont {Marco}\
  \bibnamefont {Barbieri}}, \ and\ \bibinfo {author} {\bibfnamefont {Andrew~G}\
  \bibnamefont {White}},\ }\bibfield  {title} {\enquote {\bibinfo {title}
  {Towards quantum chemistry on a quantum computer},}\ }\href@noop {}
  {\bibfield  {journal} {\bibinfo  {journal} {Nature Chemistry}\ }\textbf
  {\bibinfo {volume} {2}},\ \bibinfo {pages} {106--111} (\bibinfo {year}
  {2010})}\BibitemShut {NoStop}%
\bibitem [{\citenamefont {Kassal}\ \emph {et~al.}(2011)\citenamefont {Kassal},
  \citenamefont {Whitfield}, \citenamefont {Perdomo-Ortiz}, \citenamefont
  {Yung},\ and\ \citenamefont {Aspuru-Guzik}}]{KWP+11}%
  \BibitemOpen
  \bibfield  {author} {\bibinfo {author} {\bibfnamefont {Ivan}\ \bibnamefont
  {Kassal}}, \bibinfo {author} {\bibfnamefont {James~D}\ \bibnamefont
  {Whitfield}}, \bibinfo {author} {\bibfnamefont {Alejandro}\ \bibnamefont
  {Perdomo-Ortiz}}, \bibinfo {author} {\bibfnamefont {Man-Hong}\ \bibnamefont
  {Yung}}, \ and\ \bibinfo {author} {\bibfnamefont {Al{\'a}n}\ \bibnamefont
  {Aspuru-Guzik}},\ }\bibfield  {title} {\enquote {\bibinfo {title} {Simulating
  chemistry using quantum computers},}\ }\href@noop {} {\bibfield  {journal}
  {\bibinfo  {journal} {Annual Review of Physical Chemistry}\ }\textbf
  {\bibinfo {volume} {62}},\ \bibinfo {pages} {185--207} (\bibinfo {year}
  {2011})}\BibitemShut {NoStop}%
\bibitem [{\citenamefont {Whitfield}\ \emph {et~al.}(2011)\citenamefont
  {Whitfield}, \citenamefont {Biamonte},\ and\ \citenamefont
  {Aspuru-Guzik}}]{WBA11}%
  \BibitemOpen
  \bibfield  {author} {\bibinfo {author} {\bibfnamefont {James~D}\ \bibnamefont
  {Whitfield}}, \bibinfo {author} {\bibfnamefont {Jacob}\ \bibnamefont
  {Biamonte}}, \ and\ \bibinfo {author} {\bibfnamefont {Al{\'a}n}\ \bibnamefont
  {Aspuru-Guzik}},\ }\bibfield  {title} {\enquote {\bibinfo {title} {Simulation
  of electronic structure hamiltonians using quantum computers},}\ }\href@noop
  {} {\bibfield  {journal} {\bibinfo  {journal} {Molecular Physics}\ }\textbf
  {\bibinfo {volume} {109}},\ \bibinfo {pages} {735--750} (\bibinfo {year}
  {2011})}\BibitemShut {NoStop}%
\bibitem [{\citenamefont {Jones}\ \emph {et~al.}(2012)\citenamefont {Jones},
  \citenamefont {Whitfield}, \citenamefont {McMahon}, \citenamefont {Yung},
  \citenamefont {Van~Meter}, \citenamefont {Aspuru-Guzik},\ and\ \citenamefont
  {Yamamoto}}]{JWM+12}%
  \BibitemOpen
  \bibfield  {author} {\bibinfo {author} {\bibfnamefont {N~Cody}\ \bibnamefont
  {Jones}}, \bibinfo {author} {\bibfnamefont {James~D}\ \bibnamefont
  {Whitfield}}, \bibinfo {author} {\bibfnamefont {Peter~L}\ \bibnamefont
  {McMahon}}, \bibinfo {author} {\bibfnamefont {Man-Hong}\ \bibnamefont
  {Yung}}, \bibinfo {author} {\bibfnamefont {Rodney}\ \bibnamefont
  {Van~Meter}}, \bibinfo {author} {\bibfnamefont {Al{\'a}n}\ \bibnamefont
  {Aspuru-Guzik}}, \ and\ \bibinfo {author} {\bibfnamefont {Yoshihisa}\
  \bibnamefont {Yamamoto}},\ }\bibfield  {title} {\enquote {\bibinfo {title}
  {Faster quantum chemistry simulation on fault-tolerant quantum computers},}\
  }\href@noop {} {\bibfield  {journal} {\bibinfo  {journal} {New Journal of
  Physics}\ }\textbf {\bibinfo {volume} {14}},\ \bibinfo {pages} {115023}
  (\bibinfo {year} {2012})}\BibitemShut {NoStop}%
\bibitem [{\citenamefont {Peruzzo}\ \emph {et~al.}(2014)\citenamefont
  {Peruzzo}, \citenamefont {McClean}, \citenamefont {Shadbolt}, \citenamefont
  {Yung}, \citenamefont {Zhou}, \citenamefont {Love}, \citenamefont
  {Aspuru-Guzik},\ and\ \citenamefont {O’Brien}}]{PMS+14}%
  \BibitemOpen
  \bibfield  {author} {\bibinfo {author} {\bibfnamefont {Alberto}\ \bibnamefont
  {Peruzzo}}, \bibinfo {author} {\bibfnamefont {Jarrod}\ \bibnamefont
  {McClean}}, \bibinfo {author} {\bibfnamefont {Peter}\ \bibnamefont
  {Shadbolt}}, \bibinfo {author} {\bibfnamefont {Man-Hong}\ \bibnamefont
  {Yung}}, \bibinfo {author} {\bibfnamefont {Xiao-Qi}\ \bibnamefont {Zhou}},
  \bibinfo {author} {\bibfnamefont {Peter~J}\ \bibnamefont {Love}}, \bibinfo
  {author} {\bibfnamefont {Al{\'a}n}\ \bibnamefont {Aspuru-Guzik}}, \ and\
  \bibinfo {author} {\bibfnamefont {Jeremy~L}\ \bibnamefont {O’Brien}},\
  }\bibfield  {title} {\enquote {\bibinfo {title} {A variational eigenvalue
  solver on a photonic quantum processor},}\ }\href@noop {} {\bibfield
  {journal} {\bibinfo  {journal} {Nature communications}\ }\textbf {\bibinfo
  {volume} {5}} (\bibinfo {year} {2014})}\BibitemShut {NoStop}%
\bibitem [{\citenamefont {Wecker}\ \emph {et~al.}(2014)\citenamefont {Wecker},
  \citenamefont {Bauer}, \citenamefont {Clark}, \citenamefont {Hastings},\ and\
  \citenamefont {Troyer}}]{WBC+14}%
  \BibitemOpen
  \bibfield  {author} {\bibinfo {author} {\bibfnamefont {Dave}\ \bibnamefont
  {Wecker}}, \bibinfo {author} {\bibfnamefont {Bela}\ \bibnamefont {Bauer}},
  \bibinfo {author} {\bibfnamefont {Bryan~K}\ \bibnamefont {Clark}}, \bibinfo
  {author} {\bibfnamefont {Matthew~B}\ \bibnamefont {Hastings}}, \ and\
  \bibinfo {author} {\bibfnamefont {Matthias}\ \bibnamefont {Troyer}},\
  }\bibfield  {title} {\enquote {\bibinfo {title} {Gate-count estimates for
  performing quantum chemistry on small quantum computers},}\ }\href@noop {}
  {\bibfield  {journal} {\bibinfo  {journal} {Physical Review A}\ }\textbf
  {\bibinfo {volume} {90}},\ \bibinfo {pages} {022305} (\bibinfo {year}
  {2014})}\BibitemShut {NoStop}%
\bibitem [{\citenamefont {McClean}\ \emph
  {et~al.}(2014{\natexlab{a}})\citenamefont {McClean}, \citenamefont {Babbush},
  \citenamefont {Love},\ and\ \citenamefont {Aspuru-Guzik}}]{MBL+14}%
  \BibitemOpen
  \bibfield  {author} {\bibinfo {author} {\bibfnamefont {Jarrod~R}\
  \bibnamefont {McClean}}, \bibinfo {author} {\bibfnamefont {Ryan}\
  \bibnamefont {Babbush}}, \bibinfo {author} {\bibfnamefont {Peter~J}\
  \bibnamefont {Love}}, \ and\ \bibinfo {author} {\bibfnamefont {Al{\'a}n}\
  \bibnamefont {Aspuru-Guzik}},\ }\bibfield  {title} {\enquote {\bibinfo
  {title} {Exploiting locality in quantum computation for quantum chemistry},}\
  }\href@noop {} {\bibfield  {journal} {\bibinfo  {journal} {The journal of
  physical chemistry letters}\ }\textbf {\bibinfo {volume} {5}},\ \bibinfo
  {pages} {4368--4380} (\bibinfo {year} {2014}{\natexlab{a}})}\BibitemShut
  {NoStop}%
\bibitem [{\citenamefont {Hastings}\ \emph {et~al.}(2015)\citenamefont
  {Hastings}, \citenamefont {Wecker}, \citenamefont {Bauer},\ and\
  \citenamefont {Troyer}}]{HWB15}%
  \BibitemOpen
  \bibfield  {author} {\bibinfo {author} {\bibfnamefont {Matthew~B}\
  \bibnamefont {Hastings}}, \bibinfo {author} {\bibfnamefont {Dave}\
  \bibnamefont {Wecker}}, \bibinfo {author} {\bibfnamefont {Bela}\ \bibnamefont
  {Bauer}}, \ and\ \bibinfo {author} {\bibfnamefont {Matthias}\ \bibnamefont
  {Troyer}},\ }\bibfield  {title} {\enquote {\bibinfo {title} {Improving
  quantum algorithms for quantum chemistry},}\ }\href@noop {} {\bibfield
  {journal} {\bibinfo  {journal} {Quantum Information \& Computation}\ }\textbf
  {\bibinfo {volume} {15}},\ \bibinfo {pages} {1--21} (\bibinfo {year}
  {2015})}\BibitemShut {NoStop}%
\bibitem [{\citenamefont {Poulin}\ \emph {et~al.}(2015)\citenamefont {Poulin},
  \citenamefont {Hastings}, \citenamefont {Wecker}, \citenamefont {Wiebe},
  \citenamefont {Doberty},\ and\ \citenamefont {Troyer}}]{PHW+15}%
  \BibitemOpen
  \bibfield  {author} {\bibinfo {author} {\bibfnamefont {David}\ \bibnamefont
  {Poulin}}, \bibinfo {author} {\bibfnamefont {Matthew~B}\ \bibnamefont
  {Hastings}}, \bibinfo {author} {\bibfnamefont {Dave}\ \bibnamefont {Wecker}},
  \bibinfo {author} {\bibfnamefont {Nathan}\ \bibnamefont {Wiebe}}, \bibinfo
  {author} {\bibfnamefont {Andrew~C}\ \bibnamefont {Doberty}}, \ and\ \bibinfo
  {author} {\bibfnamefont {Matthias}\ \bibnamefont {Troyer}},\ }\bibfield
  {title} {\enquote {\bibinfo {title} {The trotter step size required for
  accurate quantum simulation of quantum chemistry},}\ }\href@noop {}
  {\bibfield  {journal} {\bibinfo  {journal} {Quantum Information \&
  Computation}\ }\textbf {\bibinfo {volume} {15}},\ \bibinfo {pages} {361--384}
  (\bibinfo {year} {2015})}\BibitemShut {NoStop}%
\bibitem [{\citenamefont {Wecker}\ \emph
  {et~al.}(2015{\natexlab{a}})\citenamefont {Wecker}, \citenamefont
  {Hastings},\ and\ \citenamefont {Troyer}}]{WHT15}%
  \BibitemOpen
  \bibfield  {author} {\bibinfo {author} {\bibfnamefont {Dave}\ \bibnamefont
  {Wecker}}, \bibinfo {author} {\bibfnamefont {Matthew~B}\ \bibnamefont
  {Hastings}}, \ and\ \bibinfo {author} {\bibfnamefont {Matthias}\ \bibnamefont
  {Troyer}},\ }\bibfield  {title} {\enquote {\bibinfo {title} {Progress towards
  practical quantum variational algorithms},}\ }\href@noop {} {\bibfield
  {journal} {\bibinfo  {journal} {Physical Review A}\ }\textbf {\bibinfo
  {volume} {92}},\ \bibinfo {pages} {042303} (\bibinfo {year}
  {2015}{\natexlab{a}})}\BibitemShut {NoStop}%
\bibitem [{\citenamefont {Babbush}\ \emph {et~al.}(2015)\citenamefont
  {Babbush}, \citenamefont {McClean}, \citenamefont {Wecker}, \citenamefont
  {Aspuru-Guzik},\ and\ \citenamefont {Wiebe}}]{BMW+15}%
  \BibitemOpen
  \bibfield  {author} {\bibinfo {author} {\bibfnamefont {Ryan}\ \bibnamefont
  {Babbush}}, \bibinfo {author} {\bibfnamefont {Jarrod}\ \bibnamefont
  {McClean}}, \bibinfo {author} {\bibfnamefont {Dave}\ \bibnamefont {Wecker}},
  \bibinfo {author} {\bibfnamefont {Al{\'a}n}\ \bibnamefont {Aspuru-Guzik}}, \
  and\ \bibinfo {author} {\bibfnamefont {Nathan}\ \bibnamefont {Wiebe}},\
  }\bibfield  {title} {\enquote {\bibinfo {title} {Chemical basis of
  trotter-suzuki errors in quantum chemistry simulation},}\ }\href@noop {}
  {\bibfield  {journal} {\bibinfo  {journal} {Physical Review A}\ }\textbf
  {\bibinfo {volume} {91}},\ \bibinfo {pages} {022311} (\bibinfo {year}
  {2015})}\BibitemShut {NoStop}%
\bibitem [{\citenamefont {Bauer}\ \emph {et~al.}(2015)\citenamefont {Bauer},
  \citenamefont {Wecker}, \citenamefont {Millis}, \citenamefont {Hastings},\
  and\ \citenamefont {Troyer}}]{BWM+15}%
  \BibitemOpen
  \bibfield  {author} {\bibinfo {author} {\bibfnamefont {Bela}\ \bibnamefont
  {Bauer}}, \bibinfo {author} {\bibfnamefont {Dave}\ \bibnamefont {Wecker}},
  \bibinfo {author} {\bibfnamefont {Andrew~J}\ \bibnamefont {Millis}}, \bibinfo
  {author} {\bibfnamefont {Matthew~B}\ \bibnamefont {Hastings}}, \ and\
  \bibinfo {author} {\bibfnamefont {Matthias}\ \bibnamefont {Troyer}},\
  }\bibfield  {title} {\enquote {\bibinfo {title} {Hybrid quantum-classical
  approach to correlated materials},}\ }\href@noop {} {\bibfield  {journal}
  {\bibinfo  {journal} {arXiv preprint arXiv:1510.03859}\ } (\bibinfo {year}
  {2015})}\BibitemShut {NoStop}%
\bibitem [{\citenamefont {Zhang}\ \emph {et~al.}(2015)\citenamefont {Zhang},
  \citenamefont {Morrison}, \citenamefont {Kaiser},\ and\ \citenamefont
  {Rees}}]{4wes}%
  \BibitemOpen
  \bibfield  {author} {\bibinfo {author} {\bibfnamefont {L.~M.}\ \bibnamefont
  {Zhang}}, \bibinfo {author} {\bibfnamefont {C.~N.}\ \bibnamefont {Morrison}},
  \bibinfo {author} {\bibfnamefont {J.~T.}\ \bibnamefont {Kaiser}}, \ and\
  \bibinfo {author} {\bibfnamefont {D.~C}\ \bibnamefont {Rees}},\ }\bibfield
  {title} {\enquote {\bibinfo {title} {Nitrogenase mofe protein from
  clostridium pasteurianum at 1.08 angstrom resolution: comparison with the
  azotobacter vinelandii mofe protein},}\ }\href@noop {} {\bibfield  {journal}
  {\bibinfo  {journal} {Acta Crystallogr.}\ }\textbf {\bibinfo {volume}
  {D71}},\ \bibinfo {pages} {274--282} (\bibinfo {year} {2015})}\BibitemShut
  {NoStop}%
\bibitem [{\citenamefont {Hoffman}\ \emph {et~al.}(2014)\citenamefont
  {Hoffman}, \citenamefont {Lukoyanov}, \citenamefont {Yang}, \citenamefont
  {Dean},\ and\ \citenamefont {Seefeldt}}]{review}%
  \BibitemOpen
  \bibfield  {author} {\bibinfo {author} {\bibfnamefont {Brian~M.}\
  \bibnamefont {Hoffman}}, \bibinfo {author} {\bibfnamefont {Dmitriy}\
  \bibnamefont {Lukoyanov}}, \bibinfo {author} {\bibfnamefont {Zhi-Yong}\
  \bibnamefont {Yang}}, \bibinfo {author} {\bibfnamefont {Dennis~R.}\
  \bibnamefont {Dean}}, \ and\ \bibinfo {author} {\bibfnamefont {Lance~C.}\
  \bibnamefont {Seefeldt}},\ }\bibfield  {title} {\enquote {\bibinfo {title}
  {Mechanism of nitrogen fixation by nitrogenase: The next stage},}\
  }\href@noop {} {\bibfield  {journal} {\bibinfo  {journal} {Chem. Rev.}\
  }\textbf {\bibinfo {volume} {114}},\ \bibinfo {pages} {4041--4062} (\bibinfo
  {year} {2014})}\BibitemShut {NoStop}%
\bibitem [{\citenamefont {Spatzal}\ \emph {et~al.}(2011)\citenamefont
  {Spatzal}, \citenamefont {Aksoyoglu}, \citenamefont {Zhang}, \citenamefont
  {Andrade}, \citenamefont {Schleicher}, \citenamefont {Weber}, \citenamefont
  {Rees},\ and\ \citenamefont {Einsle}}]{science1}%
  \BibitemOpen
  \bibfield  {author} {\bibinfo {author} {\bibfnamefont {Thomas}\ \bibnamefont
  {Spatzal}}, \bibinfo {author} {\bibfnamefont {M\"uge}\ \bibnamefont
  {Aksoyoglu}}, \bibinfo {author} {\bibfnamefont {Limei}\ \bibnamefont
  {Zhang}}, \bibinfo {author} {\bibfnamefont {Susana L.~A.}\ \bibnamefont
  {Andrade}}, \bibinfo {author} {\bibfnamefont {Erik}\ \bibnamefont
  {Schleicher}}, \bibinfo {author} {\bibfnamefont {Stefan}\ \bibnamefont
  {Weber}}, \bibinfo {author} {\bibfnamefont {Douglas~C.}\ \bibnamefont
  {Rees}}, \ and\ \bibinfo {author} {\bibfnamefont {Oliver}\ \bibnamefont
  {Einsle}},\ }\bibfield  {title} {\enquote {\bibinfo {title} {Evidence for
  interstitial carbon in nitrogenase femo cofactor},}\ }\href@noop {}
  {\bibfield  {journal} {\bibinfo  {journal} {Science}\ }\textbf {\bibinfo
  {volume} {334}},\ \bibinfo {pages} {940} (\bibinfo {year}
  {2011})}\BibitemShut {NoStop}%
\bibitem [{\citenamefont {Lancaster}\ \emph {et~al.}(2011)\citenamefont
  {Lancaster}, \citenamefont {Roemelt}, \citenamefont {Ettenhuber},
  \citenamefont {Hu}, \citenamefont {Ribbe}, \citenamefont {Neese},
  \citenamefont {Bergmann},\ and\ \citenamefont {DeBeer}}]{science2}%
  \BibitemOpen
  \bibfield  {author} {\bibinfo {author} {\bibfnamefont {Kyle~M.}\ \bibnamefont
  {Lancaster}}, \bibinfo {author} {\bibfnamefont {Michael}\ \bibnamefont
  {Roemelt}}, \bibinfo {author} {\bibfnamefont {Patrick}\ \bibnamefont
  {Ettenhuber}}, \bibinfo {author} {\bibfnamefont {Yilin}\ \bibnamefont {Hu}},
  \bibinfo {author} {\bibfnamefont {Markus~W.}\ \bibnamefont {Ribbe}}, \bibinfo
  {author} {\bibfnamefont {Frank}\ \bibnamefont {Neese}}, \bibinfo {author}
  {\bibfnamefont {Uwe}\ \bibnamefont {Bergmann}}, \ and\ \bibinfo {author}
  {\bibfnamefont {Serena}\ \bibnamefont {DeBeer}},\ }\bibfield  {title}
  {\enquote {\bibinfo {title} {X-ray emission spectroscopy evidences a central
  carbon in the nitrogenase iron-molybdenum cofactor},}\ }\href@noop {}
  {\bibfield  {journal} {\bibinfo  {journal} {Science}\ }\textbf {\bibinfo
  {volume} {334}},\ \bibinfo {pages} {974--977} (\bibinfo {year}
  {2011})}\BibitemShut {NoStop}%
\bibitem [{\citenamefont {Yandulov}\ and\ \citenamefont
  {Schrock}(2003)}]{schrock}%
  \BibitemOpen
  \bibfield  {author} {\bibinfo {author} {\bibfnamefont {Dmitry~V.}\
  \bibnamefont {Yandulov}}\ and\ \bibinfo {author} {\bibfnamefont {Richard~R.}\
  \bibnamefont {Schrock}},\ }\bibfield  {title} {\enquote {\bibinfo {title}
  {Catalytic reduction of dinitrogen to ammonia at a single molybdenum
  center},}\ }\href@noop {} {\bibfield  {journal} {\bibinfo  {journal}
  {Science}\ }\textbf {\bibinfo {volume} {301}},\ \bibinfo {pages} {76--78}
  (\bibinfo {year} {2003})}\BibitemShut {NoStop}%
\bibitem [{\citenamefont {Arashiba}\ \emph {et~al.}(2011)\citenamefont
  {Arashiba}, \citenamefont {Miyake},\ and\ \citenamefont
  {Nishibayashi}}]{nishiba}%
  \BibitemOpen
  \bibfield  {author} {\bibinfo {author} {\bibfnamefont {K.}~\bibnamefont
  {Arashiba}}, \bibinfo {author} {\bibfnamefont {Y.}~\bibnamefont {Miyake}}, \
  and\ \bibinfo {author} {\bibfnamefont {Y.}~\bibnamefont {Nishibayashi}},\
  }\bibfield  {title} {\enquote {\bibinfo {title} {A molybdenum complex bearing
  pnp-type pincer ligands leads to the catalytic reduction of dinitrogen into
  ammonia},}\ }\href@noop {} {\bibfield  {journal} {\bibinfo  {journal} {Nature
  Chem.}\ }\textbf {\bibinfo {volume} {3}},\ \bibinfo {pages} {120--125}
  (\bibinfo {year} {2011})}\BibitemShut {NoStop}%
\bibitem [{\citenamefont {Anderson}\ \emph {et~al.}(2013)\citenamefont
  {Anderson}, \citenamefont {Rittle},\ and\ \citenamefont {Peters}}]{peters}%
  \BibitemOpen
  \bibfield  {author} {\bibinfo {author} {\bibfnamefont {John~S.}\ \bibnamefont
  {Anderson}}, \bibinfo {author} {\bibfnamefont {Jonathan}\ \bibnamefont
  {Rittle}}, \ and\ \bibinfo {author} {\bibfnamefont {Jonas~C.}\ \bibnamefont
  {Peters}},\ }\bibfield  {title} {\enquote {\bibinfo {title} {Catalytic
  conversion of nitrogen to ammonia by an iron model complex},}\ }\href@noop {}
  {\bibfield  {journal} {\bibinfo  {journal} {Nature}\ }\textbf {\bibinfo
  {volume} {501}},\ \bibinfo {pages} {84--87} (\bibinfo {year}
  {2013})}\BibitemShut {NoStop}%
\bibitem [{\citenamefont {Bergeler}\ \emph {et~al.}(2015)\citenamefont
  {Bergeler}, \citenamefont {Simm}, \citenamefont {Proppe},\ and\ \citenamefont
  {Reiher}}]{heurex}%
  \BibitemOpen
  \bibfield  {author} {\bibinfo {author} {\bibfnamefont {Maike}\ \bibnamefont
  {Bergeler}}, \bibinfo {author} {\bibfnamefont {Gregor~N.}\ \bibnamefont
  {Simm}}, \bibinfo {author} {\bibfnamefont {Jonny}\ \bibnamefont {Proppe}}, \
  and\ \bibinfo {author} {\bibfnamefont {Markus}\ \bibnamefont {Reiher}},\
  }\bibfield  {title} {\enquote {\bibinfo {title} {Heuristics-guided
  exploration of reaction mechanisms},}\ }\href@noop {} {\bibfield  {journal}
  {\bibinfo  {journal} {J. Chem. Theory Comput.}\ }\textbf {\bibinfo {volume}
  {11}},\ \bibinfo {pages} {5712--5722} (\bibinfo {year} {2015})}\BibitemShut
  {NoStop}%
\bibitem [{\citenamefont {Olsen}\ and\ \citenamefont {Kongsted}(2011)}]{kong}%
  \BibitemOpen
  \bibfield  {author} {\bibinfo {author} {\bibfnamefont {Jógvan
  Magnus~Haugaard}\ \bibnamefont {Olsen}}\ and\ \bibinfo {author}
  {\bibfnamefont {Jacob}\ \bibnamefont {Kongsted}},\ }\bibfield  {title}
  {\enquote {\bibinfo {title} {Molecular properties through polarizable
  embedding},}\ }\href@noop {} {\bibfield  {journal} {\bibinfo  {journal} {Adv.
  Quantum Chem.}\ }\textbf {\bibinfo {volume} {61}},\ \bibinfo {pages}
  {107--143} (\bibinfo {year} {2011})}\BibitemShut {NoStop}%
\bibitem [{\citenamefont {Jacob}\ and\ \citenamefont
  {Neugebauer}(2014)}]{jone}%
  \BibitemOpen
  \bibfield  {author} {\bibinfo {author} {\bibfnamefont {Christoph~R.}\
  \bibnamefont {Jacob}}\ and\ \bibinfo {author} {\bibfnamefont {Johannes}\
  \bibnamefont {Neugebauer}},\ }\bibfield  {title} {\enquote {\bibinfo {title}
  {Subsystem density-functional theory},}\ }\href@noop {} {\bibfield  {journal}
  {\bibinfo  {journal} {WIREs Comput. Molec. Sci.}\ }\textbf {\bibinfo {volume}
  {4}},\ \bibinfo {pages} {325--362} (\bibinfo {year} {2014})}\BibitemShut
  {NoStop}%
\bibitem [{\citenamefont {Warshel}(2014)}]{warshel}%
  \BibitemOpen
  \bibfield  {author} {\bibinfo {author} {\bibfnamefont {Arieh}\ \bibnamefont
  {Warshel}},\ }\bibfield  {title} {\enquote {\bibinfo {title} {Multiscale
  modeling of biological functions: From enzymes to molecular machines (nobel
  lecture)},}\ }\href@noop {} {\bibfield  {journal} {\bibinfo  {journal}
  {Angew. Chem. Int. Ed.}\ }\textbf {\bibinfo {volume} {53}},\ \bibinfo {pages}
  {10020--10031} (\bibinfo {year} {2014})}\BibitemShut {NoStop}%
\bibitem [{\citenamefont {Shedge}\ and\ \citenamefont {Zhou}(2015)}]{weso}%
  \BibitemOpen
  \bibfield  {author} {\bibinfo {author} {\bibfnamefont {Tomasz A.
  Wesolowski~Sapana}\ \bibnamefont {Shedge}}\ and\ \bibinfo {author}
  {\bibfnamefont {Xiuwen}\ \bibnamefont {Zhou}},\ }\bibfield  {title} {\enquote
  {\bibinfo {title} {Frozen-density embedding strategy for multilevel
  simulations of electronic structure},}\ }\href@noop {} {\bibfield  {journal}
  {\bibinfo  {journal} {Chem. Rev.}\ }\textbf {\bibinfo {volume} {115}},\
  \bibinfo {pages} {5891--5928} (\bibinfo {year} {2015})}\BibitemShut {NoStop}%
\bibitem [{\citenamefont {Wouters}\ \emph {et~al.}(2016)\citenamefont
  {Wouters}, \citenamefont {Jim\'enez-Hoyos}, \citenamefont {Sun},\ and\
  \citenamefont {Chan}}]{garnet}%
  \BibitemOpen
  \bibfield  {author} {\bibinfo {author} {\bibfnamefont {Sebastian}\
  \bibnamefont {Wouters}}, \bibinfo {author} {\bibfnamefont {Carlos~A.}\
  \bibnamefont {Jim\'enez-Hoyos}}, \bibinfo {author} {\bibfnamefont {Qiming}\
  \bibnamefont {Sun}}, \ and\ \bibinfo {author} {\bibfnamefont {Garnet
  Kin-Lic}\ \bibnamefont {Chan}},\ }\bibfield  {title} {\enquote {\bibinfo
  {title} {A practical guide to density matrix embedding theory in quantum
  chemistry},}\ }\href@noop {} {\bibfield  {journal} {\bibinfo  {journal}
  {arXiv}\ ,\ \bibinfo {pages} {1603.08443}} (\bibinfo {year}
  {2016})}\BibitemShut {NoStop}%
\bibitem [{\citenamefont {Olsson}\ \emph {et~al.}(2006)\citenamefont {Olsson},
  \citenamefont {Mavri},\ and\ \citenamefont {Warshel}}]{warshel2}%
  \BibitemOpen
  \bibfield  {author} {\bibinfo {author} {\bibfnamefont {M.~H.}\ \bibnamefont
  {Olsson}}, \bibinfo {author} {\bibfnamefont {J.}~\bibnamefont {Mavri}}, \
  and\ \bibinfo {author} {\bibfnamefont {A.}~\bibnamefont {Warshel}},\
  }\bibfield  {title} {\enquote {\bibinfo {title} {Transition state theory can
  be used in studies of enzyme catalysis: lessons from simulations of
  tunnelling and dynamical effects in lipoxygenase and other systems},}\
  }\href@noop {} {\bibfield  {journal} {\bibinfo  {journal} {Philos. Trans.
  Roy. Soc. Lond. B}\ }\textbf {\bibinfo {volume} {361}},\ \bibinfo {pages}
  {1417--1432} (\bibinfo {year} {2006})}\BibitemShut {NoStop}%
\bibitem [{\citenamefont {Glowacki}\ \emph {et~al.}(2012)\citenamefont
  {Glowacki}, \citenamefont {Harvey},\ and\ \citenamefont
  {Mulholland}}]{ockham}%
  \BibitemOpen
  \bibfield  {author} {\bibinfo {author} {\bibfnamefont {David~R.}\
  \bibnamefont {Glowacki}}, \bibinfo {author} {\bibfnamefont {Jeremy~N.}\
  \bibnamefont {Harvey}}, \ and\ \bibinfo {author} {\bibfnamefont {Adrian~J.}\
  \bibnamefont {Mulholland}},\ }\bibfield  {title} {\enquote {\bibinfo {title}
  {Taking ockham's razor to enzyme dynamics and catalysis},}\ }\href@noop {}
  {\bibfield  {journal} {\bibinfo  {journal} {Nature Chem.}\ }\textbf {\bibinfo
  {volume} {4}},\ \bibinfo {pages} {169--176} (\bibinfo {year}
  {2012})}\BibitemShut {NoStop}%
\bibitem [{\citenamefont {Helgaker}\ \emph {et~al.}(2000)\citenamefont
  {Helgaker}, \citenamefont {J{\o}rgensen},\ and\ \citenamefont
  {Olsen}}]{helgaker}%
  \BibitemOpen
  \bibfield  {author} {\bibinfo {author} {\bibfnamefont {Trygve}\ \bibnamefont
  {Helgaker}}, \bibinfo {author} {\bibfnamefont {Poul}\ \bibnamefont
  {J{\o}rgensen}}, \ and\ \bibinfo {author} {\bibfnamefont {Jeppe}\
  \bibnamefont {Olsen}},\ }\href@noop {} {\emph {\bibinfo {title} {Molecular
  Electronic-Structure Theory}}}\ (\bibinfo  {publisher} {{John Wiley \&
  Sons}},\ \bibinfo {year} {2000})\BibitemShut {NoStop}%
\bibitem [{\citenamefont {Pulay}\ and\ \citenamefont {Hamilton}(1988)}]{pulay}%
  \BibitemOpen
  \bibfield  {author} {\bibinfo {author} {\bibfnamefont {Peter}\ \bibnamefont
  {Pulay}}\ and\ \bibinfo {author} {\bibfnamefont {Tracy~P.}\ \bibnamefont
  {Hamilton}},\ }\bibfield  {title} {\enquote {\bibinfo {title} {Uhf natural
  orbitals for defining and starting mc-scf calculations},}\ }\href@noop {}
  {\bibfield  {journal} {\bibinfo  {journal} {J. Chem. Phys.}\ }\textbf
  {\bibinfo {volume} {88}},\ \bibinfo {pages} {4926--4933} (\bibinfo {year}
  {1988})}\BibitemShut {NoStop}%
\bibitem [{\citenamefont {Stein}\ and\ \citenamefont {Reiher}(2016)}]{stein}%
  \BibitemOpen
  \bibfield  {author} {\bibinfo {author} {\bibfnamefont {Christopher~J.}\
  \bibnamefont {Stein}}\ and\ \bibinfo {author} {\bibfnamefont {Markus}\
  \bibnamefont {Reiher}},\ }\bibfield  {title} {\enquote {\bibinfo {title}
  {Automated selection of active orbital spaces},}\ }\href@noop {} {\bibfield
  {journal} {\bibinfo  {journal} {J. Chem. Theory Comput.}\ }\textbf {\bibinfo
  {volume} {12}},\ \bibinfo {pages} {1760--1771} (\bibinfo {year}
  {2016})}\BibitemShut {NoStop}%
\bibitem [{\citenamefont {Aquilante}\ \emph {et~al.}(2016)\citenamefont
  {Aquilante}, \citenamefont {Autschbach}, \citenamefont {Carlson},
  \citenamefont {Chibotaru}, \citenamefont {Delcey}, \citenamefont {De~Vico},
  \citenamefont {Fdez.~Galv\'an}, \citenamefont {Ferr\'e}, \citenamefont
  {Frutos}, \citenamefont {Gagliardi}, \citenamefont {Garavelli}, \citenamefont
  {Giussani}, \citenamefont {Hoyer}, \citenamefont {Li~Manni}, \citenamefont
  {Lischka}, \citenamefont {Ma}, \citenamefont {Malmqvist}, \citenamefont
  {M\"uller}, \citenamefont {Nenov}, \citenamefont {Olivucci}, \citenamefont
  {Pedersen}, \citenamefont {Peng}, \citenamefont {Plasser}, \citenamefont
  {Pritchard}, \citenamefont {Reiher}, \citenamefont {Rivalta}, \citenamefont
  {Schapiro}, \citenamefont {Segarra-Martí}, \citenamefont {Stenrup},
  \citenamefont {Truhlar}, \citenamefont {Ungur}, \citenamefont {Valentini},
  \citenamefont {Vancoillie}, \citenamefont {Veryazov}, \citenamefont
  {Vysotskiy}, \citenamefont {Weingart}, \citenamefont {Zapata},\ and\
  \citenamefont {Lindh}}]{molcas8}%
  \BibitemOpen
  \bibfield  {author} {\bibinfo {author} {\bibfnamefont {Francesco}\
  \bibnamefont {Aquilante}}, \bibinfo {author} {\bibfnamefont {Jochen}\
  \bibnamefont {Autschbach}}, \bibinfo {author} {\bibfnamefont {Rebecca~K.}\
  \bibnamefont {Carlson}}, \bibinfo {author} {\bibfnamefont {Liviu~F.}\
  \bibnamefont {Chibotaru}}, \bibinfo {author} {\bibfnamefont {Mickael~G.}\
  \bibnamefont {Delcey}}, \bibinfo {author} {\bibfnamefont {Luca}\ \bibnamefont
  {De~Vico}}, \bibinfo {author} {\bibfnamefont {Ignacio}\ \bibnamefont
  {Fdez.~Galv\'an}}, \bibinfo {author} {\bibfnamefont {Nicolas}\ \bibnamefont
  {Ferr\'e}}, \bibinfo {author} {\bibfnamefont {Luis~Manuel}\ \bibnamefont
  {Frutos}}, \bibinfo {author} {\bibfnamefont {Laura}\ \bibnamefont
  {Gagliardi}}, \bibinfo {author} {\bibfnamefont {Marco}\ \bibnamefont
  {Garavelli}}, \bibinfo {author} {\bibfnamefont {Angelo}\ \bibnamefont
  {Giussani}}, \bibinfo {author} {\bibfnamefont {Chad~E.}\ \bibnamefont
  {Hoyer}}, \bibinfo {author} {\bibfnamefont {Giovanni}\ \bibnamefont
  {Li~Manni}}, \bibinfo {author} {\bibfnamefont {Hans}\ \bibnamefont
  {Lischka}}, \bibinfo {author} {\bibfnamefont {Dongxia}\ \bibnamefont {Ma}},
  \bibinfo {author} {\bibfnamefont {Per~{\AA}ke}\ \bibnamefont {Malmqvist}},
  \bibinfo {author} {\bibfnamefont {Thomas}\ \bibnamefont {M\"uller}}, \bibinfo
  {author} {\bibfnamefont {Artur}\ \bibnamefont {Nenov}}, \bibinfo {author}
  {\bibfnamefont {Massimo}\ \bibnamefont {Olivucci}}, \bibinfo {author}
  {\bibfnamefont {Thomas~Bondo}\ \bibnamefont {Pedersen}}, \bibinfo {author}
  {\bibfnamefont {Daoling}\ \bibnamefont {Peng}}, \bibinfo {author}
  {\bibfnamefont {Felix}\ \bibnamefont {Plasser}}, \bibinfo {author}
  {\bibfnamefont {Ben}\ \bibnamefont {Pritchard}}, \bibinfo {author}
  {\bibfnamefont {Markus}\ \bibnamefont {Reiher}}, \bibinfo {author}
  {\bibfnamefont {Ivan}\ \bibnamefont {Rivalta}}, \bibinfo {author}
  {\bibfnamefont {Igor}\ \bibnamefont {Schapiro}}, \bibinfo {author}
  {\bibfnamefont {Javier}\ \bibnamefont {Segarra-Martí}}, \bibinfo {author}
  {\bibfnamefont {Michael}\ \bibnamefont {Stenrup}}, \bibinfo {author}
  {\bibfnamefont {Donald~G.}\ \bibnamefont {Truhlar}}, \bibinfo {author}
  {\bibfnamefont {Liviu}\ \bibnamefont {Ungur}}, \bibinfo {author}
  {\bibfnamefont {Alessio}\ \bibnamefont {Valentini}}, \bibinfo {author}
  {\bibfnamefont {Steven}\ \bibnamefont {Vancoillie}}, \bibinfo {author}
  {\bibfnamefont {Valera}\ \bibnamefont {Veryazov}}, \bibinfo {author}
  {\bibfnamefont {Victor~P.}\ \bibnamefont {Vysotskiy}}, \bibinfo {author}
  {\bibfnamefont {Oliver}\ \bibnamefont {Weingart}}, \bibinfo {author}
  {\bibfnamefont {Felipe}\ \bibnamefont {Zapata}}, \ and\ \bibinfo {author}
  {\bibfnamefont {Roland}\ \bibnamefont {Lindh}},\ }\bibfield  {title}
  {\enquote {\bibinfo {title} {Molcas 8: New capabilities for
  multiconfigurational quantum chemical calculations across the periodic
  table},}\ }\href@noop {} {\bibfield  {journal} {\bibinfo  {journal} {J.
  Comput. Chem.}\ ,\ \bibinfo {pages} {506--541}} (\bibinfo {year}
  {2016})}\BibitemShut {NoStop}%
\bibitem [{\citenamefont {White}(1992)}]{white}%
  \BibitemOpen
  \bibfield  {author} {\bibinfo {author} {\bibfnamefont {Steven~R.}\
  \bibnamefont {White}},\ }\bibfield  {title} {\enquote {\bibinfo {title}
  {Density matrix formulation for quantum renormalization groups},}\
  }\href@noop {} {\bibfield  {journal} {\bibinfo  {journal} {Phys. Rev. Lett.}\
  }\textbf {\bibinfo {volume} {69}},\ \bibinfo {pages} {2863--2866} (\bibinfo
  {year} {1992})}\BibitemShut {NoStop}%
\bibitem [{\citenamefont {Bofill}\ and\ \citenamefont {Pulay}(1989)}]{pulay2}%
  \BibitemOpen
  \bibfield  {author} {\bibinfo {author} {\bibfnamefont {Josep~M.}\
  \bibnamefont {Bofill}}\ and\ \bibinfo {author} {\bibfnamefont {Peter}\
  \bibnamefont {Pulay}},\ }\bibfield  {title} {\enquote {\bibinfo {title} {The
  unrestricted natural orbital-complete active space (uno-cas) method: An
  inexpensive alternative to the complete active space-self-consistent-field
  (cas-scf) method},}\ }\href@noop {} {\bibfield  {journal} {\bibinfo
  {journal} {J. Chem. Phys.}\ }\textbf {\bibinfo {volume} {90}},\ \bibinfo
  {pages} {3637--3646} (\bibinfo {year} {1989})}\BibitemShut {NoStop}%
\bibitem [{\citenamefont {Fromager}\ \emph {et~al.}(2007)\citenamefont
  {Fromager}, \citenamefont {Toulouse},\ and\ \citenamefont {Jensen}}]{jensen}%
  \BibitemOpen
  \bibfield  {author} {\bibinfo {author} {\bibfnamefont {E.}~\bibnamefont
  {Fromager}}, \bibinfo {author} {\bibfnamefont {J.}~\bibnamefont {Toulouse}},
  \ and\ \bibinfo {author} {\bibfnamefont {H.~J.~{\AA}.}\ \bibnamefont
  {Jensen}},\ }\bibfield  {title} {\enquote {\bibinfo {title} {{On the
  universality of the long-/short-range separation in multiconfigurational
  density-functional theory}},}\ }\href@noop {} {\bibfield  {journal} {\bibinfo
   {journal} {J. Chem. Phys.}\ }\textbf {\bibinfo {volume} {126}},\ \bibinfo
  {pages} {074111} (\bibinfo {year} {2007})}\BibitemShut {NoStop}%
\bibitem [{\citenamefont {Hedeg\r{a}rd}\ \emph {et~al.}(2015)\citenamefont
  {Hedeg\r{a}rd}, \citenamefont {Knecht}, \citenamefont {Kielberg},
  \citenamefont {Jensen},\ and\ \citenamefont {Reiher}}]{hedegard}%
  \BibitemOpen
  \bibfield  {author} {\bibinfo {author} {\bibfnamefont {Erik~Donovan}\
  \bibnamefont {Hedeg\r{a}rd}}, \bibinfo {author} {\bibfnamefont {Stefan}\
  \bibnamefont {Knecht}}, \bibinfo {author} {\bibfnamefont {Jesper~Skau}\
  \bibnamefont {Kielberg}}, \bibinfo {author} {\bibfnamefont {Hans
  J{\"o}rgen~Aagaard}\ \bibnamefont {Jensen}}, \ and\ \bibinfo {author}
  {\bibfnamefont {Markus}\ \bibnamefont {Reiher}},\ }\bibfield  {title}
  {\enquote {\bibinfo {title} {Density matrix renormalization group with
  efficient dynamical electron correlation through range separation},}\
  }\href@noop {} {\bibfield  {journal} {\bibinfo  {journal} {J. Chem. Phys.}\
  }\textbf {\bibinfo {volume} {142}},\ \bibinfo {pages} {224108} (\bibinfo
  {year} {2015})}\BibitemShut {NoStop}%
\bibitem [{\citenamefont {Abrams}\ and\ \citenamefont {Lloyd}(1997)}]{AL97}%
  \BibitemOpen
  \bibfield  {author} {\bibinfo {author} {\bibfnamefont {Daniel~S}\
  \bibnamefont {Abrams}}\ and\ \bibinfo {author} {\bibfnamefont {Seth}\
  \bibnamefont {Lloyd}},\ }\bibfield  {title} {\enquote {\bibinfo {title}
  {Simulation of many-body fermi systems on a universal quantum computer},}\
  }\href@noop {} {\bibfield  {journal} {\bibinfo  {journal} {Physical Review
  Letters}\ }\textbf {\bibinfo {volume} {79}},\ \bibinfo {pages} {2586}
  (\bibinfo {year} {1997})}\BibitemShut {NoStop}%
\bibitem [{\citenamefont {Berry}\ \emph {et~al.}(2007)\citenamefont {Berry},
  \citenamefont {Ahokas}, \citenamefont {Cleve},\ and\ \citenamefont
  {Sanders}}]{BACS07}%
  \BibitemOpen
  \bibfield  {author} {\bibinfo {author} {\bibfnamefont {Dominic~W}\
  \bibnamefont {Berry}}, \bibinfo {author} {\bibfnamefont {Graeme}\
  \bibnamefont {Ahokas}}, \bibinfo {author} {\bibfnamefont {Richard}\
  \bibnamefont {Cleve}}, \ and\ \bibinfo {author} {\bibfnamefont {Barry~C}\
  \bibnamefont {Sanders}},\ }\bibfield  {title} {\enquote {\bibinfo {title}
  {Efficient quantum algorithms for simulating sparse hamiltonians},}\
  }\href@noop {} {\bibfield  {journal} {\bibinfo  {journal} {Communications in
  Mathematical Physics}\ }\textbf {\bibinfo {volume} {270}},\ \bibinfo {pages}
  {359--371} (\bibinfo {year} {2007})}\BibitemShut {NoStop}%
\bibitem [{\citenamefont {Childs}\ and\ \citenamefont {Wiebe}(2012)}]{CW12}%
  \BibitemOpen
  \bibfield  {author} {\bibinfo {author} {\bibfnamefont {Andrew~M}\
  \bibnamefont {Childs}}\ and\ \bibinfo {author} {\bibfnamefont {Nathan}\
  \bibnamefont {Wiebe}},\ }\bibfield  {title} {\enquote {\bibinfo {title}
  {Hamiltonian simulation using linear combinations of unitary operations},}\
  }\href@noop {} {\bibfield  {journal} {\bibinfo  {journal} {Quantum
  Information \& Computation}\ }\textbf {\bibinfo {volume} {12}},\ \bibinfo
  {pages} {901--924} (\bibinfo {year} {2012})}\BibitemShut {NoStop}%
\bibitem [{\citenamefont {Berry}\ \emph {et~al.}(2015)\citenamefont {Berry},
  \citenamefont {Childs}, \citenamefont {Cleve}, \citenamefont {Kothari},\ and\
  \citenamefont {Somma}}]{berry2015simulating}%
  \BibitemOpen
  \bibfield  {author} {\bibinfo {author} {\bibfnamefont {Dominic~W}\
  \bibnamefont {Berry}}, \bibinfo {author} {\bibfnamefont {Andrew~M}\
  \bibnamefont {Childs}}, \bibinfo {author} {\bibfnamefont {Richard}\
  \bibnamefont {Cleve}}, \bibinfo {author} {\bibfnamefont {Robin}\ \bibnamefont
  {Kothari}}, \ and\ \bibinfo {author} {\bibfnamefont {Rolando~D}\ \bibnamefont
  {Somma}},\ }\bibfield  {title} {\enquote {\bibinfo {title} {Simulating
  hamiltonian dynamics with a truncated taylor series},}\ }\href@noop {}
  {\bibfield  {journal} {\bibinfo  {journal} {Physical review letters}\
  }\textbf {\bibinfo {volume} {114}},\ \bibinfo {pages} {090502} (\bibinfo
  {year} {2015})}\BibitemShut {NoStop}%
\bibitem [{\citenamefont {Fowler}\ \emph {et~al.}(2012)\citenamefont {Fowler},
  \citenamefont {Mariantoni}, \citenamefont {Martinis},\ and\ \citenamefont
  {Cleland}}]{fowler2012surface}%
  \BibitemOpen
  \bibfield  {author} {\bibinfo {author} {\bibfnamefont {Austin~G}\
  \bibnamefont {Fowler}}, \bibinfo {author} {\bibfnamefont {Matteo}\
  \bibnamefont {Mariantoni}}, \bibinfo {author} {\bibfnamefont {John~M}\
  \bibnamefont {Martinis}}, \ and\ \bibinfo {author} {\bibfnamefont {Andrew~N}\
  \bibnamefont {Cleland}},\ }\bibfield  {title} {\enquote {\bibinfo {title}
  {Surface codes: Towards practical large-scale quantum computation},}\
  }\href@noop {} {\bibfield  {journal} {\bibinfo  {journal} {Physical Review
  A}\ }\textbf {\bibinfo {volume} {86}},\ \bibinfo {pages} {032324} (\bibinfo
  {year} {2012})}\BibitemShut {NoStop}%
\bibitem [{\citenamefont {Nielsen}\ and\ \citenamefont {Chuang}(2010)}]{NC00}%
  \BibitemOpen
  \bibfield  {author} {\bibinfo {author} {\bibfnamefont {Michael~A}\
  \bibnamefont {Nielsen}}\ and\ \bibinfo {author} {\bibfnamefont {Isaac~L}\
  \bibnamefont {Chuang}},\ }\href@noop {} {\emph {\bibinfo {title} {Quantum
  computation and quantum information}}}\ (\bibinfo  {publisher} {Cambridge
  university press},\ \bibinfo {year} {2010})\BibitemShut {NoStop}%
\bibitem [{\citenamefont {Bravyi}\ and\ \citenamefont
  {Kitaev}(2005)}]{bravyi2005universal}%
  \BibitemOpen
  \bibfield  {author} {\bibinfo {author} {\bibfnamefont {Sergey}\ \bibnamefont
  {Bravyi}}\ and\ \bibinfo {author} {\bibfnamefont {Alexei}\ \bibnamefont
  {Kitaev}},\ }\bibfield  {title} {\enquote {\bibinfo {title} {Universal
  quantum computation with ideal {Clifford} gates and noisy ancillas},}\
  }\href@noop {} {\bibfield  {journal} {\bibinfo  {journal} {Physical Review
  A}\ }\textbf {\bibinfo {volume} {71}},\ \bibinfo {pages} {022316} (\bibinfo
  {year} {2005})}\BibitemShut {NoStop}%
\bibitem [{\citenamefont {Bocharov}\ \emph {et~al.}(2015)\citenamefont
  {Bocharov}, \citenamefont {Roetteler},\ and\ \citenamefont
  {Svore}}]{bocharov2015efficient}%
  \BibitemOpen
  \bibfield  {author} {\bibinfo {author} {\bibfnamefont {Alex}\ \bibnamefont
  {Bocharov}}, \bibinfo {author} {\bibfnamefont {Martin}\ \bibnamefont
  {Roetteler}}, \ and\ \bibinfo {author} {\bibfnamefont {Krysta~M}\
  \bibnamefont {Svore}},\ }\bibfield  {title} {\enquote {\bibinfo {title}
  {Efficient synthesis of probabilistic quantum circuits with fallback},}\
  }\href@noop {} {\bibfield  {journal} {\bibinfo  {journal} {Physical Review
  A}\ }\textbf {\bibinfo {volume} {91}},\ \bibinfo {pages} {052317} (\bibinfo
  {year} {2015})}\BibitemShut {NoStop}%
\bibitem [{\citenamefont {Selinger}(2015)}]{selinger2015efficient}%
  \BibitemOpen
  \bibfield  {author} {\bibinfo {author} {\bibfnamefont {Peter}\ \bibnamefont
  {Selinger}},\ }\bibfield  {title} {\enquote {\bibinfo {title} {Efficient
  {Clifford}+ t approximation of single-qubit operators},}\ }\href@noop {}
  {\bibfield  {journal} {\bibinfo  {journal} {Quantum Information \&
  Computation}\ }\textbf {\bibinfo {volume} {15}},\ \bibinfo {pages} {159--180}
  (\bibinfo {year} {2015})}\BibitemShut {NoStop}%
\bibitem [{\citenamefont {Sarma}\ \emph {et~al.}(2015)\citenamefont {Sarma},
  \citenamefont {Freedman},\ and\ \citenamefont {Nayak}}]{topo}%
  \BibitemOpen
  \bibfield  {author} {\bibinfo {author} {\bibfnamefont {Sankar~Das}\
  \bibnamefont {Sarma}}, \bibinfo {author} {\bibfnamefont {Michael}\
  \bibnamefont {Freedman}}, \ and\ \bibinfo {author} {\bibfnamefont {Chetan}\
  \bibnamefont {Nayak}},\ }\bibfield  {title} {\enquote {\bibinfo {title}
  {Majorana zero modes and topological quantum computation},}\ }\href@noop {}
  {\bibfield  {journal} {\bibinfo  {journal} {arXiv preprint arXiv:1501.02813}\
  } (\bibinfo {year} {2015})}\BibitemShut {NoStop}%
\bibitem [{\citenamefont {Bocharov}\ \emph {et~al.}(2013)\citenamefont
  {Bocharov}, \citenamefont {Gurevich},\ and\ \citenamefont
  {Svore}}]{bocharov2013efficient}%
  \BibitemOpen
  \bibfield  {author} {\bibinfo {author} {\bibfnamefont {Alex}\ \bibnamefont
  {Bocharov}}, \bibinfo {author} {\bibfnamefont {Yuri}\ \bibnamefont
  {Gurevich}}, \ and\ \bibinfo {author} {\bibfnamefont {Krysta~M}\ \bibnamefont
  {Svore}},\ }\bibfield  {title} {\enquote {\bibinfo {title} {Efficient
  decomposition of single-qubit gates into v basis circuits},}\ }\href@noop {}
  {\bibfield  {journal} {\bibinfo  {journal} {Physical Review A}\ }\textbf
  {\bibinfo {volume} {88}},\ \bibinfo {pages} {012313} (\bibinfo {year}
  {2013})}\BibitemShut {NoStop}%
\bibitem [{\citenamefont {Forest}\ \emph {et~al.}(2015)\citenamefont {Forest},
  \citenamefont {Gosset}, \citenamefont {Kliuchnikov},\ and\ \citenamefont
  {McKinnon}}]{forest2015exact}%
  \BibitemOpen
  \bibfield  {author} {\bibinfo {author} {\bibfnamefont {Simon}\ \bibnamefont
  {Forest}}, \bibinfo {author} {\bibfnamefont {David}\ \bibnamefont {Gosset}},
  \bibinfo {author} {\bibfnamefont {Vadym}\ \bibnamefont {Kliuchnikov}}, \ and\
  \bibinfo {author} {\bibfnamefont {David}\ \bibnamefont {McKinnon}},\
  }\bibfield  {title} {\enquote {\bibinfo {title} {Exact synthesis of
  single-qubit unitaries over {Clifford}-cyclotomic gate sets},}\ }\href@noop
  {} {\bibfield  {journal} {\bibinfo  {journal} {Journal of Mathematical
  Physics}\ }\textbf {\bibinfo {volume} {56}},\ \bibinfo {pages} {082201}
  (\bibinfo {year} {2015})}\BibitemShut {NoStop}%
\bibitem [{\citenamefont {Dawson}\ and\ \citenamefont
  {Nielsen}(2005)}]{dawson2005solovay}%
  \BibitemOpen
  \bibfield  {author} {\bibinfo {author} {\bibfnamefont {Christopher~M}\
  \bibnamefont {Dawson}}\ and\ \bibinfo {author} {\bibfnamefont {Michael~A}\
  \bibnamefont {Nielsen}},\ }\bibfield  {title} {\enquote {\bibinfo {title}
  {The solovay-kitaev algorithm},}\ }\href@noop {} {\bibfield  {journal}
  {\bibinfo  {journal} {arXiv preprint quant-ph/0505030}\ } (\bibinfo {year}
  {2005})}\BibitemShut {NoStop}%
\bibitem [{\citenamefont {Kliuchnikov}\ \emph {et~al.}(2013)\citenamefont
  {Kliuchnikov}, \citenamefont {Maslov},\ and\ \citenamefont
  {Mosca}}]{kliuchnikov2013fast}%
  \BibitemOpen
  \bibfield  {author} {\bibinfo {author} {\bibfnamefont {Vadym}\ \bibnamefont
  {Kliuchnikov}}, \bibinfo {author} {\bibfnamefont {Dmitri}\ \bibnamefont
  {Maslov}}, \ and\ \bibinfo {author} {\bibfnamefont {Michele}\ \bibnamefont
  {Mosca}},\ }\bibfield  {title} {\enquote {\bibinfo {title} {Fast and
  efficient exact synthesis of single-qubit unitaries generated by {Clifford}
  and t gates},}\ }\href@noop {} {\bibfield  {journal} {\bibinfo  {journal}
  {Quantum Information \& Computation}\ }\textbf {\bibinfo {volume} {13}},\
  \bibinfo {pages} {607--630} (\bibinfo {year} {2013})}\BibitemShut {NoStop}%
\bibitem [{\citenamefont {Ross}\ and\ \citenamefont
  {Selinger}(2014)}]{ross2014optimal}%
  \BibitemOpen
  \bibfield  {author} {\bibinfo {author} {\bibfnamefont {Neil~J}\ \bibnamefont
  {Ross}}\ and\ \bibinfo {author} {\bibfnamefont {Peter}\ \bibnamefont
  {Selinger}},\ }\bibfield  {title} {\enquote {\bibinfo {title} {Optimal
  ancilla-free {Clifford}+ t approximation of z-rotations},}\ }\href@noop {}
  {\bibfield  {journal} {\bibinfo  {journal} {arXiv preprint arXiv:1403.2975}\
  } (\bibinfo {year} {2014})}\BibitemShut {NoStop}%
\bibitem [{\citenamefont {Wiebe}\ and\ \citenamefont
  {Kliuchnikov}(2013)}]{wiebe2013floating}%
  \BibitemOpen
  \bibfield  {author} {\bibinfo {author} {\bibfnamefont {Nathan}\ \bibnamefont
  {Wiebe}}\ and\ \bibinfo {author} {\bibfnamefont {Vadym}\ \bibnamefont
  {Kliuchnikov}},\ }\bibfield  {title} {\enquote {\bibinfo {title} {Floating
  point representations in quantum circuit synthesis},}\ }\href@noop {}
  {\bibfield  {journal} {\bibinfo  {journal} {New Journal of Physics}\ }\textbf
  {\bibinfo {volume} {15}},\ \bibinfo {pages} {093041} (\bibinfo {year}
  {2013})}\BibitemShut {NoStop}%
\bibitem [{\citenamefont {Seeley}\ \emph {et~al.}(2012)\citenamefont {Seeley},
  \citenamefont {Richard},\ and\ \citenamefont {Love}}]{seeley2012bravyi}%
  \BibitemOpen
  \bibfield  {author} {\bibinfo {author} {\bibfnamefont {Jacob~T}\ \bibnamefont
  {Seeley}}, \bibinfo {author} {\bibfnamefont {Martin~J}\ \bibnamefont
  {Richard}}, \ and\ \bibinfo {author} {\bibfnamefont {Peter~J}\ \bibnamefont
  {Love}},\ }\bibfield  {title} {\enquote {\bibinfo {title} {The bravyi-kitaev
  transformation for quantum computation of electronic structure},}\
  }\href@noop {} {\bibfield  {journal} {\bibinfo  {journal} {The Journal of
  chemical physics}\ }\textbf {\bibinfo {volume} {137}},\ \bibinfo {pages}
  {224109} (\bibinfo {year} {2012})}\BibitemShut {NoStop}%
\bibitem [{\citenamefont {Fowler}\ \emph {et~al.}(2009)\citenamefont {Fowler},
  \citenamefont {Stephens},\ and\ \citenamefont
  {Groszkowski}}]{fowler2009high}%
  \BibitemOpen
  \bibfield  {author} {\bibinfo {author} {\bibfnamefont {Austin~G}\
  \bibnamefont {Fowler}}, \bibinfo {author} {\bibfnamefont {Ashley~M}\
  \bibnamefont {Stephens}}, \ and\ \bibinfo {author} {\bibfnamefont {Peter}\
  \bibnamefont {Groszkowski}},\ }\bibfield  {title} {\enquote {\bibinfo {title}
  {High-threshold universal quantum computation on the surface code},}\
  }\href@noop {} {\bibfield  {journal} {\bibinfo  {journal} {Physical Review
  A}\ }\textbf {\bibinfo {volume} {80}},\ \bibinfo {pages} {052312} (\bibinfo
  {year} {2009})}\BibitemShut {NoStop}%
\bibitem [{\citenamefont {Barends}\ \emph {et~al.}(2014)\citenamefont
  {Barends}, \citenamefont {Kelly}, \citenamefont {Megrant}, \citenamefont
  {Veitia}, \citenamefont {Sank}, \citenamefont {Jeffrey}, \citenamefont
  {White}, \citenamefont {Mutus}, \citenamefont {Fowler}, \citenamefont
  {Campbell} \emph {et~al.}}]{barends2014superconducting}%
  \BibitemOpen
  \bibfield  {author} {\bibinfo {author} {\bibfnamefont {R}~\bibnamefont
  {Barends}}, \bibinfo {author} {\bibfnamefont {J}~\bibnamefont {Kelly}},
  \bibinfo {author} {\bibfnamefont {A}~\bibnamefont {Megrant}}, \bibinfo
  {author} {\bibfnamefont {A}~\bibnamefont {Veitia}}, \bibinfo {author}
  {\bibfnamefont {D}~\bibnamefont {Sank}}, \bibinfo {author} {\bibfnamefont
  {E}~\bibnamefont {Jeffrey}}, \bibinfo {author} {\bibfnamefont
  {TC}~\bibnamefont {White}}, \bibinfo {author} {\bibfnamefont {J}~\bibnamefont
  {Mutus}}, \bibinfo {author} {\bibfnamefont {AG}~\bibnamefont {Fowler}},
  \bibinfo {author} {\bibfnamefont {B}~\bibnamefont {Campbell}},  \emph
  {et~al.},\ }\bibfield  {title} {\enquote {\bibinfo {title} {Superconducting
  quantum circuits at the surface code threshold for fault tolerance},}\
  }\href@noop {} {\bibfield  {journal} {\bibinfo  {journal} {Nature}\ }\textbf
  {\bibinfo {volume} {508}},\ \bibinfo {pages} {500--503} (\bibinfo {year}
  {2014})}\BibitemShut {NoStop}%
\bibitem [{\citenamefont {Paetznick}\ and\ \citenamefont
  {Reichardt}(2013)}]{PR13}%
  \BibitemOpen
  \bibfield  {author} {\bibinfo {author} {\bibfnamefont {Adam}\ \bibnamefont
  {Paetznick}}\ and\ \bibinfo {author} {\bibfnamefont {Ben~W}\ \bibnamefont
  {Reichardt}},\ }\bibfield  {title} {\enquote {\bibinfo {title} {Universal
  fault-tolerant quantum computation with only transversal gates and error
  correction},}\ }\href@noop {} {\bibfield  {journal} {\bibinfo  {journal}
  {Physical review letters}\ }\textbf {\bibinfo {volume} {111}},\ \bibinfo
  {pages} {090505} (\bibinfo {year} {2013})}\BibitemShut {NoStop}%
\bibitem [{\citenamefont {Bombin}\ and\ \citenamefont
  {Martin-Delgado}(2009)}]{bombin2009quantum}%
  \BibitemOpen
  \bibfield  {author} {\bibinfo {author} {\bibfnamefont {H}~\bibnamefont
  {Bombin}}\ and\ \bibinfo {author} {\bibfnamefont {MA}~\bibnamefont
  {Martin-Delgado}},\ }\bibfield  {title} {\enquote {\bibinfo {title} {Quantum
  measurements and gates by code deformation},}\ }\href@noop {} {\bibfield
  {journal} {\bibinfo  {journal} {Journal of Physics A: Mathematical and
  Theoretical}\ }\textbf {\bibinfo {volume} {42}},\ \bibinfo {pages} {095302}
  (\bibinfo {year} {2009})}\BibitemShut {NoStop}%
\bibitem [{\citenamefont {Bravyi}\ and\ \citenamefont
  {Cross}(2015)}]{bravyi2015doubled}%
  \BibitemOpen
  \bibfield  {author} {\bibinfo {author} {\bibfnamefont {Sergey}\ \bibnamefont
  {Bravyi}}\ and\ \bibinfo {author} {\bibfnamefont {Andrew}\ \bibnamefont
  {Cross}},\ }\bibfield  {title} {\enquote {\bibinfo {title} {Doubled color
  codes},}\ }\href@noop {} {\bibfield  {journal} {\bibinfo  {journal} {arXiv
  preprint arXiv:1509.03239}\ } (\bibinfo {year} {2015})}\BibitemShut {NoStop}%
\bibitem [{\citenamefont {Jones}(2013)}]{jones2013multilevel}%
  \BibitemOpen
  \bibfield  {author} {\bibinfo {author} {\bibfnamefont {Cody}\ \bibnamefont
  {Jones}},\ }\bibfield  {title} {\enquote {\bibinfo {title} {Multilevel
  distillation of magic states for quantum computing},}\ }\href@noop {}
  {\bibfield  {journal} {\bibinfo  {journal} {Physical Review A}\ }\textbf
  {\bibinfo {volume} {87}},\ \bibinfo {pages} {042305} (\bibinfo {year}
  {2013})}\BibitemShut {NoStop}%
\bibitem [{\citenamefont {Wiseman}\ and\ \citenamefont
  {Killip}(1997)}]{wiseman1997adaptive}%
  \BibitemOpen
  \bibfield  {author} {\bibinfo {author} {\bibfnamefont {HM}~\bibnamefont
  {Wiseman}}\ and\ \bibinfo {author} {\bibfnamefont {RB}~\bibnamefont
  {Killip}},\ }\bibfield  {title} {\enquote {\bibinfo {title} {Adaptive
  single-shot phase measurements: A semiclassical approach},}\ }\href@noop {}
  {\bibfield  {journal} {\bibinfo  {journal} {Physical Review A}\ }\textbf
  {\bibinfo {volume} {56}},\ \bibinfo {pages} {944} (\bibinfo {year}
  {1997})}\BibitemShut {NoStop}%
\bibitem [{\citenamefont {Van~Dam}\ \emph {et~al.}(2007)\citenamefont
  {Van~Dam}, \citenamefont {D'Ariano}, \citenamefont {Ekert}, \citenamefont
  {Macchiavello},\ and\ \citenamefont {Mosca}}]{van2007optimal}%
  \BibitemOpen
  \bibfield  {author} {\bibinfo {author} {\bibfnamefont {Wim}\ \bibnamefont
  {Van~Dam}}, \bibinfo {author} {\bibfnamefont {G~Mauro}\ \bibnamefont
  {D'Ariano}}, \bibinfo {author} {\bibfnamefont {Artur}\ \bibnamefont {Ekert}},
  \bibinfo {author} {\bibfnamefont {Chiara}\ \bibnamefont {Macchiavello}}, \
  and\ \bibinfo {author} {\bibfnamefont {Michele}\ \bibnamefont {Mosca}},\
  }\bibfield  {title} {\enquote {\bibinfo {title} {Optimal phase estimation in
  quantum networks},}\ }\href@noop {} {\bibfield  {journal} {\bibinfo
  {journal} {Journal of Physics A: Mathematical and Theoretical}\ }\textbf
  {\bibinfo {volume} {40}},\ \bibinfo {pages} {7971} (\bibinfo {year}
  {2007})}\BibitemShut {NoStop}%
\bibitem [{\citenamefont {Berry}\ \emph {et~al.}(2009)\citenamefont {Berry},
  \citenamefont {Higgins}, \citenamefont {Bartlett}, \citenamefont {Mitchell},
  \citenamefont {Pryde},\ and\ \citenamefont {Wiseman}}]{berry2009perform}%
  \BibitemOpen
  \bibfield  {author} {\bibinfo {author} {\bibfnamefont {Dominic~W}\
  \bibnamefont {Berry}}, \bibinfo {author} {\bibfnamefont {Brendon~L}\
  \bibnamefont {Higgins}}, \bibinfo {author} {\bibfnamefont {Stephen~D}\
  \bibnamefont {Bartlett}}, \bibinfo {author} {\bibfnamefont {Morgan~W}\
  \bibnamefont {Mitchell}}, \bibinfo {author} {\bibfnamefont {Geoff~J}\
  \bibnamefont {Pryde}}, \ and\ \bibinfo {author} {\bibfnamefont {Howard~M}\
  \bibnamefont {Wiseman}},\ }\bibfield  {title} {\enquote {\bibinfo {title}
  {How to perform the most accurate possible phase measurements},}\ }\href@noop
  {} {\bibfield  {journal} {\bibinfo  {journal} {Physical Review A}\ }\textbf
  {\bibinfo {volume} {80}},\ \bibinfo {pages} {052114} (\bibinfo {year}
  {2009})}\BibitemShut {NoStop}%
\bibitem [{\citenamefont {Wiebe}\ and\ \citenamefont
  {Granade}(2015)}]{wiebe2015efficient}%
  \BibitemOpen
  \bibfield  {author} {\bibinfo {author} {\bibfnamefont {Nathan}\ \bibnamefont
  {Wiebe}}\ and\ \bibinfo {author} {\bibfnamefont {Christopher~E}\ \bibnamefont
  {Granade}},\ }\bibfield  {title} {\enquote {\bibinfo {title} {Efficient
  bayesian phase estimation},}\ }\href@noop {} {\bibfield  {journal} {\bibinfo
  {journal} {arXiv preprint arXiv:1508.00869}\ } (\bibinfo {year}
  {2015})}\BibitemShut {NoStop}%
\bibitem [{\citenamefont {Wiebe}\ \emph {et~al.}(2014)\citenamefont {Wiebe},
  \citenamefont {Granade}, \citenamefont {Ferrie},\ and\ \citenamefont
  {Cory}}]{wiebe2014hamiltonian}%
  \BibitemOpen
  \bibfield  {author} {\bibinfo {author} {\bibfnamefont {Nathan}\ \bibnamefont
  {Wiebe}}, \bibinfo {author} {\bibfnamefont {Christopher}\ \bibnamefont
  {Granade}}, \bibinfo {author} {\bibfnamefont {Christopher}\ \bibnamefont
  {Ferrie}}, \ and\ \bibinfo {author} {\bibfnamefont {DG}~\bibnamefont
  {Cory}},\ }\bibfield  {title} {\enquote {\bibinfo {title} {Hamiltonian
  learning and certification using quantum resources},}\ }\href@noop {}
  {\bibfield  {journal} {\bibinfo  {journal} {Physical review letters}\
  }\textbf {\bibinfo {volume} {112}},\ \bibinfo {pages} {190501} (\bibinfo
  {year} {2014})}\BibitemShut {NoStop}%
\bibitem [{\citenamefont {Hentschel}\ and\ \citenamefont
  {Sanders}(2010)}]{hentschel2010machine}%
  \BibitemOpen
  \bibfield  {author} {\bibinfo {author} {\bibfnamefont {Alexander}\
  \bibnamefont {Hentschel}}\ and\ \bibinfo {author} {\bibfnamefont {Barry~C}\
  \bibnamefont {Sanders}},\ }\bibfield  {title} {\enquote {\bibinfo {title}
  {Machine learning for precise quantum measurement},}\ }\href@noop {}
  {\bibfield  {journal} {\bibinfo  {journal} {Physical review letters}\
  }\textbf {\bibinfo {volume} {104}},\ \bibinfo {pages} {063603} (\bibinfo
  {year} {2010})}\BibitemShut {NoStop}%
\bibitem [{\citenamefont {Granade}\ \emph {et~al.}(2012)\citenamefont
  {Granade}, \citenamefont {Ferrie}, \citenamefont {Wiebe},\ and\ \citenamefont
  {Cory}}]{granade2012robust}%
  \BibitemOpen
  \bibfield  {author} {\bibinfo {author} {\bibfnamefont {Christopher~E}\
  \bibnamefont {Granade}}, \bibinfo {author} {\bibfnamefont {Christopher}\
  \bibnamefont {Ferrie}}, \bibinfo {author} {\bibfnamefont {Nathan}\
  \bibnamefont {Wiebe}}, \ and\ \bibinfo {author} {\bibfnamefont {David~G}\
  \bibnamefont {Cory}},\ }\bibfield  {title} {\enquote {\bibinfo {title}
  {Robust online hamiltonian learning},}\ }\href@noop {} {\bibfield  {journal}
  {\bibinfo  {journal} {New Journal of Physics}\ }\textbf {\bibinfo {volume}
  {14}},\ \bibinfo {pages} {103013} (\bibinfo {year} {2012})}\BibitemShut
  {NoStop}%
\bibitem [{\citenamefont {Bonato}\ \emph {et~al.}(2015)\citenamefont {Bonato},
  \citenamefont {Blok}, \citenamefont {Dinani}, \citenamefont {Berry},
  \citenamefont {Markham}, \citenamefont {Twitchen},\ and\ \citenamefont
  {Hanson}}]{bonato2015optimized}%
  \BibitemOpen
  \bibfield  {author} {\bibinfo {author} {\bibfnamefont {Cristian}\
  \bibnamefont {Bonato}}, \bibinfo {author} {\bibfnamefont {Machiel~S}\
  \bibnamefont {Blok}}, \bibinfo {author} {\bibfnamefont {Hossein~T}\
  \bibnamefont {Dinani}}, \bibinfo {author} {\bibfnamefont {Dominic~W}\
  \bibnamefont {Berry}}, \bibinfo {author} {\bibfnamefont {Matthew~L}\
  \bibnamefont {Markham}}, \bibinfo {author} {\bibfnamefont {Daniel~J}\
  \bibnamefont {Twitchen}}, \ and\ \bibinfo {author} {\bibfnamefont {Ronald}\
  \bibnamefont {Hanson}},\ }\bibfield  {title} {\enquote {\bibinfo {title}
  {Optimized quantum sensing with a single electron spin using real-time
  adaptive measurements},}\ }\href@noop {} {\bibfield  {journal} {\bibinfo
  {journal} {Nature nanotechnology}\ } (\bibinfo {year} {2015})}\BibitemShut
  {NoStop}%
\bibitem [{\citenamefont {Wecker}\ \emph
  {et~al.}(2015{\natexlab{b}})\citenamefont {Wecker}, \citenamefont {Hastings},
  \citenamefont {Wiebe}, \citenamefont {Clark}, \citenamefont {Nayak},\ and\
  \citenamefont {Troyer}}]{WHW+15}%
  \BibitemOpen
  \bibfield  {author} {\bibinfo {author} {\bibfnamefont {Dave}\ \bibnamefont
  {Wecker}}, \bibinfo {author} {\bibfnamefont {Matthew~B}\ \bibnamefont
  {Hastings}}, \bibinfo {author} {\bibfnamefont {Nathan}\ \bibnamefont
  {Wiebe}}, \bibinfo {author} {\bibfnamefont {Bryan~K}\ \bibnamefont {Clark}},
  \bibinfo {author} {\bibfnamefont {Chetan}\ \bibnamefont {Nayak}}, \ and\
  \bibinfo {author} {\bibfnamefont {Matthias}\ \bibnamefont {Troyer}},\
  }\bibfield  {title} {\enquote {\bibinfo {title} {Solving strongly correlated
  electron models on a quantum computer},}\ }\href@noop {} {\bibfield
  {journal} {\bibinfo  {journal} {Physical Review A}\ }\textbf {\bibinfo
  {volume} {92}},\ \bibinfo {pages} {062318} (\bibinfo {year}
  {2015}{\natexlab{b}})}\BibitemShut {NoStop}%
\bibitem [{\citenamefont {Raeisi}\ \emph {et~al.}(2012)\citenamefont {Raeisi},
  \citenamefont {Wiebe},\ and\ \citenamefont {Sanders}}]{raeisi2012quantum}%
  \BibitemOpen
  \bibfield  {author} {\bibinfo {author} {\bibfnamefont {Sadegh}\ \bibnamefont
  {Raeisi}}, \bibinfo {author} {\bibfnamefont {Nathan}\ \bibnamefont {Wiebe}},
  \ and\ \bibinfo {author} {\bibfnamefont {Barry~C}\ \bibnamefont {Sanders}},\
  }\bibfield  {title} {\enquote {\bibinfo {title} {Quantum-circuit design for
  efficient simulations of many-body quantum dynamics},}\ }\href@noop {}
  {\bibfield  {journal} {\bibinfo  {journal} {New Journal of Physics}\ }\textbf
  {\bibinfo {volume} {14}},\ \bibinfo {pages} {103017} (\bibinfo {year}
  {2012})}\BibitemShut {NoStop}%
\bibitem [{\citenamefont {Garcíaa}\ \emph {et~al.}(1995)\citenamefont
  {Garcíaa}, \citenamefont {Castell}, \citenamefont {Caballol},\ and\
  \citenamefont {Malrieu}}]{malrieu}%
  \BibitemOpen
  \bibfield  {author} {\bibinfo {author} {\bibfnamefont {V.~M.}\ \bibnamefont
  {Garcíaa}}, \bibinfo {author} {\bibfnamefont {O.}~\bibnamefont {Castell}},
  \bibinfo {author} {\bibfnamefont {R.}~\bibnamefont {Caballol}}, \ and\
  \bibinfo {author} {\bibfnamefont {J.~P.}\ \bibnamefont {Malrieu}},\
  }\bibfield  {title} {\enquote {\bibinfo {title} {An iterative
  difference-dedicated configuration interaction. proposal and test studies},}\
  }\href@noop {} {\bibfield  {journal} {\bibinfo  {journal} {Chem. Phys.
  Lett.}\ }\textbf {\bibinfo {volume} {238}},\ \bibinfo {pages} {222--229}
  (\bibinfo {year} {1995})}\BibitemShut {NoStop}%
\bibitem [{\citenamefont {Legeza}\ and\ \citenamefont
  {S{\'o}lyom}()}]{legeza2004}%
  \BibitemOpen
  \bibfield  {author} {\bibinfo {author} {\bibfnamefont {{\"O}.}~\bibnamefont
  {Legeza}}\ and\ \bibinfo {author} {\bibfnamefont {J.}~\bibnamefont
  {S{\'o}lyom}},\ }\href
  {http://www.itp.uni-hannover.de/~jeckelm/dmrg/workshop/proceedings.html} {\
  \textbf {\bibinfo {volume} {International Workshop on Recent Progress and
  Prospects in Density-Matrix Renormalization}}},\ \bibinfo {note} {lorentz
  Center, Leiden University, The Netherlands, \textbf{2004}}\BibitemShut
  {NoStop}%
\bibitem [{\citenamefont {Legeza}()}]{legeza2010}%
  \BibitemOpen
  \bibfield  {author} {\bibinfo {author} {\bibfnamefont {{\"O}.}~\bibnamefont
  {Legeza}},\ }\href@noop {} {\ \textbf {\bibinfo {volume} {CECAM workshop for
  tensor network methods for quantum chemistry}}},\ \bibinfo {note} {eTH
  Z{\"u}rich, \textbf{2010}}\BibitemShut {NoStop}%
\bibitem [{\citenamefont {Keller}\ and\ \citenamefont
  {Reiher}(2014)}]{keller2014determining}%
  \BibitemOpen
  \bibfield  {author} {\bibinfo {author} {\bibfnamefont {Sebastian~F}\
  \bibnamefont {Keller}}\ and\ \bibinfo {author} {\bibfnamefont {Markus}\
  \bibnamefont {Reiher}},\ }\bibfield  {title} {\enquote {\bibinfo {title}
  {Determining factors for the accuracy of dmrg in chemistry},}\ }\href@noop {}
  {\bibfield  {journal} {\bibinfo  {journal} {CHIMIA International Journal for
  Chemistry}\ }\textbf {\bibinfo {volume} {68}},\ \bibinfo {pages} {200--203}
  (\bibinfo {year} {2014})}\BibitemShut {NoStop}%
\bibitem [{\citenamefont {Wu}\ \emph {et~al.}(2002)\citenamefont {Wu},
  \citenamefont {Byrd},\ and\ \citenamefont {Lidar}}]{wu2002polynomial}%
  \BibitemOpen
  \bibfield  {author} {\bibinfo {author} {\bibfnamefont {L-A}\ \bibnamefont
  {Wu}}, \bibinfo {author} {\bibfnamefont {MS}~\bibnamefont {Byrd}}, \ and\
  \bibinfo {author} {\bibfnamefont {DA}~\bibnamefont {Lidar}},\ }\bibfield
  {title} {\enquote {\bibinfo {title} {Polynomial-time simulation of pairing
  models on a quantum computer},}\ }\href@noop {} {\bibfield  {journal}
  {\bibinfo  {journal} {Physical Review Letters}\ }\textbf {\bibinfo {volume}
  {89}},\ \bibinfo {pages} {057904} (\bibinfo {year} {2002})}\BibitemShut
  {NoStop}%
\bibitem [{\citenamefont {Cheung}\ \emph {et~al.}(2011)\citenamefont {Cheung},
  \citenamefont {H{\o}yer},\ and\ \citenamefont {Wiebe}}]{cheung2011improved}%
  \BibitemOpen
  \bibfield  {author} {\bibinfo {author} {\bibfnamefont {Donny}\ \bibnamefont
  {Cheung}}, \bibinfo {author} {\bibfnamefont {Peter}\ \bibnamefont
  {H{\o}yer}}, \ and\ \bibinfo {author} {\bibfnamefont {Nathan}\ \bibnamefont
  {Wiebe}},\ }\bibfield  {title} {\enquote {\bibinfo {title} {Improved error
  bounds for the adiabatic approximation},}\ }\href@noop {} {\bibfield
  {journal} {\bibinfo  {journal} {Journal of Physics A: Mathematical and
  Theoretical}\ }\textbf {\bibinfo {volume} {44}},\ \bibinfo {pages} {415302}
  (\bibinfo {year} {2011})}\BibitemShut {NoStop}%
\bibitem [{\citenamefont {Wiebe}\ \emph {et~al.}(2010)\citenamefont {Wiebe},
  \citenamefont {Berry}, \citenamefont {H{\o}yer},\ and\ \citenamefont
  {Sanders}}]{WBH+10}%
  \BibitemOpen
  \bibfield  {author} {\bibinfo {author} {\bibfnamefont {Nathan}\ \bibnamefont
  {Wiebe}}, \bibinfo {author} {\bibfnamefont {Dominic}\ \bibnamefont {Berry}},
  \bibinfo {author} {\bibfnamefont {Peter}\ \bibnamefont {H{\o}yer}}, \ and\
  \bibinfo {author} {\bibfnamefont {Barry~C}\ \bibnamefont {Sanders}},\
  }\bibfield  {title} {\enquote {\bibinfo {title} {Higher order decompositions
  of ordered operator exponentials},}\ }\href@noop {} {\bibfield  {journal}
  {\bibinfo  {journal} {Journal of Physics A: Mathematical and Theoretical}\
  }\textbf {\bibinfo {volume} {43}},\ \bibinfo {pages} {065203} (\bibinfo
  {year} {2010})}\BibitemShut {NoStop}%
\bibitem [{\citenamefont {Wiebe}\ \emph {et~al.}(2011)\citenamefont {Wiebe},
  \citenamefont {Berry}, \citenamefont {H{\o}yer},\ and\ \citenamefont
  {Sanders}}]{wiebe2011simulating}%
  \BibitemOpen
  \bibfield  {author} {\bibinfo {author} {\bibfnamefont {Nathan}\ \bibnamefont
  {Wiebe}}, \bibinfo {author} {\bibfnamefont {Dominic~W}\ \bibnamefont
  {Berry}}, \bibinfo {author} {\bibfnamefont {Peter}\ \bibnamefont {H{\o}yer}},
  \ and\ \bibinfo {author} {\bibfnamefont {Barry~C}\ \bibnamefont {Sanders}},\
  }\bibfield  {title} {\enquote {\bibinfo {title} {Simulating quantum dynamics
  on a quantum computer},}\ }\href@noop {} {\bibfield  {journal} {\bibinfo
  {journal} {Journal of Physics A: Mathematical and Theoretical}\ }\textbf
  {\bibinfo {volume} {44}},\ \bibinfo {pages} {445308} (\bibinfo {year}
  {2011})}\BibitemShut {NoStop}%
\bibitem [{\citenamefont {McClean}\ \emph
  {et~al.}(2014{\natexlab{b}})\citenamefont {McClean}, \citenamefont {Babbush},
  \citenamefont {Love},\ and\ \citenamefont
  {Aspuru-Guzik}}]{mcclean2014exploiting}%
  \BibitemOpen
  \bibfield  {author} {\bibinfo {author} {\bibfnamefont {Jarrod~R}\
  \bibnamefont {McClean}}, \bibinfo {author} {\bibfnamefont {Ryan}\
  \bibnamefont {Babbush}}, \bibinfo {author} {\bibfnamefont {Peter~J}\
  \bibnamefont {Love}}, \ and\ \bibinfo {author} {\bibfnamefont {Al{\'a}n}\
  \bibnamefont {Aspuru-Guzik}},\ }\bibfield  {title} {\enquote {\bibinfo
  {title} {Exploiting locality in quantum computation for quantum chemistry},}\
  }\href@noop {} {\bibfield  {journal} {\bibinfo  {journal} {The journal of
  physical chemistry letters}\ }\textbf {\bibinfo {volume} {5}},\ \bibinfo
  {pages} {4368--4380} (\bibinfo {year} {2014}{\natexlab{b}})}\BibitemShut
  {NoStop}%
\bibitem [{\citenamefont {Leigh}(2002)}]{leig02}%
  \BibitemOpen
  \bibinfo {editor} {\bibfnamefont {G.~Jeffery}\ \bibnamefont {Leigh}},\ ed.,\
  \href@noop {} {\emph {\bibinfo {title} {{Nitrogen Fixation at the
  Millenium}}}}\ (\bibinfo  {publisher} {Elsevier Science},\ \bibinfo {address}
  {Amsterdam},\ \bibinfo {year} {2002})\BibitemShut {NoStop}%
\bibitem [{\citenamefont {Kim}\ and\ \citenamefont
  {Rees}(1992)}]{science_1992_257_1677}%
  \BibitemOpen
  \bibfield  {author} {\bibinfo {author} {\bibfnamefont {J.}~\bibnamefont
  {Kim}}\ and\ \bibinfo {author} {\bibfnamefont {D.~C.}\ \bibnamefont {Rees}},\
  }\bibfield  {title} {\enquote {\bibinfo {title} {{Structural Models for the
  Metal Centers in the Nitrogenase Molybdenum-Iron Protein}},}\ }\href@noop {}
  {\bibfield  {journal} {\bibinfo  {journal} {Science}\ }\textbf {\bibinfo
  {volume} {257}},\ \bibinfo {pages} {1677--1682} (\bibinfo {year}
  {1992})}\BibitemShut {NoStop}%
\bibitem [{\citenamefont {Einsle}\ \emph {et~al.}(2002)\citenamefont {Einsle},
  \citenamefont {Tezcan}, \citenamefont {Andrade}, \citenamefont {Schmid},
  \citenamefont {Yoshida}, \citenamefont {Howard},\ and\ \citenamefont
  {Rees}}]{eins02}%
  \BibitemOpen
  \bibfield  {author} {\bibinfo {author} {\bibfnamefont {Oliver}\ \bibnamefont
  {Einsle}}, \bibinfo {author} {\bibfnamefont {F.~Akif}\ \bibnamefont
  {Tezcan}}, \bibinfo {author} {\bibfnamefont {Susana L.~A.}\ \bibnamefont
  {Andrade}}, \bibinfo {author} {\bibfnamefont {Benedikt}\ \bibnamefont
  {Schmid}}, \bibinfo {author} {\bibfnamefont {Mika}\ \bibnamefont {Yoshida}},
  \bibinfo {author} {\bibfnamefont {James~B.}\ \bibnamefont {Howard}}, \ and\
  \bibinfo {author} {\bibfnamefont {Douglas~C.}\ \bibnamefont {Rees}},\
  }\bibfield  {title} {\enquote {\bibinfo {title} {{Nitrogenase MoFe-Protein at
  1.16 \AA Resolution: A Central Ligand in the FeMo-Cofactor}},}\ }\href@noop
  {} {\bibfield  {journal} {\bibinfo  {journal} {Science}\ }\textbf {\bibinfo
  {volume} {297}},\ \bibinfo {pages} {1696--1700} (\bibinfo {year}
  {2002})}\BibitemShut {NoStop}%
\bibitem [{\citenamefont {Ribbe}\ \emph {et~al.}(2014)\citenamefont {Ribbe},
  \citenamefont {Hu}, \citenamefont {Hodgson},\ and\ \citenamefont
  {Hedman}}]{ribbe}%
  \BibitemOpen
  \bibfield  {author} {\bibinfo {author} {\bibfnamefont {M.~W.}\ \bibnamefont
  {Ribbe}}, \bibinfo {author} {\bibfnamefont {Y.}~\bibnamefont {Hu}}, \bibinfo
  {author} {\bibfnamefont {K.~O.}\ \bibnamefont {Hodgson}}, \ and\ \bibinfo
  {author} {\bibfnamefont {B}~\bibnamefont {Hedman}},\ }\bibfield  {title}
  {\enquote {\bibinfo {title} {Biosynthesis of nitrogenase metalloclusters},}\
  }\href@noop {} {\bibfield  {journal} {\bibinfo  {journal} {Chem. Rev.}\
  }\textbf {\bibinfo {volume} {114}},\ \bibinfo {pages} {4063--4080} (\bibinfo
  {year} {2014})}\BibitemShut {NoStop}%
\bibitem [{\citenamefont {Liu}\ \emph {et~al.}(2015)\citenamefont {Liu},
  \citenamefont {Liu},\ and\ \citenamefont {Hall}}]{HallHyd}%
  \BibitemOpen
  \bibfield  {author} {\bibinfo {author} {\bibfnamefont {Caiping}\ \bibnamefont
  {Liu}}, \bibinfo {author} {\bibfnamefont {Tianbiao}\ \bibnamefont {Liu}}, \
  and\ \bibinfo {author} {\bibfnamefont {Michael~B.}\ \bibnamefont {Hall}},\
  }\bibfield  {title} {\enquote {\bibinfo {title} {{Influence of the Density
  Functional and Basis Set on the Relative Stabilities of Oxygenated Isomers of
  Diiron Models for the Active Site of [FeFe]-Hydrogenase}},}\ }\href@noop {}
  {\bibfield  {journal} {\bibinfo  {journal} {J. Chem. Theory Comput.}\
  }\textbf {\bibinfo {volume} {11}},\ \bibinfo {pages} {205--214} (\bibinfo
  {year} {2015})}\BibitemShut {NoStop}%
\bibitem [{\citenamefont {Jacob}\ and\ \citenamefont {Reiher}(2012)}]{jaco12}%
  \BibitemOpen
  \bibfield  {author} {\bibinfo {author} {\bibfnamefont {C.~R.}\ \bibnamefont
  {Jacob}}\ and\ \bibinfo {author} {\bibfnamefont {M.}~\bibnamefont {Reiher}},\
  }\bibfield  {title} {\enquote {\bibinfo {title} {{Spin in density-functional
  theory}},}\ }\href@noop {} {\bibfield  {journal} {\bibinfo  {journal} {Int.
  J. Quantum Chem.}\ }\textbf {\bibinfo {volume} {112}},\ \bibinfo {pages}
  {3661--3684} (\bibinfo {year} {2012})}\BibitemShut {NoStop}%
\bibitem [{\citenamefont {Ahlrichs}\ \emph {et~al.}(1989)\citenamefont
  {Ahlrichs}, \citenamefont {B\"{a}r}, \citenamefont {H\"{a}ser}, \citenamefont
  {Horn},\ and\ \citenamefont {K\"{o}lmel}}]{turbomole}%
  \BibitemOpen
  \bibfield  {author} {\bibinfo {author} {\bibfnamefont {Reinhart}\
  \bibnamefont {Ahlrichs}}, \bibinfo {author} {\bibfnamefont {Michael}\
  \bibnamefont {B\"{a}r}}, \bibinfo {author} {\bibfnamefont {Marco}\
  \bibnamefont {H\"{a}ser}}, \bibinfo {author} {\bibfnamefont {Hans}\
  \bibnamefont {Horn}}, \ and\ \bibinfo {author} {\bibfnamefont {Christoph}\
  \bibnamefont {K\"{o}lmel}},\ }\bibfield  {title} {\enquote {\bibinfo {title}
  {{Electronic Structure Calculations on Workstation Computers: The Program
  System \textsc{Turbomole}}},}\ }\href@noop {} {\bibfield  {journal} {\bibinfo
   {journal} {Chem. Phys. Lett.}\ }\textbf {\bibinfo {volume} {162}},\ \bibinfo
  {pages} {165--169} (\bibinfo {year} {1989})}\BibitemShut {NoStop}%
\bibitem [{\citenamefont {Lee}\ \emph {et~al.}(1988)\citenamefont {Lee},
  \citenamefont {Yang},\ and\ \citenamefont {Parr}}]{lyp}%
  \BibitemOpen
  \bibfield  {author} {\bibinfo {author} {\bibfnamefont {Chengteh}\
  \bibnamefont {Lee}}, \bibinfo {author} {\bibfnamefont {Weitao}\ \bibnamefont
  {Yang}}, \ and\ \bibinfo {author} {\bibfnamefont {Robert~G.}\ \bibnamefont
  {Parr}},\ }\bibfield  {title} {\enquote {\bibinfo {title} {Development of the
  colle-salvetti correlation-energy formula into a functional of the electron
  density},}\ }\href@noop {} {\bibfield  {journal} {\bibinfo  {journal} {Phys.
  Rev. B}\ }\textbf {\bibinfo {volume} {37}},\ \bibinfo {pages} {785--789}
  (\bibinfo {year} {1988})}\BibitemShut {NoStop}%
\bibitem [{\citenamefont {Becke}(1988)}]{becke88}%
  \BibitemOpen
  \bibfield  {author} {\bibinfo {author} {\bibfnamefont {A.~D.}\ \bibnamefont
  {Becke}},\ }\bibfield  {title} {\enquote {\bibinfo {title}
  {{Density-Functional Exchange-Energy Approximation with Correct Asymptotic
  Behavior}},}\ }\href@noop {} {\bibfield  {journal} {\bibinfo  {journal}
  {Phys. Rev. A}\ }\textbf {\bibinfo {volume} {38}},\ \bibinfo {pages}
  {3098--3100} (\bibinfo {year} {1988})}\BibitemShut {NoStop}%
\bibitem [{\citenamefont {Becke}(1993)}]{becke93}%
  \BibitemOpen
  \bibfield  {author} {\bibinfo {author} {\bibfnamefont {Axel~D.}\ \bibnamefont
  {Becke}},\ }\bibfield  {title} {\enquote {\bibinfo {title}
  {Density-functional thermochemistry. iii. the role of exact exchange},}\
  }\href@noop {} {\bibfield  {journal} {\bibinfo  {journal} {J. Chem. Phys.}\
  }\textbf {\bibinfo {volume} {98}},\ \bibinfo {pages} {5648--5652} (\bibinfo
  {year} {1993})}\BibitemShut {NoStop}%
\bibitem [{\citenamefont {Stephens}\ \emph {et~al.}(1994)\citenamefont
  {Stephens}, \citenamefont {Devlin}, \citenamefont {Chabalowski},\ and\
  \citenamefont {Frisch}}]{stephens}%
  \BibitemOpen
  \bibfield  {author} {\bibinfo {author} {\bibfnamefont {P.~J.}\ \bibnamefont
  {Stephens}}, \bibinfo {author} {\bibfnamefont {F.~J.}\ \bibnamefont
  {Devlin}}, \bibinfo {author} {\bibfnamefont {C.~F.}\ \bibnamefont
  {Chabalowski}}, \ and\ \bibinfo {author} {\bibfnamefont {M.~J.}\ \bibnamefont
  {Frisch}},\ }\bibfield  {title} {\enquote {\bibinfo {title} {Ab initio
  calculation of vibrational absorption and circular dichroism spectra using
  density functional force fields},}\ }\href@noop {} {\bibfield  {journal}
  {\bibinfo  {journal} {J. Phys. Chem.}\ }\textbf {\bibinfo {volume} {98}},\
  \bibinfo {pages} {11623--11627} (\bibinfo {year} {1994})}\BibitemShut
  {NoStop}%
\bibitem [{\citenamefont {Weigend}\ and\ \citenamefont
  {Ahlrichs}(2005)}]{phys_chem_chem_phys_2005_7_3297}%
  \BibitemOpen
  \bibfield  {author} {\bibinfo {author} {\bibfnamefont {F.}~\bibnamefont
  {Weigend}}\ and\ \bibinfo {author} {\bibfnamefont {R.}~\bibnamefont
  {Ahlrichs}},\ }\bibfield  {title} {\enquote {\bibinfo {title} {{Balanced
  Basis Sets of Split Valence, Triple Zeta Valence and Quadruple Zeta Valence
  Quality for H to Rn: Design and Assessment of Accuracy}},}\ }\href@noop {}
  {\bibfield  {journal} {\bibinfo  {journal} {Phys. Chem. Chem. Phys.}\
  }\textbf {\bibinfo {volume} {7}},\ \bibinfo {pages} {3297--3305} (\bibinfo
  {year} {2005})}\BibitemShut {NoStop}%
\bibitem [{\citenamefont {Andrae}\ \emph {et~al.}(1990)\citenamefont {Andrae},
  \citenamefont {H\"{a}u\ss{}ermann}, \citenamefont {Dolg}, \citenamefont
  {Stoll},\ and\ \citenamefont {Preu\ss{}}}]{theor_chim_acta_1990_77_123}%
  \BibitemOpen
  \bibfield  {author} {\bibinfo {author} {\bibfnamefont {D.}~\bibnamefont
  {Andrae}}, \bibinfo {author} {\bibfnamefont {U.}~\bibnamefont
  {H\"{a}u\ss{}ermann}}, \bibinfo {author} {\bibfnamefont {M.}~\bibnamefont
  {Dolg}}, \bibinfo {author} {\bibfnamefont {H.}~\bibnamefont {Stoll}}, \ and\
  \bibinfo {author} {\bibfnamefont {H.}~\bibnamefont {Preu\ss{}}},\ }\bibfield
  {title} {\enquote {\bibinfo {title} {{Energy-adjusted {\it ab initio}
  Pseudopotentials for the Second and Third Row Transition Elements}},}\
  }\href@noop {} {\bibfield  {journal} {\bibinfo  {journal} {Theor. Chim.
  Acta}\ }\textbf {\bibinfo {volume} {77}},\ \bibinfo {pages} {123--141}
  (\bibinfo {year} {1990})}\BibitemShut {NoStop}%
\bibitem [{\citenamefont {Keller}\ \emph {et~al.}(2015)\citenamefont {Keller},
  \citenamefont {Boguslawski}, \citenamefont {Janowski}, \citenamefont
  {Reiher},\ and\ \citenamefont {Pulay}}]{bogus15}%
  \BibitemOpen
  \bibfield  {author} {\bibinfo {author} {\bibfnamefont {Sebastian}\
  \bibnamefont {Keller}}, \bibinfo {author} {\bibfnamefont {Katharina}\
  \bibnamefont {Boguslawski}}, \bibinfo {author} {\bibfnamefont {Tomasz}\
  \bibnamefont {Janowski}}, \bibinfo {author} {\bibfnamefont {Markus}\
  \bibnamefont {Reiher}}, \ and\ \bibinfo {author} {\bibfnamefont {Peter}\
  \bibnamefont {Pulay}},\ }\bibfield  {title} {\enquote {\bibinfo {title}
  {Selection of active spaces for multiconfigurational wavefunctions},}\
  }\href@noop {} {\bibfield  {journal} {\bibinfo  {journal} {J. Chem. Phys}\
  }\textbf {\bibinfo {volume} {142}},\ \bibinfo {pages} {244104} (\bibinfo
  {year} {2015})}\BibitemShut {NoStop}%
\end{thebibliography}
\end{document}